\documentstyle[12pt]{article}
\def\ie{\hbox{\it i.e.}{}}      
\def\eg{\hbox{\it e.g.}{}}

\catcode`\@=11

\newcommand{\be}{\begin{equation}}
\newcommand{\ee}{\end{equation}}
\newcommand{\ba}{\begin{eqnarray}}
\newcommand{\ea}{\end{eqnarray}}
\def\NPB#1#2#3{{\it Nucl.~Phys.} {\bf{B#1}} (19#2) #3}
\def\PLB#1#2#3{{\it Phys.~Lett.} {\bf{B#1}} (19#2) #3}
\def\PRD#1#2#3{{\it Phys.~Rev.} {\bf{D#1}} (19#2) #3}
\def\PRL#1#2#3{{\it Phys.~Rev.~Lett.} {\bf{#1}} (19#2) #3}

\def\PR#1#2#3{{\it Phys.~Rep.} {\bf#1} (19#2) #3}

\def\RMP#1#2#3{{\it Rev.~Mod.~Phys.} {\bf#1} (19#2) #3}

\def\JHEP#1#2#3{{\it J. High Energy Phys.} {\bf#1} (19#2) #3}

\def\darr#1{\raise1.5ex\hbox{$\leftrightarrow$}\mkern-16.5mu #1}
\def\part{\partial}

\def\a{\alpha}
\def\b{\beta}
\def\g{\gamma}

\def\d{\delta}
\def\e{\epsilon}

\def\m\mu 
\def\n{\nu}

\def\@normalsize{\@setsize\normalsize{15pt}\xiipt\@xiipt
\abovedisplayskip 14pt plus3pt minus3pt%
\belowdisplayskip \abovedisplayskip
\abovedisplayshortskip  \z@ plus3pt%
\belowdisplayshortskip  7pt plus3.5pt minus0pt}
\def\small{\@setsize\small{13.6pt}\xipt\@xipt
\abovedisplayskip 13pt plus3pt minus3pt%
\belowdisplayskip \abovedisplayskip
\abovedisplayshortskip  \z@ plus3pt%
\belowdisplayshortskip  7pt plus3.5pt minus0pt
\def\@listi{\parsep 4.5pt plus 2pt minus 1pt
            \itemsep \parsep
            \topsep 9pt plus 3pt minus 3pt}}

\def\underline#1{\relax\ifmmode\@@underline#1\else
        $\@@underline{\hbox{#1}}$\relax\fi}
\@twosidetrue
\relax

\catcode`@=12

\catcode`\@=11

\def\section{\@startsection{section}{1}{\z@}{3.5ex plus 1ex minus
   .2ex}{2.3ex plus .2ex}{\large\bf}}

\def\thesubsection{\Roman{section}-\arabic{subsection}}

\def\ps@headings{\def\@oddfoot{}\def\@evenfoot{}
\def\@oddhead{\hbox{}\hfill
        \makebox[.5\textwidth]{\raggedright\ignorespaces --\thepage{}--
        \hfill }}
\def\@evenhead{\@oddhead}
\def\subsectionmark##1{\markboth{##1}{}} }
\renewcommand{\subsection}[1]{\addtocounter{subsection}{1}
\vspace{2.5mm}\par\noindent {\em \thesubsection . #1}\par
 \vspace{0.5mm} }

\ps@headings

\catcode`\@=12

\relax

\def\figcap{\section*{Figure Captions\markboth
        {FIGURECAPTIONS}{FIGURECAPTIONS}}\list
        {Fig. \arabic{enumi}:\hfill}{\settowidth\labelwidth{Fig. 999:}
        \leftmargin\labelwidth
        \advance\leftmargin\labelsep\usecounter{enumi}}}
 \relax
\def\tablecap{\section*{Table Captions\markboth
        {TABLECAPTIONS}{TABLECAPTIONS}}\list
        {Table \arabic{enumi}:\hfill}{\settowidth\labelwidth{Table
999:}
        \leftmargin\labelwidth
        \advance\leftmargin\labelsep\usecounter{enumi}}}
 \relax
\def\reflist{\section*{References\markboth
        {REFLIST}{REFLIST}}\list
        {[\arabic{enumi}]\hfill}{\settowidth\labelwidth{[999]}
        \leftmargin\labelwidth
        \advance\leftmargin\labelsep\usecounter{enumi}}}
 \relax

\catcode`\@=11

\def\marginnote#1{}
\newcount\hour
\newcount\minute
\newtoks\amorpm
\hour=\time\divide\hour by60
\minute=\time{\multiply\hour by60 \global\advance\minute by-
\hour}
\edef\standardtime{{\ifnum\hour<12 \global\amorpm={am}%
    \else\global\amorpm={pm}\advance\hour by-12 \fi
    \ifnum\hour=0 \hour=12 \fi
    \number\hour:\ifnum\minute<100\fi\number\minute\the\amorpm}}
\edef\militarytime{\number\hour:\ifnum\minute<100\fi\number\minute}
\def\draftlabel#1{{\@bsphack\if@filesw {\let\thepage\relax
  \xdef\@gtempa{\write\@auxout{\string
    \newlabel{#1}{{\@currentlabel}{\thepage}}}}}\@gtempa
    \if@nobreak \ifvmode\nobreak\fi\fi\fi\@esphack}
     \gdef\@eqnlabel{#1}}
\def\@eqnlabel{}
\def\@vacuum{}
\def\draftmarginnote#1{\marginpar{\raggedright\scriptsize\tt#1}}
\def\draft{\oddsidemargin -.5truein
        \def\@oddfoot{\sl preliminary draft \hfil
        \rm\thepage\hfil\sl\today\quad\militarytime}
        \let\@evenfoot\@oddfoot \overfullrule 3pt
        \let\label=\draftlabel
        \let\marginnote=\draftmarginnote
   
\def\@eqnnum{(\theequation)\rlap{\kern\marginparsep\tt\@eqnlabel}%
\global\let\@eqnlabel\@vacuum}  }
\def\preprint{\twocolumn\sloppy\flushbottom\parindent 1em
        \leftmargini 2em\leftmarginv .5em\leftmarginvi .5em
        \oddsidemargin -.5in    \evensidemargin -.5in
        \columnsep 15mm \footheight 0pt
        \textwidth 250mmin      \topmargin  -.4in
        \headheight 12pt \topskip .4in
        \textheight 175mm
        \footskip 0pt
        
\def\@oddhead{\thepage\hfil\addtocounter{page}{1}\thepage}
        \let\@evenhead\@oddhead \def\@oddfoot{} \def\@evenfoot{}  }
\def\titlepage{\@restonecolfalse\if@twocolumn\@restonecoltrue\onecolumn
     \else \newpage \fi \thispagestyle{empty}\c@page\z@
        \def\thefootnote{\fnsymbol{footnote}} }
\def\endtitlepage{\if@restonecol\twocolumn \else  \fi
        \def\thefootnote{\arabic{footnote}}
        \setcounter{footnote}{0}}  
\catcode`@=12
\relax

\def\ps@headings{\def\@oddfoot{}\def\@evenfoot{}
\def\@oddhead{\hbox{}\hfill
        \makebox[.5\textwidth]{\raggedright\ignorespaces --\thepage{}--
        \hfill }}
\def\@evenhead{\@oddhead}
\def\subsectionmark##1{\markboth{##1}{}} }

\ps@headings

\relax

\long\def\@caption#1[#2]#3{\par\addcontentsline{\csname
  ext@#1\endcsname}{#1}{\protect\numberline{\csname
  the#1\endcsname}{\ignorespaces #2}}\begingroup
    \small
    \@parboxrestore
    \@makecaption{\csname fnum@#1\endcsname}{\ignorespaces #3}\par
  \endgroup}
\catcode`@=12

\input epsf

\def\firstpage#1#2#3#4#5#6{
\begin{document}


\begin{titlepage}
\nopagebreak
\title{\begin{flushright}
        \vspace*{-1.8in}
        {\normalsize LPT-ORSAY 00/52}\\[-10mm]
        {\normalsize hep-ph/0006190}\\[-4mm]
\end{flushright}
\vfill {#3}}
\author{\large #4 \\[1.0cm] #5}
\maketitle
\vskip -9mm     
\nopagebreak 
\begin{abstract} {\noindent #6}
\end{abstract}
\vfill
\begin{flushleft}
\rule{16.1cm}{0.2mm}\\[-4mm]
$^{\dagger}${\small This review is based on the Th{\`e}se
d'Habilitation of the author.}\\
$^{\ddagger}${\small Unit{\'e} mixte de recherche du CNRS (UMR 8627).}\\
\today
\end{flushleft}
\thispagestyle{empty}
\end{titlepage}}

\date{}
\firstpage{3118}{IC/95/34} {\huge\bf Theory and Phenomenology 
of \\ Type I strings and M-theory$^\dagger$}  
{Emilian Dudas$^{\,a}$} 
{\small\sl $^a$  Laboratoire de Physique Th\'eorique$^\ddagger$,
B{\^a}t. 210, Univ. Paris-Sud, F-91405 Orsay, FRANCE}
{ The physical motivations and the basic construction rules for Type I
strings and M-theory compactifications are reviewed in light of the
recent developments. The first part contains the basic
theoretical ingredients needed for building four-dimensional
supersymmetric models, models with broken supersymmetry and for 
computing low-energy actions and
quantum corrections to them. The second part contains some
phenomenological applications to brane world scenarios with low values
of the string scale and large extra dimensions.}
\setcounter{page}{0}
{\huge {\bf TABLE OF CONTENTS}}

{\bf 1. Introduction}

{\large \bf Theoretical aspects} 

{\bf 2. From heterotic strings to Type I strings and M-theory}

{\bf 3. Building blocks for Type I strings}

{\bf 4. M-theory}

{\bf 5. Type I compactifications to four-dimensions}

{\bf 6. Effective Lagrangian and quantum corrections in Type I strings}

{\bf 7. String mechanisms for breaking supersymmetry, brane-antibrane systems}

\vskip .5cm
{\large \bf Phenomenology of low-scale strings}

{\bf 8. Large extra dimensions, mm and TeV dimensions}

{\bf 9. Gauge coupling unification}

{\bf 10. Supersymmetry breaking}

{\bf 11. Bulk physics: Neutrino and axion masses with large extra dimensions}

{\bf 12. Low-scale string predictions for accelerators}

{\bf 13. Conclusions}

{\bf 14. Appendix}

\vfill
\eject 
\section{Introduction}

Since the discovery of the anomaly cancellation for superstrings in ten
dimensions (10d) \cite{green}, the construction of the heterotic strings
\cite{ghmr} and the seminal
papers on the compactification to four dimensions \cite{chsw},
\cite{dhvw}, string theory has become the best candidate for a fundamental
quantum theory of all interactions including Einstein gravity. The 
theory contains only one free dimensionful parameter, the string scale $M_s$, 
while the four-dimensional (4d) gauge group and the matter content are manifestations
of the geometric properties of the internal space that, however, we are
unable to select in a unique fashion. The Standard
Model hopefully would correspond to a particular internal space or vacuum 
configuration chosen by nature by some still unknown mechanism. 
The 4d low-energy couplings
depend only on the string scale and on various vacuum expectation values (vev's)
of fields describing the string coupling constant, the size and the shape of the
internal manifold. There is therefore, in principle, the hope to
understand the empirically observed pattern of the parameters in the
Standard Model. A long activity in heterotic strings \cite{abk},
\cite{ibanez} was devoted to this program \cite{pheno},
in the hope that these rather tight constraints would determine in some way the
right vacuum describing our world. Despite serious insights into the
structure and the phenomeneological properties of 4d models
\cite{abk, ibanez}, no unique candidate having as low-energy limit
the Standard Model emerged. Moreover, there were (and there still are)
conceptual problems to be solved, as for example the large degeneracy of
the string vacua and the related
problems of spacetime supersymmetry breaking and dilaton stabilization.   
Most of these problems asked for a better understanding of the strong
coupling limit of string theory, of which very little was known for a
long time. The other string theories were, for a long time, discarded as
inappropriate for phenomenological purposes. Type II strings were considered unable
to produce a realistic gauge group, while Type I strings, despite
serious advances made over the years \cite{sagnotti1,ps,bs,io}, that
revealed striking differences with respect to heterotic strings, were less studied
and their consistency rules not widely known as for heterotic vacua.
   
By the middle of the last decade, 
it became clear that all known string theories are actually related by
various dualities to each other and to a mysterious eleven
dimensional theory, provisionally called M-theory \cite{M-theory}. It
therefore became  possible to obtain some
nonperturbative string results, at least for theories with enough supersymmetry.
Moreover, the discovery and the study of D-branes \cite{polchinski} put the
duality predictions on a firmer quantitative basis and, on the other
hand, was an important step in unravelling the geometric structure
underlying the consistency conditions of Type I vacua, stimulating a new
activity in this field \cite{gp}.   
The first chiral 4d Type I model was proposed \cite{abpss}, 
and efforts on 4d model building allowed a
better understanding of supersymmetric 4d vacua \cite{bl} and of their
gauge and gravitational anomaly cancellation mechanisms \cite{iru},
similar to the 6d generalized Green-Schwarz mechanism discovered by
Sagnotti \cite{sagnotti2}. 
The presence of D-branes in Type I models led to new mechanisms for
breaking supersymmetry by compactification \cite{ads1, adds}, by internal
magnetic fields \cite{ft,bachas1} or directly on
some (anti)branes \cite{ads2,au,aadds}, providing perturbatively stable 
non-BPS analogs of Type IIB configurations \cite{sen}. 
 
On the phenomenological side, the M-theory compactification of Horava
and Witten \cite{hw}, with a fundamental scale $M_{11} \sim 2 \times 10^{16}$ GeV, 
provided a framework \cite{witten1} for the perturbative MSSM unification of gauge
couplings \cite{unif}, and stimulated studies of gaugino condensation 
\cite{horava}, of 4d compactifications  \cite{bd,dg,ovrut} and of supersymmetry
breaking along the new (eleventh) dimension \cite{aq,dg}. Moreover, it
was noticed \cite{lykken} that in Type I strings the string
scale can be lowered all the way down to the TeV range. Similar ideas appeared 
for lowering the fundamental Planck scale in theories with
(sub)millimeter gravitational dimensions \cite{add}, as an alternative
solution to the gauge hierarchy problem, and, simultaneously, a new
way for lowering the GUT scale in theories with large (TeV) dimensions
\cite{ddg} was proposed. The new emerging picture found a simple realization in a
perturbative Type I setting \cite{aahdd} with low  
string scale (in the TeV range) and became the subject of an intense
activity, mostly on the phenomenological side, but also on the theoretical side. 

The goal of the present paper is to review the ideas which led to this 
new picture and to present a comprehensive introduction to the basic 
string tools necessary for understanding the 
corresponding physics. The convention for the metric signature throughout the
paper is $(-,+, \cdots +)$, ten-dimensional (10d) indices are denoted by
$A,B, \cdots$, eleven dimensional indices by $I,J, \cdots $,
five-dimensional (5d) indices by $M,N, \cdots$ and four-dimensional
indices (4d) by $\mu,\nu, \cdots$. 
   
\section{From heterotic strings to Type I strings and M-theory}

To date, the (super)strings are the only known consistent quantum theories
including Einstein gravity. They are therefore promising candidates for a unifying
picture of elementary particles and fundamental interactions.

It has been known for a long time that there are five consistent
(anomaly-free) superstring theories in 10d, namely:

- The heterotic closed string theories, with gauge groups
$SO(32)$ and $E_8 \times E_8$ and
${\cal N}=1$ spacetime supersymmetry, that after a toroidal
compactification corresponds
to ${\cal N}=4$ supersymmetry in four dimensions. There are also
nonsupersymmetric heterotic vacua, in particular a non-tachyonic one
based on the gauge group $SO(16) \times SO(16)$ \cite{dh,agmv}.

- The (non-chiral) Type IIA and (chiral) Type IIB closed string theories, with
${\cal N}=2$ spacetime supersymmetry, that after a toroidal compactification
corresponds to ${\cal N}=8$ supersymmetry in four dimensions. Different modular
invariant GSO projections in 10d give rise to nonsupersymmetric theories,
called 0A and 0B \cite{sw}.

- The Type I open string theory, with gauge group $SO(32)$ and ${\cal N}=1$
supersymmetry. In this case the (Chan-Paton) gauge quantum numbers sit at the ends of the   
string and allow, in more general cases, to construct the gauge groups
$O(n)$, $USp(n)$ and $U(n)$ \cite{grinstein}. This theory can be defined as a projection
(orientifold) of the Type IIB string \cite{sagnotti1}. Analogously, 
(nonsupersymmetric)
orientifolds of Type 0A and 0B can be constructed \cite{bs}, in
particular a nontachyonic 0B orientifold with gauge group $U(32)$ \cite{augusto}.

\noindent The massless modes of the above superstring theories and their interactions are
described by effective 10d supergravity theories, namely:

- The low energy limit of the the two heterotic theories {\it and} of the Type I open 
string are described by the
ten dimensional ${\cal N}=1$ (or $(1,0)$) supergravity coupled to the
super Yang-Mills system based
on the gauge groups $SO(32)$ and $E_8 \times E_8$, respectively.

- The low energy limit of the Type II strings is described by the ${\cal
N}=2$ Type IIA supergravity (with $(1,1)$ supersymmetry) and Type IIB 
supergravity (with $(2,0)$ supersymmetry).

\noindent The common features of all the effective ten dimensional superstring theories 
is the
presence of supersymmetric multiplets containing in the bosonic sector the graviton
$g_{\mu \nu}$, the dilaton $\phi$ and an antisymmetric tensor $B_{\mu \nu}$.
The string coupling constant is a dynamical variable
$\lambda = \exp ({\phi})$, and the only free parameter is the string length
$\alpha'=1/M_s^2$, where $M_s$ is the string mass scale.

The 4d theories are defined after a compactification 
similar to the old Kaluza-Klein scenario. Typically, the
ten dimensional spacetime is decomposed as $M_{10}=M_4 \times K_6$, 
where $M_4$ is our four dimensional Minkowski spacetime and $K_6$ is a compact
manifold whose volume $V$ traditionally defines the compactification scale $M_c$
\be
V = M_c^{-6} \equiv M_{GUT}^{-6} \ , \label{1.1}
\ee
the scale of the Kaluza-Klein mass excitations in the internal space.
The compactification scale was also identified above with the grand unified
scale $M_{GUT}$ in the string unification picture, because the field theory
description breaks down above $M_c$. However, we will see later that
(\ref{1.1}) can be substantially altered in some string models.

The massless fields in a toroidal
compactification are the zero modes of the 10d fields, that in more
general settings depend on the topology of the compact space $K_6$. If we denote by
$i,j$ six dimensional internal indices, then we have, for example, the following
decompositions:
\ba
g_{AB} \ : \ \ g_{\mu \nu} \ \ g_{ij} \ \ g_{\mu i} \ , \nonumber \\
B_{AC} \ : \ \ B_{\mu \nu} \ \ B_{ij} \ \ B_{\mu i} \ \nonumber 
\ , 
\ea
where in 4d $g_{\mu \nu}$ is the graviton, $g_{\mu i}$, $B_{\mu i}$
are gauge fields and $g_{ij}$ are scalars describing the shape of the compact
space. On the other hand, $B_{\mu \nu}$ and $B_{ij}$ are pseudoscalar, 
axion-type fields.

Four dimensional string couplings and scales are predicted in terms of the
string mass scale $M_s$ and of various dynamical fields: dilaton, volume of compact
space, etc. In contrast to the usual GUT models, which do not incorporate gravity
and thus make no predictions for Newton's constant,  the perturbative
string models do make a definite prediction for the gravitational coupling strength.  
Since the length scale of string theory $\sqrt{\alpha'}$,
the volume $V$ of the internal manifold and the expectation value of the dilaton 
field $\phi$ are not directly
known from experiment, one might naively think that by adjusting 
$\alpha'$, $V$, and $\langle\phi\rangle$ one could fit to any desired
values the Newton's constant, the GUT scale $M_{GUT}$, 
and the GUT coupling constant $\alpha_{GUT}$.
However,  this is not true for the weakly
coupled heterotic strings.  In 10d, the low energy supergravity
effective action looks like
\be
S_{eff}= 
\int d^{10}x \sqrt g \ e^{-2\phi} \left({4\over (\alpha')^{4}} R
- {1\over (\alpha')^{3}} {\rm tr} F^2 + \cdots \right) \ , \label{1.2}
\ee
where $R$ is the scalar curvature and ${\rm tr} F^2$ is the Yang-Mills kinetic term.
After compactification on an internal manifold 
of volume $V$ (in the string metric), one gets a
four-dimensional effective action that looks like
\be
S_{eff}= \int d^4x \sqrt g \ e^{-2\phi} V
\left({4\over(\alpha')^{4}}R
- {1\over(\alpha')^{3}}{\rm tr} F^2+\cdots\right) \ .  \label{1.3}
\ee
Notice that the same function $Ve^{-2\phi}$ multiplies both $R$ and
${\rm tr} F^2$. From (\ref{1.3}), defining the heterotic scale
$M_H={\alpha'}^{-1/2}$, one thus gets
\be
M_H = {({\alpha_{GUT} \over 8})}^{1/2} M_P\ , \ 
\lambda_H  = 2 ({\alpha_{GUT}} V)^{1/2} M_H^3 \ , \label{1.03}
\ee
where $M_P=G_N^{-1/2}$ is the Planck mass. Then $M_H \sim 5
\times 10^{17}$ GeV, and
therefore there is some (slight) discrepancy between the GUT scale
$M_{GUT}$ and the string scale
$M_H$.  Indeed, from (\ref{1.1}) and (\ref{1.03}) we find $M_{GUT}/M_H =
(4 \alpha_{GUT} / \lambda_H^2)^{1/6}$ which asks, in order to find
$M_{GUT} \sim 2-3 \times 10^{16}$ GeV, for a very large string 
coupling $\lambda_H$.
The problem might be alleviated
by considering an anisotropic Calabi-Yau space and a lot of effort in this
direction was made over the years \cite{pheno}.

The above picture evolved considerably in the last few years. 
First of all, it was a 
puzzle that the heterotic $SO(32)$ and the Type I ten dimensional strings share
the same low-energy theory. Indeed, the two low-energy actions coincide if the
following identifications are made
\be
{\lambda}_I = {1 \over \lambda_H} \ \ , \ \ M_I = {M_H \over \sqrt{\lambda_H}}
\ , \label{1.4}
\ee
where $M_I,M_H$ are the heterotic and Type I string scales and
$\lambda_I, \lambda_H$ are the corresponding string couplings. A natural
conjecture was made, that the two string theories are dual (in the
weak-coupling strong-coupling sense) to each other
\cite{witten2,pw}. New arguments 
in favor of this duality came soon:

- The heterotic $SO(32)$ string can be obtained as a soliton solution of the
Type I string \cite{atish}.

- There is a precise mapping of BPS states (and their masses) betwwen the two
theories. If we compactify, for example, both theories to nine dimensions
on a circle of radius $R_I (R_H)$ in Type I (heterotic) units,
we can relate states with the masses
\be
{\cal M}_I^2 = l^2 R_I^2M_I^4 + {m^2 R_I^2 M_I^4 \over \lambda_I^2} + 
{n^2 \over R_I^2}
\leftrightarrow  
{\cal M}_H^2 = m^2 R_H^2M_H^4 + {l^2 R_H^2 M_H^4 \over \lambda_H^2} + 
{n^2 \over R_H^2}
\ , \label{1.5}
\ee
where $n,l$ ($n,m$) are Kaluza-Klein and winding numbers on Type I (heterotic)
side. It is interesting to notice in this formula how perturbative heterotic
winding states ($m$) become non perturbative on the Type I side. An important 
role in checking dualities in various dimensions is played by extended objects
called Dirichlet (D) branes \cite{polchinski}, which correspond on the
heterotic side to non perturbative states.

A second, far more surprising conclusion was reached in studying the strong
coupling limit of the ten dimensional Type IIA string. It was already known
that a simple truncation of eleven dimensional supergravity
\cite{cjs} on a circle
of radius $R_{11}$ gives the Type IIA supergravity in 10d, and that the 
Type IIA string coupling $\lambda$ is related to the radius by \cite{witten2} 
\be
M_{11} R_{11}= {\lambda}^{2/3} \ . \label{1.6}
\ee
On the other hand, if we consider Kaluza-Klein masses of the
compactified 11d supergravity and map them in Type IIA string units, we find
\be
m_n = {n \over R_{11}} \leftrightarrow m_n = {n \over \lambda} M_{IIA}
\ . \label{1.7}
\ee
Therefore, on Type IIA side, they can be interpreted as non perturbative,
and, with a bit more effort, BPS D0 brane states. The natural conclusion is that in
the strong coupling limit $\lambda \rightarrow \infty$ of the Type IIA
string,
a new dimension reveals itself ($R_{11} \rightarrow \infty$ using
(\ref{1.6})) and the low energy theory becomes the uncompactified
11d supergravity \cite{ht}, \cite{witten2}! As there is
no known quantum theory whose low energy limit describes the 11d
supergravity, a new name was invented for this underlying structure, 
the M-theory \cite{M-theory}.

Soon after, Horava and Witten gave convincing arguments that the
11d supergravity compactified on a line segment $S^1/Z_2$
(or, equivalently, on a circle with opposite points identified) should describe the strong
coupling limit of the $E_8 \times E_8$ heterotic string \cite{hw}. They argued
that the two gauge
factors sit at the ends of the interval, very much like the gauge quantum
numbers of open strings are sitting at their ends. The basic 
argument is that only half (one Majorana-Weyl) of the original
(Majorana) 11d gravitino lives on the boundary. This would 
produce gravitational anomalies unless $248$ new Majorana-Weyl fermions 
appear at each end. This is exactly the dimension of the gauge group $E_8$.

The compactification pattern of this theory down to 4d is 
different according to the relative value of the eleventh radius compared to the other 
radii, that are denoted collectively $R$ in the following. Assuming for simplicity an
isotropic compact space, there are two distinct compactification patterns
\ba
R_{11}<R \ : \ 11d \rightarrow 10d \rightarrow 4d \ , \nonumber \\
R_{11}>R \ : \ 11d \rightarrow 5d \rightarrow 4d \ . \label{1.8}
\ea
In the strong coupling limit $R_{11}>R$, there is therefore an energy range where the
spacetime is effectively five dimensional.
        
Finally, let us notice that in ten dimensions the Type IIB string is 
conjectured to be self-dual in the sense of an $SL(2,Z)$ strong-weak coupling
S-duality. Moreover, the $SO(32)$ heterotic 
string compactified on a circle of radius $R$ is T-dual to 
$E_8 \times E_8$ heterotic string compactified on a circle of radius
$1/R$ and similarly Type IIA and Type IIB strings are T-dual to each
other. By combining all
the above information one can build a whole web of dualities, which
becomes richer
and richer when new space dimensions are compactified \cite{lelonde}.  
     
In the light of the new picture described above, let us
see what changes in the strong coupling regime and let us investigate whether, for a
string scale of the order of the GUT scale, the gauge unification
problem has a natural solution in a region of {\it large} string coupling constant.  
The behavior is completely
different depending on whether one considers the $SO(32)$ or the $E_8\times E_8$
heterotic string.  

\noindent Let us first consider the  strongly-coupled $SO(32)$ heterotic
string, equivalent to the weakly-coupled Type I string.  We repeat the above 
discussion, using
the Type I dilaton $\phi_I$, metric $g_I$, and scalar curvature $R_I$.
The analog of (\ref{1.2}) is 
\be
L_{eff}= \int d^{10}x \sqrt {g_I} 
\left(e^{-2\phi_I}{4\over (\alpha')^{4}}R_I
- e^{-\phi_I}{1\over (\alpha')^{3}}{\rm tr} F^2+\dots\right) \ . \label{1.9}
\ee
Contrary to the heterotic string case, the gravitational
and gauge actions multiply different functions of $\phi_I$, 
$e^{-2\phi_I}$ and $e^{-\phi_I}$, since the first is
generated by a world-sheet path integral on the sphere, while the second 
arises from the disk. The analog of (\ref{1.3}) is then
\be
L_{eff}= \int d^4x\sqrt g_I V \left({4e^{-2\phi_I}
\over(\alpha')^{4}}R
- {e^{-\phi_I}\over(\alpha')^{3}}{\rm tr} F^2+\dots\right) \ . \label{1.10} 
\ee
The 4d quantites can be expressed as
\be
M_I = ({2 \over {\alpha}_{GUT}^2 M_P^2})^{1/4} V^{-1/4} \ , \
\lambda_I =  4 \alpha_{GUT} M_I^6 V \ . \label{1.11}
\ee                       
Hence
\be 
M_I = ({ \alpha_{GUT} \lambda_I \over 8})^{1/2} M_P \ , \label{1.12}
\ee
showing that after taking $\alpha_{GUT}$ from experiment
one can make $M_I$ as small as one wishes simply by taking $e^{\phi_I}$ to
be small, that is, by taking the Type I superstring to be weakly
coupled\footnote{In this case, however, $M_I^6 V << 1$ and a better 
physical picture is obtained by performing T-dualities, thus generating 
lower-dimensional branes.}. 
In particular, as mentioned in the Introduction, $M_I$ can be lowered
down to the weak scale \cite{lykken}. In this case the unification
picture is completely different \cite{ddg,bachas2}, as we will see in
the following sections. 

We will now argue that the $E_8\times E_8$ heterotic string
has a similar strong coupling behavior: one retains the standard
GUT relations among the gauge couplings, but loosing the prediction
for Newton's constant, which can thus be considerably below
the weak coupling bound. 

At strong coupling, the ten-dimensional $E_8\times E_8$ heterotic string 
becomes $M$-theory on $R^{10}\times S^1/Z_2$ \cite{hw}.
The gravitational field propagates in the bulk of the eleventh dimension,
while the $E_8\times E_8$ gauge fields live at the 
$Z_2$ fixed points $0$ and $\pi R_{11}$.
We write $M^{11}$ for $R^{10}\times S^1$ and $M^{10}_i$, $i=1,2$ 
for the two fixed (hyper)planes.
The gauge and gravitational kinetic energies take the form
\be
L= {1\over 2\kappa_{11}^2}\int_{M^{11}} d^{11}x \sqrt g R -\sum_i
{3^{1/3} \over 4\pi
(2 \pi \kappa_{11}^2)^{2/3}}\int_{M^{10}_i}d^{10}x\sqrt g \ 
{\rm tr} F_i^2 \ , \label{1.13}
\ee
where $\kappa_{11} $ is here the eleven-dimensional gravitational 
coupling and  $F_i$, for
$i=1,2$, is the field strength of the $i^{th}$ $E_8$, which propagates
on the fixed plane $M^{10}_i$.

Now compactify to four/five dimensions on a compact manifold whose volume
(in the eleven-dimensional metric, from now on) is $V$.
Let $S^1$ have a radius $R_{11}$, or a circumference $2\pi R_{11}$, and define the
eleven dimensional scale $M_{11}=2 \pi (4 \pi \kappa_{11}^2)^{-1/9}$.  
Upon reducing (\ref{1.13}) down to 4d, one can express $M_{11}$ and $R_{11}$
in terms of four-dimensional parameters
\be
M_{11}=(2 \alpha_{GUT} V)^{-1/6} \ , \nonumber \\
R_{11}^{-1} = ({2 \over \alpha_{GUT}})^{3/2} M_P^{-2} V^{-1/2} \ . \label{1.14}
\ee
{}From the first relation we find that $M_{11} \sim M_{GUT}$, and therefore
the Horava-Witten theory can accomodate a traditional MSSM
unification-type scenario with fundamental scale $M_{11} \sim 10^{16}$
GeV. The second one, for
$V=M_{GUT}^{-6}$, gives $R_{11}^{-1} \sim 10^{13}-10^{15}$ GeV.
This is again a sensible result, since $R_{11}$ has to be large compared
to the eleven-dimensional Planck scale in order to have a reliable field-theory
description of the theory.

\section{Building blocks for Type I strings}

Type I strings  describe the dynamics of open and closed superstrings.
Denoting by $0 \leq \sigma \leq \pi$ the coordinate describing the
open string at a given time, the two ends $\sigma =0, \pi$ contain the gauge
group (Chan-Paton) degrees of freedom and the corresponding charged
matter fields. The open string quantum states can be conveniently
described by matrices 
\be
|k;a> = \sum_{i,j=1}^N \lambda_{i,j}^a |k;i,j> \ , \label{2.01}
\ee
where $i,j=1 \cdots N$ denote Chan-Paton
indices and $k$ other internal quantum numbers. At the ends of open
strings, we must add boundary conditions, which for string coordinates
can be of two types 
\be
{\partial X^{\mu} \over \partial \sigma}|_{\sigma = 0, \pi} = 0 \quad ,
\ (N) \quad , \ X^{\mu}|_{\sigma = 0, \pi} = {\rm cst} \quad , \ (D) 
\ , \label{2.1}
\ee
where the two different possibilities denote the Neumann (N) and Dirichlet
(D) strings. As will be explained later on, even if we start
with a theory containing only Neumann strings, the Dirichlet strings can
arise after performing various T-duality operations or on
orbifolds containing $Z_2$-type elements.
Since by joining two open strings one can create a closed string, propagation
of closed strings must be added for consistency. The corresponding
quantum fluctuations produce the closed (gravitational-type) spectrum of
the theory, neutral under the Chan-Paton gauge group, that always
contains the gravitational (super)multiplet. 
The string oscillators are defined as Fourier modes of the string
coordinates. For closed coordinates, the expansion reads
\be
X_c^{\mu} = x^{\mu} + 2 \alpha' p^{\mu} \tau + {i \over 2} \sqrt{2
\alpha'} \sum_{n \not=0} {1 \over n} \left[ \alpha_n^{\mu} 
e^{-2in(\tau-\sigma)}
+ {\tilde \alpha}_n^{\mu} e^{-2in(\tau+\sigma)} \right] \ .
\label{2.08}
\ee 
The usual canonical quantization gives the commutators for the left movers 
$ [{\alpha}_m^{\mu}, {\alpha}_n^{\mu}]=m \delta_{m+n} \eta^{\mu \nu}$
and similarly for the right movers.  
For open strings with Neumann boundary conditions, for example,
the oscillator expansion reads
\be
X_o^{\mu} = x^{\mu} + 2 \alpha' p^{\mu} \tau + i  \sqrt{2
\alpha'} \sum_{n \not=0} {1 \over n} \left[ \alpha_n^{\mu} e^{-i n \tau}
\cos{n \sigma} \right] \ . \label{2.09}
\ee  
Type I can be seen as a projection (or {\it orientifold}) of Type IIB
theory, obtained by projecting the Type IIB spectrum by the involution
$\Omega$, exchanging the left and right closed oscillators 
$ \alpha_m^{\mu}, {\tilde \alpha}_m^{\mu}$ and acting
on the open-strings ones by phases
\be
{\rm closed :} \ \Omega \quad : \quad \alpha_m^{\mu} \leftrightarrow {\tilde
\alpha}_m^{\mu} \quad ,\quad {\rm open :} \ \Omega \quad : 
\quad \alpha_m^{\mu} \rightarrow \pm (-1)^m \alpha_m^{\mu} \ . \label{2.2}
\ee 
In addition, $\Omega$ acts on the zero-modes (compactification lattice) of
closed strings by interchanging left and right momenta ${\bf p}_L
\leftrightarrow {\bf p}_R$.

The parent Type IIB string contains D(-1),D1,D3,D5,D7 (and D9) branes,
coupling electrically or magnetically to the various RR forms present in
the massless spectrum. Out of them, the D1, D5 and D9 branes are
invariant under $\Omega$ and therefore are present in the Type I theory,
as (sub)spaces on which open string ends can terminate.
In some sense, open strings can be considered as twisted states of the $\Omega$
involution \cite{sagnotti1}, in
analogy with twisted states in orbifold compactifications of closed strings.
 
The perturbative, topological expansion in Type I strings involves
two-dimensional surfaces with holes $h$, boundaries $b$ and crosscaps
$c$. Each surface has an associated factor
$\lambda_I^{- \chi}$, where
\be
\chi = 2 - 2h - b - c \label{2.02}
\ee
is the Euler genus of the corresponding surface. 
Tree-level diagrams include, in addition to the
sphere with genus $\chi =2$, the disk with one boundary $\chi=1$, where open string
vertex operators can be attached, and the projective plane $RP^2$ with
one crosscap ($\chi=1)$ . One-loop diagrams include, in addition to the usual
torus ${\cal T}$ with one handle, the Klein bottle ${\cal K}$ with two
crosscaps, the annulus ${\cal A}$ with two boundaries
and the M{\"o}bius ${\cal M}$ with one boundary and one crosscap, all of them
having $\chi=0$ .
The last two diagrams allow the propagation of
open strings with Chan-Paton charges $|k;ij>$ in the annulus and $|k;ii>$ in
the M{\"o}bius, containing the gauge group and the charged matter degrees 
of freedom. On the other hand, the torus and the Klein bottle describe 
the propagation of closed string degrees of freedom. 

One-loop string diagrams may be constructed as generalizations of the 
one-loop vacuum energy in field-theory. In d noncompact dimensions, the
vacuum energy contribution of a real boson of mass $m$ is
\ba
\Gamma &=& {1 \over 2} \int {d^d p \over (2 \pi)^d} \ln (p^2+m^2) = 
-{1 \over 2} \int_0^{\infty} {dt \over t} \int {d^d p \over (2 \pi)^d}
e^{-(p^2+m^2) t } \nonumber \\
&=& -{1 \over 2 (4 \pi)^{d/2}} \int_0^{\infty} {dt \over t^{1+d/2}} e^{-t m^2} \ ,
\label{2.03}
\ea
where we introduced a Schwinger proper-time parameter through the
identity
\be
\ln {A \over B} = - \int_0^{\infty} {dt \over t} (e^{-t A}-e^{-t B})
\label{2.04}
\ee
and where we also neglected in (\ref{2.03}) an (infinite) irrelevant
mass-independent term.
The result (\ref{2.03}) readily generalizes to the case of more
particles in the loop with mass operator $m$ and different spin, as
\be
\Gamma=  -{1 \over 2 (4 \pi)^{d/2}} \ {\rm Str} \ \int_0^{\infty} {dt \over t^{1+d/2}}
e^{-t m^2} \ , \label{2.05}
\ee
where ${\rm Str}$ takes into account the multiplicities of particles and
their spin and reduces in 4d to the usual definition 
${\rm Str} \ m^{2k}= \sum_J (-1)^{2J} (2J+1) \ {\rm
tr} \ m_J^{2k}$, where $m_J$ denotes the mass matrix of particles of spin $J$.

The generalization of (\ref{2.05}) to the Type IIB torus
partition function in $d$ noncompact (and $10-d$ compact) dimensions is
(keeping only internal metric moduli here for simplicity)
\ba 
&&{\cal T}= Tr \ {1+(-1)^G \over 2} {1+(-1)^{\bar G} \over 2} \ {\cal P} \ 
q^{L_0} {\bar q}^{\bar L_0} = \nonumber \\ 
&&{1 \over ({4 \pi^2 {\alpha'}})^{d \over 2}} \sum_{rs}\, X_{rs}\, 
\int_F {d^2 \tau \over ({\rm Im \ \tau})^{1+{d \over 2}}} 
\chi_r(\tau)\, \chi_s({\bar \tau} ) \ \Gamma_{rs}^{(10-d,10-d)} (\tau,
{\bar \tau}, g_{i{\bar j}} ) \quad ,
\label{2.3}
\ea 
where $q=exp(2 \pi i \tau)$ and $\tau$ is the modular parameter of the
torus, $L_0,{\bar L}_0$ are Virasoro operators for the left and the
right movers, $(-1)^G$  ($(-1)^{\bar G}$) is the world-sheet left
(right) fermion number implementing the GSO projection and ${\cal P}$ an
operator needed in orbifold
compactifications (see Section 5) in order to project onto physical states.
In (\ref{2.3}), the $\chi$'s are a set of modular functions of the underlying conformal
field theory, $\Gamma_{rs}^{(10-d,10-d)}$ is the contribution from the
compactification lattice depending on the compact metric components
$g_{i{\bar j}}$ and $X$ is a matrix of integers. The integral in (\ref{2.3}) is
performed over the fundamental region 
\be
F \ : \ {\rm Im} \tau \geq 0 \ , \ -{1 \over 2} \leq \tau_1 \leq {1
\over 2} \ , \ |\tau| \geq 1 \ , \label{2.030}
\ee
and the ${\rm Im} \ \tau$ factors come
from integrating over noncompact momenta. The typical form of the
characters is 
\be
\chi_r = q^{h_r -{c \over 24}} \sum_{n=0}^{\infty} d_n^r q^n \ , \label{2.031} 
\ee
where $h_r$ is the conformal weight, $c$ is the central
charge of the conformal field theory and the $d_n^r$ are positive integers.
 
Let us start with a brief review of the algorithm used in the
following. This was introduced in \cite{ps,bs}, and
developed further in \cite{pss}. The starting point consists in
adding to the (halved) torus amplitude the Klein-bottle
${\cal K}$.  This completes the projection induced by $\Omega$, 
and is a linear combination
of the diagonal contributions to the torus amplitude, with argument $q
{\bar q}$. Then one obtains 
\footnote{As discussed in \cite{pss}, 
in general one has the option of modifying eq. (\ref{2.4}), altering $X_{ii}$
by signs $\epsilon_i$. These
turn sectors symmetrized under left-right interchange into
antisymmetrized ones,
and vice-versa, and are in general constrained by compatibility with the
fusion rules. This freedom, which has the spacetime interpretation of
flipping the RR charge of some orientifold planes, will turn out to be
crucial later on.}
\be 
{\cal K}= Tr \ {\Omega \over 2} {1+(-1)^G \over 2} \ {\cal P} \ 
e^{-4 \pi \tau_2 L_0} = {1 \over 2 ({4 \pi^2 {\alpha'}})^{d \over 2}} 
\ \int_0^{\infty} {d \tau_2 \over \tau_2^{1+{d \over 2}}} 
\sum_{r}\, X_{rr} \, \chi_r(2i\tau_2) \ \Gamma_{{\cal K},r}^{(10-d)} (
i \tau_2 , g_{i{\bar j}} ) \ , \label{2.4}
\ee 
with $\tau_2$ the proper time for the closed string and  
$\Gamma_{{\cal K},r}^{(10-d)} (i \tau_2 , g_{i{\bar j}} )$ the contribution of
the compactification lattice. In order to identify the
corresponding open sector, it is useful to perform the $S$ modular transformation
induced by 
\ba 
{\cal K}:\qquad\qquad 2\tau_2 \ &{{{}\atop\longrightarrow}\atop {{}\atop S}}& 
\ {1\over 2\tau_2}\equiv l\ ,
\label{2.5}
\ea
thus turning the direct-channel Klein-bottle amplitude
${\cal K}$ into the transverse-channel amplitude. The latter describes
the propagation of the closed spectrum on a cylinder of length $l$ terminating at two
crosscaps\footnote{The crosscap, or real projective plane, is a non-orientable surface
that may be defined starting from a 2-sphere and identifying antipodal points.}, and
has the generic form
\be 
{\cal K}={1 \over 2 ({4 \pi^2 {\alpha'}})^{d \over 2}} 
\ \int_0^{\infty} dl \sum_{r}\, \Gamma_r^2 \, \chi_r(il) \ 
{\tilde \Gamma}_{{\cal K},r}^{(10-d)} (i l , g_{i{\bar j}} ) \equiv 
{1 \over 2 ({4 \pi^2 })^{d \over 2}{\alpha'}^5} \int_0^{\infty} 
dl \ {\tilde{\cal K}}\ ,
\label{2.6}
\ee 
where ${\tilde \Gamma}_{{\cal K},r}^{(10-d)} (i l, g_{i{\bar j}} )$ is the
Poisson transform of $\Gamma_{{\cal K},r}$ and the coefficients
$\Gamma_r$ can be related to the one-point functions of the
closed-string fields in the presence of a
crosscap. Alternatively, in a spacetime language, the $\Omega$
involution has fixed (hyper)surfaces called orientifold (O) planes, carrying RR
charge. The Klein bottle amplitude is then interpreted as describing the
closed string propagation
starting and ending on orientifold (O) planes. Since the modulus of the
Klein amplitude is $0 \leq \tau_2 < \infty $, the $\tau_2$ integral is not
cut in the ultraviolet (UV) and is generically UV
divergent. Physically, this divergence is related to the presence of an
uncanceled RR flux from the O planes, which asks for the
introduction of D branes and corresponding open strings. It will be
important later on to distinguish between several types of
O-planes. First of all, in supersymmetric models there are $O_{+}$
planes carrying negative RR charge and $O_{-}$ planes carrying positive
RR charge and also flipped NS-NS couplings, in order to preserve
supersymmetry. In nonsupersymmetric models, there can exist ${\bar O}_{+}$
planes with flipped RR charge compared to their supersymmetric $O_{+}$
cousins, but with the same NS-NS couplings, therefore breaking
supersymmetry. Analogously, we can define  ${\bar O}_{-}$ planes,
starting from  $O_{-}$ planes and flipping only the RR charge. We will
exemplify later on in detail the couplings of these four different
types of O-planes to supergravity fields in different models.

The open strings may be deduced from the closed-string spectrum in a
similar fashion. A very important property of one-loop open string
amplitudes is that they all have a dual interpretation as tree-level
closed string propagation (see Figure 1).  
First, the direct-channel annulus amplitude
may be deduced from the transverse-channel boundary-to-boundary
amplitude. This has the general form \cite{ps} (see also \cite{io})
\be 
{\cal A}= \ {1 \over ({8 \pi^2 {\alpha'}})^{d \over 2}}
\int_0^{\infty} dl \sum_{r}\, B_r^2 \, \chi_r(il) \ {\tilde \Gamma}_{{\cal
A},r}^{(10-d)} (il, g_{i{\bar j}}) \equiv {1 \over 
({8 \pi^2 })^{d \over 2} {\alpha'}^5}
\int_0^{\infty} dl \ {\tilde{\cal A}} \ ,
\label{2.7}
\ee 
where the coefficients $B_r$ can be related to the one-point functions of
closed-string fields on the disk and on the $RP^2$ crosscap. In a spacetime
interpretation, the annulus amplitudes describe open strings with ends stuck
on D branes.
\begin{figure}
\vspace{4 cm}
\special{hscale=60 vscale=60 voffset=0 hoffset=120
psfile=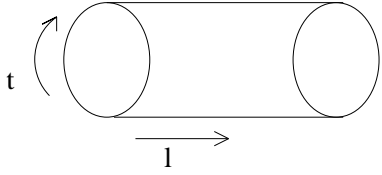}
\caption{The annulus amplitude has a dual interpretation of one-loop
open string propagation with vertical time $t$ and tree-level closed string 
propagation with horizontal time $l$.}
\end{figure}
The relevant $S$ modular
transformation now maps the closed string proper time $l$ on the tube to the
open-string proper time $t$ on the annulus, according to
\ba 
{\cal A}:\qquad\qquad l \ &{{{}\atop\longrightarrow}\atop {{}\atop S}}&  
{1\over l}\equiv {t\over 2} \ .
\label{2.8}
\ea
The direct-channel annulus amplitude then takes the form
\be 
{\cal A} \!=\! {1 \over 2} Tr {1 + (-1)^G \over 2} \ 
{\cal P} \ e^{- \pi t L_0} \!=\! {1 \over 2 ({8 \pi^2 {\alpha'}})^{d \over 2}} 
\int_0^{\infty} {d t \over t^{1+{d \over 2}}} 
\sum_{r,a,b} A^r_{ab} \ n_a\ n_b\ 
\chi_r \left({it \over 2}\right)\ \Gamma_{{\cal A},r}^{(10-d)} ({it 
\over 2}, g_{i{\bar j}}) \ , \label{2.9}
\ee 
where $L_0$ in (\ref{2.9}) is the Virasoro operator in the open sector,
the $n$'s are integers that have the interpretation of Chan-Paton
multiplicities for the boundaries (D branes) and the $A^r$ are a set of
matrices with integer
elements. These matrices are obtained solving diophantine equations determined by the
condition that the modular transform of eq. (\ref{2.9}) involves only integer
coefficients, while the Chan-Paton multiplicities arise as free parameters of the
solution. Supersymmetric models contain only D-branes, i.e. objects
carrying positive RR charges. Nonsupersymmetric models ask generically
also for {\it antibranes}, objects carrying negative RR charges but with
NS-NS couplings identical to those of branes. 

Finally, the transverse-channel M{\"o}bius amplitude $\tilde{\cal M}$ describes the
propagation of closed strings between D branes and O planes
(or boundaries and crosscaps, in worldsheet language), and is determined by
factorization from
$\tilde{\cal K}$ and $\tilde{\cal A}$. It contains the characters common to the two
expressions, with coefficients that are geometric means of those present in
$\tilde{\cal K}$ and $\tilde{\cal A}$ \cite{ps}, \cite{bs}. Thus
\be 
{\cal M}= \ - 2  {1 \over ({8 \pi^2 {\alpha'}})^{d \over 2}} 
\int_0^{\infty}
dl \sum_{r}\, B_r \ \Gamma_r \ {\hat\chi}_r \ (il+\frac{1}{2}) \ 
{\tilde \Gamma}_{{\cal M},r}^{(10-d)} (il, g_{i{\bar j}}) \equiv
{1 \over ({8 \pi^2 })^{d \over 2 {\alpha'}^5}} \int_0^{\infty} 
dl \ {\tilde{\cal M}} \ ,
\label{2.10}
\ee 
where the hatted characters form a real basis and are obtained by the redefinitions
\be 
{\hat\chi}_r \ (il+\frac{1}{2})=e^{-i\pi h_r} \chi_r \ (il+\frac{1}{2})\ .
\label{2.11}
\ee 
The direct-channel M{\"o}bius amplitude can then be related
to $\tilde{\cal M}$ by a
modular $P$ transformation and by the redefinition (\ref{2.11})
\ba
{\cal M}:\qquad\qquad {it \over 2}+{1 \over 2} \ 
&{{{}\atop\longrightarrow}\atop {{}\atop P}}&  
{i \over 2t}+{1 \over 2} \equiv il + {1 \over 2} \ . \label{2.12}
\ea
This is realized on the hatted characters by the sequence
$P=T^{1/2} ST^2ST^{1/2}$, with S the matrix that implements the
transformation $\tau \rightarrow -1/\tau$ and $T$ the diagonal 
matrix that implements the transformation
$\tau\to\tau+1$. The direct-channel M{\"o}bius amplitude then takes the form
\ba 
&& {\cal M} =  Tr {\Omega \over 2} {1+(-1)^G \over 2} 
\ {\cal P} \ (-e^{-\pi t})^{L_0} = \nonumber \\
&& -  {1 \over 2 ({8 \pi^2 {\alpha'}})^{d \over 2}}
\int_0^{\infty} {d t \over t^{1+{d \over 2}}}
\sum_{r,a} \ M^r_{a} \ n_a\  {\hat\chi}_r \left({it
\over 2}+\frac{1}{2}\right) \ \Gamma_{{\cal M},r}^{(10-d)} ({i t \over
2}, g_{i{\bar j}}) \ ,
\label{2.13}
\ea 
where by consistency the integer coefficients $M^r_{a}$ satisfy 
constraints \cite{stanev} that make $\cal M$ the $\Omega$ projection 
of $\cal A$. The full one-loop vacuum amplitude is 
\be 
\int\left({1\over 2}{\cal T}(\tau,{\bar \tau})+{\cal K}(2i\tau_2)
+{\cal A}({it\over 2})+{\cal M}({it\over 2}+{1\over 2})\right)\ ,
\label{2.15}
\ee 
where the different measures of integration are left implicit.
In the remainder of this paper, we shall often omit the dependence on 
world-sheet modular parameters.

It is often convenient for a spacetime particle
interpretation to write the partition functions with the help of $SO(2n)$ characters 
\ba 
O_{2n} &=& {1 \over 2 \eta^n} ( \theta_3^n + \theta_4^n) \ , \qquad\quad 
V_{2n}={1 \over 2 \eta^n} (
\theta_3^n - \theta_4^n) \ , \nonumber \\ S_{2n} &=& {1 \over 2 \eta^n} ( \theta_2^n +
i^n
\theta_1^n) \ , \qquad C_{2n}={1 \over 2 \eta^n} ( \theta_2^n - i^n \theta_1^n) \ ,
\label{2.16}
\ea 
where the $\theta_i$ are the four Jacobi theta-functions with (half)integer
characteristics. In a spacetime interpretation, at the lowest level 
$O_{2n}$ represents a
scalar, $V_{2n}$ represents a vector, while  $S_{2n}$, $C_{2n}$ represent spinors of
opposite chiralities. In order to link the direct and transverse
channels, one needs the transformation matrices $S$ and $P$ for the
level-one $SO(2n)$ characters (\ref{2.16}). These may be simply deduced
from the corresponding transformation properties of the Jacobi theta
functions, and are
\be  
S_{(2n)} =  {1 \over 2}
\left(
\begin{array}{cccc}  1 & 1 & 1 & 1 \\ 1 & 1 & -1 & -1 \\ 1 & -1 & i^{-n} & -i^{-n} \\ 1
& -1 & -i^{-n} & i^{-n}
\end{array}
\right) \ , \  P_{(2n)} = 
\left(
\begin{array}{cccc}  c & s & 0 & 0 \\  s & -c & 0 & 0 \\ 0 & 0 & \zeta c & i \zeta s \\
0 & 0 & i
\zeta s & \zeta c
\end{array}
\right) \ ,\label{2.17}
\ee   
where $c= \cos ({n \pi /4})$, $s= \sin ({n \pi /4})$ and $\zeta= e^{-i{n
\pi/4}}$ \cite{bs}. 

The absence of UV divergences ($l \rightarrow \infty$ limit) in the
above amplitudes asks for
constraints on the Chan-Paton factors, called {\it tadpole consistency
conditions} \cite{pc}. They are equivalent to the absence of
tree-level one-point functions for some closed string fields and ensure
that the total RR charge in the theory is zero. In the notations used
here, they read
\be
B_r = \Gamma_r \ , \label{2.170}
\ee
and generically determine the Chan-Paton multiplicity, that in
ten dimensions equals $N=32$. The tadpoles for RR fields can be
related \cite{pc} to inconsistencies in the field equations of 
RR forms (often reflected in the presence of gauge and gravitational 
anomalies). Indeed, D branes and O planes are electric and magnetic
sources for RR forms. The Bianchi identities and field equations
for a form of order $n$ then read (in the  language of differential
forms )
\be
d H_{n+1}= * J_{8-n} \quad , \quad d * H_{n+1}= * J_{n} \ , \label{2.171}
\ee
where the subscript on the electric and magnetic sources denotes their rank. The field
equations are globally consistent if
\be
\int_{C_m} * J_{10-m} = 0 \ , \label{2.172}
\ee 
for all closed (sub)manifolds $C_m$. In particular, in a compact space the RR
flux must be zero, and this gives nontrivial constraints on the spectrum
of D branes in the theory. 

The situation is different for NS-NS tadpoles. Indeed, suppose there
is a dilaton tadpole, of the type
$\exp(-\Phi)$ in the string frame, generated by the presence of  (anti)brane-(anti)orientifold Dp-Op systems.  
The dilaton classical field equation reads
\be
\partial_A ( \sqrt{g} \ g^{AB} \partial_B \Phi) = \sum_i \alpha_i  \sqrt{g} \ e^{(p-3) \Phi \over 4} 
\delta^{(9-p)} (y-y_i) \ , \label{2.0172}
\ee 
where $A,B = 1 \cdots 10$ and $y_i$ denote the position of the brane-orientifold planes in the space transverse
to the brane. The uncancelled dilaton tadpole means explicitly 
\be
\sum_i \alpha_i \not = 0 \ , \ {\rm while} \   
\sum_i \alpha_i  \int_{\cal C} \sqrt{g} \ e^{(p-3) \Phi \over 4} 
\delta^{(9-p)} (y-y_i) =0 \ . \label{2.0173}
\ee     
The first inequality means that, around the {\it flat vacuum}, the r.h.s. source in
(\ref{2.0172}) does not integrate to zero
and violates the integrability condition coming from the l.h.s. of
(\ref{2.0172}). As stressed in \cite{fs}, however,
this simply means that the real background is {\it not} the flat
background, but a curved one. This explains the second equality 
in (\ref{2.0173}), where ${\cal C}$ is any closed curve or
(hyper)surface in the internal space. An explicit
example of such a Type I background was recently given in \cite{dm4}.

In order to describe some simple Type I examples, let us consider
two 10d orientifolds of Type IIB. 

i) Supersymmetric SO(32)  

The Type IIB torus amplitude reads
\be 
{\cal T}={1 \over ({4 \pi^2 {\alpha'}})^{5}} 
\int_F {d^2 \tau \over ({\rm Im} \ \tau)^{6}} |(V_8-S_8) {1 \over \eta^8}|^2 \quad ,
\label{2.18}
\ee 
in terms of the characters introduced in (\ref{2.16}). In (\ref{2.18}), the
characters $V_8-S_8$ describe the contribution of the (worldsheet, left
and right) fermionic coordinates ($\Psi^{\mu}$, ${\tilde \Psi}^{\mu}$) to
the partition function. Moreover, $1/ \eta^8$ denotes
the contribution of the eight transverse bosons $X^{\mu}$, where $\eta$
is the Dedekind modular function defined in eq. (\ref{a1}) of the
Appendix. The corresponding Klein bottle amplitude is 
\be 
{\cal K}={1 \over 2 ({4 \pi^2 {\alpha'}})^{5}} 
\ \int_0^{\infty} {d \tau_2 \over \tau_2^{6}} 
(V_8-S_8) {1 \over \eta^8} \ . \label{2.19}
\ee 
It symmetrizes the NS-NS states and antisymetrizes the RR states. In
particular, the NS-NS antisymmetric tensor is projected out of the
spectrum (still, quantized parts of it can consistently be introduced
\cite{bps}), while the RR  antisymmetric tensor survives, a general feature
in Type I models. 
 
The annulus and M{\"o}bius amplitudes in the open string channel read
\ba
{\cal A}= \ \frac{N^2}{2} \ {1 \over ({8 \pi^2 {\alpha'}})^{5}} 
\int_0^{\infty} {d t \over t^{6}} 
(V_8-S_8) {1 \over \eta^8} \ , \nonumber \\
{\cal M}= \ - \frac{N}{2} \ {1 \over ({8 \pi^2 {\alpha'}})^{5}} 
\int_0^{\infty} {d t \over t^{6}} 
(V_8-S_8) {1 \over \eta^8} \ , \label{2.20}
\ea
where $N$ is the Chan-Paton index. In order to find the massless
spectrum, we expand them in powers of the modular
parameter $q$ and retain the constant piece, obtaining 
\be
{\cal A}_0 + {\cal M}_0 \sim {N(N-1) \over 2} \int_0^{\infty} {d t \over
t^{6}} \times (8-8) \ , \label{2.21}
\ee
where the $(8-8)$ terms come from the vector $V_8$ and the spinor
$S_8$, respectively. The massless spectrum is therefore supersymmetric,
and consists of 10d vectors and Weyl spinors in the adjoint
representation of the gauge group $SO(N)$. The dimension of the group is
fixed by looking at the divergent (tadpole) piece of the amplitudes in
the transverse (closed-string) channel
\be
{\cal K}+{\cal A}+{\cal M} =  \frac{1}{2} \ {1 \over ({8 \pi^2 {\alpha'}})^{5}}
\int_0^{\infty} dl \ (32+{N^2 \over 32}-2N) \times (8-8) + \cdots \ , \label{2.22}
\ee
where the two equal terms $8-8$ come from the NS-NS and RR massless closed string
states and $\cdots$ denote exchanges of massive closed states,
with no associated IR divergences. Even if the RR and NS-NS divergent pieces
cancel each other, consistency of the theory requires cancelling each
independently, as they reflect the existence of different
couplings. This requires that $N=32$ and  therefore determines
the gauge group $SO(32)$.
The geometric interpretation of this model is that it contains 32 D9
branes and 32 O$9_{+}$ planes, that carry RR charge under an unphysical RR
10-form $A_{10}$. The effective action contains the bosonic terms
\be
S = \int d^{10} x \{ \sqrt{g} \ {\cal L}_{SUGRA}- (N-32) ( \sqrt{g}
e^{-\Phi} + A_{10}) \} + \cdots \ , \label{2.022}
\ee
clearly displaying the interaction of closed fields $g_{AB}, \Phi,
A_{10}$ with the D9 branes and the O9 planes in the model.

ii) Nonsupersymmetric USp(32)

As already explained, there is an important difference between tadpoles of RR
closed fields and tadpoles of NS-NS closed fields. While the first
signal an internal inconsistency of the theory and must therefore always
be cancelled, the latter ask for a background redefinition 
and remove flat directions, producing potentials for
the corresponding fields and leading actually to consistent models
\cite{fs}. The difference between RR and NS-NS tadpoles turns out to
play an important role in (some) models with broken supersymmetry.
Indeed, there is another consistent model in 10d described by the same
closed spectrum (\ref{2.18})-(\ref{2.19}), but with a nonsupersymmetric
open spectrum described by the Chan-Paton charge $N$. The
open string partition functions are \cite{sugimoto}
\ba
{\cal A}= \ \frac{N^2}{2} \ {1 \over ({8 \pi^2 {\alpha'}})^{5}} 
\int_0^{\infty} {d t \over t^{6}} 
 (V_8-S_8)  {1 \over \eta^8} \ , \nonumber \\
{\cal M}= \ \frac{N}{2} \ {1 \over ({8 \pi^2 {\alpha'}})^{5}} 
\int_0^{\infty} {d t \over t^{6}} 
 (V_8+S_8) {1 \over \eta^8} \ . \label{2.23}
\ea
{}From the closed string viewpoint, $V_8$ describes the NS-NS sector
(more precisely, the dilaton) and $S_8$ the RR sector. The tadpole
conditions here read
\be
{\cal K}+{\cal A}+{\cal M} =  \frac{1}{2} \ {1 \over ({8 \pi^2 {\alpha'}})^{5}}
\int_0^{\infty} dl \ \{ (N +32)^2 \times 8 - (N-32)^2 \times 8 \} + \cdots
\ . \label{2.24}
\ee
It is therefore clear that we can set to zero the RR tadpole choosing
$N=32$, but we are forced to live with a dilaton tadpole. The resulting
spectrum is nonsupersymmetric and contains the vectors of 
the gauge group $USp(32)$ and fermions in the antisymmetric (reducible) 
representation. However, the spectrum is free of gauge and gravitational
anomalies, and therefore the model appears to be consistent. This model
contains 32 ${\bar D}$9 branes and 32 O$9_{-}$ planes, such that the
total RR
charge is zero but NS-NS tadpoles are present, signaling the breaking
of supersymmetry. The effective action contains the bosonic terms
\be
S = \int d^{10} x \{ \sqrt{g} \ {\cal L}_{SUGRA}- (N+32) \sqrt{g}
e^{-\Phi} + (N-32) A_{10} \} + \cdots \ . \label{2.024}
\ee
Notice in (\ref{2.024}) the peculiar couplings of the dilaton and the
10-form to antibranes/$O_{-}$ planes, in agreement with the general
properties displayed earlier. Indeed, the coupling to the ten-form
is similar to the supersymmetric one (\ref{2.022}), modulo the
overall sign reflecting the flipped RR charge of antibranes and $O_{-}$ planes 
compared to branes and $O_{+}$ planes. The coupling to the dilaton
reflects that antibranes couple to NS-NS fields in the same way as
branes, while $O_{-}$ planes couple with a flipped sign compared to
$O_{+}$ planes. 
 
The NS-NS tadpoles generate scalar potentials for the
corresponding (closed-string) fields, in our case the (10d) dilaton.
The dilaton potential reads
\be
V \sim (N+32) e^{- \Phi } \quad , \label{2.26}
\ee
and in the Einstein frame is proportional to $(N+32) \exp(3
\Phi /2)$. It has therefore the (usual) runaway behaviour towards
zero string coupling, a feature which is common to all
perturbative constructions. However, other NS-NS fields can be
given more complicated potentials and can be stabilized in appropriate compactifications
of Type I strings with brane-antibrane systems, as we will see later on.
Due to the dilaton tadpole, the background of this model is not the 10d Minkowski space.
However, it was shown in \cite{dm4} that a background with $SO(9)$
Poincare symmetry can be explicitly
found, therefore curing the NS-NS tadpole problem. In this
background, the tenth dimension is spontaneously
compactified and the geometry is $R^9 \times S^1/Z_2$,
with localized gravity.   

There is another way to see that in 10d the only possible gauge groups
are orthogonal and symplectic. Indeed, the massless gauge bosons are
represented as $\lambda \ \alpha_{-1}^A |0>$, where $\lambda$ is the
matrix describing the Chan-Paton charges defined in (\ref{2.01}), of
size $N \times N$. The orientifold involution $\Omega$ which squares to one can have
a nontrivial action on the matrix $\lambda$
\be
\Omega \quad : \quad \lambda \rightarrow - \gamma_{\Omega} \lambda^T
\gamma_{\Omega}^{-1} \ , \label{2.25}
\ee
where $\lambda^T$ denotes the transpose of the matrix $\lambda$. The
action of $\Omega$ squares to one if $\gamma_{\Omega}
(\gamma_{\Omega}^{-1})^T=\pm I$, where $I$ is the identity matrix. Then the
gauge bosons are invariant under $\Omega$ if \cite{grinstein}

a)  $\gamma_{\Omega}= \gamma_{\Omega}^T=I$, implying $\lambda=-\lambda^T$
   and the gauge group is $SO(N)$. 

b) $\gamma_{\Omega}= - \gamma_{\Omega}^T$, implying $\lambda= \lambda^T$
   and the gauge group $USp(N)$.

The difference between the supersymmetric $SO(32)$ and nonsupersymmetric
$USp(32)$ model described previously is that in the supersymmetric case
$\Omega$ acted in the same way on NS-NS and RR states in the (transverse
channel) M{\"o}bius, while in the nonsupersymmetric case the action was
$\Omega=1$ for NS-NS states and $\Omega=-1$ for RR states. Both
possibilites are however consistent with the rules described at the
beginning of this section, namely particle interpretation and factorization.
We will interpret later the first model as containing 32 D9 branes
and the second one as containing 32 D${\bar 9}$ (anti)branes, where by
definition antibranes have reversed RR charge compared to the
corresponding branes. The $USp(32)$ model is interpreted as
containing 32 O$9_{-}$ planes of positive RR charge (instead of
the negative charged O$9_{+}$ of the supersymmetric case), asking for 32  
D${\bar 9}$ (anti)branes
in the open sector. The only change occurs in the M{\"o}bius amplitude, that
describes strings streched between (anti)branes and orientifold planes.

iii) Models with local tadpole cancellation

As we have seen in the previous models, UV divergences in the open
spectrum are related to tadpoles of massless closed fields exchanged by
the branes. Let us now compactify one dimension (the discussion easily generalizes to
more compactified dimensions). The closed string fields have
in this case a tower of winding states of mass $n R M_I^2$ that
give no additional divergences. However, in the limit $R \rightarrow 0$
all these states become massless and contribute new potential
divergences. For example, the supersymmetric $SO(32)$ model has,
in the T-dual version ($R_{\perp}=1/RM_I^2$), 32 D8 branes at the origin $y_{\perp}=0$
and 32 orientifold O8 planes equally distributed between the two
orientifold fixed planes $y_{\perp}=0, \pi R_{\perp}$. The {\it global}
RR charge is indeed cancelled, however locally there are 16 units of RR charge
at $y_{\perp}=0$ and -16 at $y_{\perp}=\pi R_{\perp}$. Consequently, the
dilaton has a variation along $y_{\perp}$ and for 
$R_{\perp} \rightarrow \infty$, the theory encounters singularities \cite{pw}.
Avoiding this pathology asks for a new condition, {\it local tadpole cancellation}
or, equivalently the local cancellation of the RR charge. In the
example at hand, the only Type I 9d model satisfying this condition is obtained 
putting, via a Wilson line,  16 D8 branes at the origin $y_{\perp}=0$
and 16 D8 branes at $y_{\perp}=\pi R_{\perp}$, thus giving the gauge group 
$SO(16) \times SO(16)$. 

This phenomenon manifests itself neatly in the one-loop vacuum amplitudes
\cite{ads1}. Indeed, let us compactify the 10d Type I string on a circle
and let us introduce a Wilson line $W=(I_{n_1}, -I_{n_2})$, which breaks the gauge
group $SO(32) \rightarrow SO(n_1) \times SO(n_2)$. The three relevant
amplitudes read, in the direct channel,
\ba
{\cal K}&=&{1 \over 2 ({4 \pi^2 {\alpha'}})^{9/2}} 
\ \int_0^{\infty} {d \tau_2 \over \tau_2^{11/2}} \sum_m P_m  
\ (V_8-S_8) {1 \over \eta^8} \ , \nonumber \\
{\cal A}&=& \ \ {1 \over ({8 \pi^2 {\alpha'}})^{9/2}} 
\int_0^{\infty} {d t \over t^{11/2}} 
(V_8-S_8) {1 \over \eta^8} \left( {n_1^2 + n_2^2 \over 2} \sum_m P_m + n_1n_2
\sum_m P_{m+1/2} \right) \ , \nonumber \\
{\cal M}&=& \ - \frac{n_1+n_2}{2} \ {1 \over ({8 \pi^2 {\alpha'}})^{9/2}} 
\int_0^{\infty} {d t \over t^{11/2}} (V_8-S_8) {1 \over \eta^8} 
\sum_m P_m \ , \label{2.27}
\ea
where the half-integer powers of $t$ and $\alpha'$ come from integrating
over the nine noncompact momenta and the lattice summations are defined
in the Appendix. The same amplitudes can be written (after S and
P transformations), in the transverse channel, 
\ba
{\tilde {\cal K}}&=& {2^5 R \over 2 {\sqrt 2}} \ \sum_n W_{2n}  
\ (V_8-S_8) {1 \over \eta^8} \ , \nonumber \\
{\tilde {\cal A}}&=& \ \ {R \over 2^5 {\sqrt 2}} 
(V_8-S_8) {1 \over \eta^8} \left( {n_1^2 + n_2^2 \over 2} \sum_n W_n + n_1n_2
\sum_n (-1)^n W_n \right) \ , \nonumber \\
{\tilde {\cal M}}&=& \ - (n_1+n_2) \ {R \over {\sqrt 2}} (V_8-S_8) 
{1 \over \eta^8} \sum_n W_{2n}  \ ,
\label{2.28}
\ea
where the various numerical coefficients in (\ref{2.28}) arise from the
S transformations in the Klein bottle and annulus amplitude and after
the P transformation in the M{\"o}bius amplitudes. The sum of the three amplitudes
\be
{\tilde {\cal K}}+{\tilde {\cal A}}+{\tilde {\cal M}}= {R \over 2 {\sqrt
2}} (V_8-S_8) {1 \over \eta^8} \{ [32 + {(n_1+n_2)^2 \over 32} -2
(n_1+n_2)] \sum_n W_{2n} + {(n_1-n_2)^2 \over 32} \sum_n W_{2n+1} \} 
\label{2.29}
\ee
tells us that for an arbitrary radius the tadpole conditions coming from
the massless states are
\be
32 + {(n_1+n_2)^2 \over 32} -2 (n_1+n_2)=0 \ , \label{2.30}
\ee
fixing the number of D9 branes $n_1+n_2=32$. However, in the $R
\rightarrow 0$ ($R_{\perp} \rightarrow \infty$) limit the odd winding
states become massless too.
Therefore the last term in (\ref{2.29}) asks for $n_1=n_2=16$ and the
gauge group is $SO(16) \times SO(16)$, as anticipated. Models with local
tadpole conditions are intimately related to M-theory compactifications
since they allow, by a suitable identification of $R_{11}$ with $R_{\perp}$,
a well-defined strong coupling heterotic (M-theory) limit
$R_{\perp} \rightarrow \infty $.
  
\section{M-theory}

The maximal supergravity theory (containing particles with spin less
than or
equal to two) was constructed long time ago by Cremmer, Julia and Scherk
\cite{cjs} in eleven dimensions and is unique. The field content
consists of the irreducible 11d
supergravity multiplet: the graviton $g_{IJ}$, the Majorana gravitino
$\psi_{I \alpha}$ and the three-form $C_{IJK}$. The bosonic part of the
action was found to be 
\be
L_{SUGRA}= {1\over 2 \kappa_{11}^2}\int_{M^{11}}d^{11}x\sqrt g
\left( R^{(11)} 
- {1\over 24} G_{IJKL}G^{IJKL}  \right) \label{eq:ar1} 
 -{\sqrt 2 \over \kappa_{11}^2} \int_{M^{11}} d^{11}x
\ C \wedge G \wedge G \ , \label{m1}
\ee
where $I,J,K,L=1 \cdots 11$ and $G$ is the field-strength of the
three-form ($G = 6 d C$ in form notation), where we use here and in the
following the usual definition
$A={1 \over p !} A_{I_1 \cdots I_p} d x^{I_1} \wedge \cdots d x^{I_p}$
for differential forms.
The role played by the 11d supergravity (SUGRA) in string theory has
been a long-standing puzzle. A hint in this direction was that the 
$S^1$ circle dimensional reduction of 11d SUGRA to 10d gives exactly the
nonchiral Type IIA SUGRA. As all the 10d SUGRA theories are low-energy
limits of the corresponding string theories, it was natural to ask for the existence
and the properties of a quantum theory containing the 11d SUGRA as
its low-energy limit. This (still unkwown) theory was called M-theory and the study 
of its connection with string theories became a central goal for the string
community in the last five years.  
 
A natural conjecture was then logically put
forward, namely that the Type IIA string in the strongly coupled regime is
described by M-theory or, equivalently, that the M-theory compactified on $S^1$ is the
10d Type IIA superstring. For example, the bosonic fields of Type IIA
SUGRA and that of the circle reduction of the
bosonic 11d SUGRA contain both the graviton $g_{AB}$, antisymmetric
tensor $B_{AC}$ ($C_{11,AC}$ in 11d SUGRA), the dilaton $\Phi$
($g_{11,11} \equiv (R_{11}M_{11})^2$ in 11d SUGRA), a one-form potential
$A_B$ ($g_{11,B}$) and a three-form potential $C_{ABC}$.  
By comparing the lagrangians of 11d SUGRA of
Newton constant $M_{11} \sim k_{11}^{-2/9}$ compactified
on a circle $S^1$ of radius $R_{11}$ and of the Type IIA string of
string scale $M_{IIA}$ and string coupling $\lambda_{IIA}$, the
following relations emerge\footnote{The second relation can
equivalently be replaced by $M_{11}=\lambda_{IIA}^{-1/3} M_{IIA}$.} 
\be 
R_{11} M_{11} = \lambda_{IIA}^{2/3} \quad , \quad  g_{AB}^M= 
\lambda_{IIA}^{-2/3} g_{AB}  \ , \label{m2}
\ee
where $g_{AB}$ and $g_{AB}^M$ are the Type IIA string and the M-theory 
metric, respectively.
These relations support the conjectured duality. In the
weak-coupling regime of the Type IIA string ($\lambda_{IIA} \rightarrow 0$), 
$R_{11} \rightarrow 0$ and therefore the low-energy limit is indeed the 10d
IIA SUGRA. On the other hand, in the strong coupling regime  
($\lambda_{IIA} \rightarrow \infty$) a new-dimension decompactifies
($R_{11} \rightarrow \infty$), and the low-energy limit of the Type IIA string
is described by 11d SUGRA. A second argument for the conjecture is motivated by
trying to identify Kaluza-Klein modes of the 11d gravitational multiplet in
string language. Using the mapping (\ref{m2}), the relation
(\ref{1.7}) can easily be proved. On the string side, these states are
interpreted as D0 branes. A nontrivial check of the duality conjecture
\cite{witten3} is that a bound state of n D0 branes has a mass n times
larger than the mass of a single D0 brane, in precise correspondence with
the Kaluza-Klein spacing on the 11d supergravity side. 
More generally, it is known that the Type IIA string contains, in addition
to the fundamental string states and to the solitonic NS fivebrane,
D0,D2,D4,D6 and D8 BPS branes, coupling (electrically or magnetically) to
the appropriate odd-rank antisymmetric
tensors present in the massless spectrum of the theory. On the other hand, the
11d SUGRA contains M2 (membranes) and
M5 (fivebranes) as classical solutions. A precise mapping between M-theory states
compactified on the circle and Type IIA branes (with the exception
of the Type IIA D8 brane) was achieved, 
and their corresponding tensions were found to be in agreement with the conjectured
duality relations (\ref{m2}).
 
There is another possible compactification of M-theory to 10d, that preserves
one-half of the original supersymmetry. Indeed, the 11d action (\ref{m1}) has
the following symmetry
\ba
&&x_{11} \rightarrow -x_{11} \ , \ \psi_{I}(-x_{11})=\Gamma_{11}
\psi_{I}(x_{11}) \ , \nonumber \\
&&g_{AB}(-x_{11}) = g_{AB}(x_{11}) \ , \ g_{11,A}(-x_{11}) = -
g_{11,A}(x_{11}) \ , \nonumber \\
&&C_{ABC}(-x_{11}) = - C_{ABC}(x_{11}) \ , \ C_{11,AB}(-x_{11}) =  C_{11,AB}(x_{11})
\ , \label{m3}
\ea
where $A,B,C=1 \cdots 10$ are ten-dimensional indices and $\Gamma_{11} =
\Gamma_1 \cdots \Gamma_{10}$. We can then compactify on an {\it orbifold},
the interval $S^1/Z_2$ obtained by identifying opposite points, of coordinates $x_{11}$
and $-x_{11}$, on the circle. The two fixed points of this operation,
$x_{11}=0$ and $x_{11} = \pi R_{11}$ play a peculiar role, as will be
seen in a moment. From a 10d viewpoint, $\Gamma_{11}$ acts
as a chiral projector and selects one-half of the original gravitino,
namely one (chiral) Majorana-Weyl spinor $\psi_{A}$ with $\Gamma^A
\psi_A$ projected out and another Majorana-Weyl spinor (of opposite chirality)
$\psi_{11}$. The two spinors can be assembled into a 10d Majorana
gravitino $\psi_A$ containing 64 degrees of freedom.
The full massless
gravitational spectrum of M-theory on $S^1/Z_2$ includes also the 10d
graviton $g_{AB}$, the dilaton $\phi$ contained in
$g_{11,11}=(R_{11}M_{11})^2=e^{4 \phi /3}$ and an antisymmetric tensor
field $C_{11,AB}$. The sum of the bosonic degrees of freedom adds up to
64, as expected by supersymmetry.  
This cannot be the end of the story, however. It is well-known that the
massless 10d gravitino gives an anomaly under the 10d diffeomorphisms,
whereas the massive Kaluza-Klein modes are nonchiral and do not
contribute to the anomaly. On the other hand, in a smooth 11d space
there is no such anomaly. A natural
possibility is that the anomaly draws its origin from the two ends of
the interval and is equally distributed between the fixed points 
$x_{11}=0$ and $x_{11}= \pi R_{11}$.       
A standard explicit computation then asks for 496 Majorana-Weyl fermions, 248 on each
of the fixed points, to cancel it. These fermions come necessarily from
super Yang-Mills vector multiplets and can be associated to the gauge group $E_8 \times
E_8$, with one gauge factor per fixed point. The bosonic part of the
Yang-Mills action is then
\be
L_{SYM} = -{1 \over \lambda_1^2} \int_{x_{11}=0} d^{10} x \sqrt{g} \ tr
F_1^2 -{1 \over \lambda_2^2} \int_{x_{11}=\pi R_{11}} d^{10} x \sqrt{g} \ tr
F_2^2 \ , \label{m4}    
\ee
where $\lambda_i$ are the two Yang-Mills couplings. 
Similarly to the weakly-coupled heterotic string, supersymmetry
invariance of the action ask for a modification of the Bianchi identity
associated to the three-form. A closer look at the SUGRA-SYM Lagrangian
requires that the modification to the Bianchi identity be concentrated on the
fixed planes and read \cite{hw}
\be
d G = {k_{11}^2 \over \sqrt{2} \lambda^2} \ \bigl\{ \ \delta(x_{11}) ({1 \over 2}
{\rm tr} R^2- {\rm tr} F_1^2)+ \delta(x_{11}-\pi R_{11}) ({1 \over 2}
{\rm tr} R^2- {\rm tr} F_2^2) \ \bigr\} \ . \label{m5} 
\ee 
A consistent M-theory compactification is obtained by using 
\be
\int_{C_5} d G = 0 \quad , \quad \int_{C_4^i} ({\rm tr} F_i^2
- {1 \over 2} {\rm tr} R^2)=m_i-{1 \over 2} p_i \ , \label{m6}
\ee   
for any closed 5-cycle $C_5$ and arbitrary 4-cycle $C_4^i$ defined at
the fixed points $x_{11}^i=0, \pi R_{11}$, where in (\ref{m6}) $m_i,p_i$
are integers. The two equations (\ref{m6}) define
thus the embedding of the spin connection into the gauge group as 
one particular solution of the equation
\be
m_1+m_2= {1 \over 2} (p_1+p_2) \quad . \label{m7}
\ee
Gauge and gravitational anomaly cancellation issues was first discussed 
by Horava and Witten \cite{hw, dealwis},  starting from a
particular solution to the Bianchi identity (\ref{m5}). 
It was later realized \cite{dm,conrad} that the original solution
\cite{hw} is not unique and that a one-parameter class a solutions exist,
parametrized by $b$ in the following.   
A critical reanalysis of anomaly cancellation appeared recently
\cite{bde}, which insists on a periodic global definition of various
M-theory fields. In particular, \cite{bde} uses a periodic generalization
of the $\epsilon (x_{11})$ function on the interval $-\pi R_{11} \le
x_{11} \le \pi R_{11}$ , whose definition and derivative are
\be
\epsilon_1(x_{11}) = {\rm sign}(x_{11})-{x_{11} \over \pi R_{11}} \quad , \quad
d \epsilon_1 = [2 \delta (x_{11}) - {1 \over \pi R_{11}}] d x_{11} \
. \label{m07}
\ee
The above defined $\epsilon_1(x_{11})$ is indeed periodic and continous
at $x_{11}=\pi R_{11}$ and has a step-type discontinuity at $x_{11}=0$.
Similarly, another function discontinous at $x_{11}=\pi R_{11}$ can be
defined by $\epsilon_2 (x_{11})=\epsilon_1 (x_{11}-\pi R_{11})$. With
these definitions, the solution to the Bianchi identity (\ref{m5}) reads
\cite{bde}
\be
G = 6 \ dC + {k_{11}^2 \over \lambda^2} \{ (b-1) \sum_{i=1}^2 \delta_i \wedge
Q_3^i + {b \over 2} \sum_{i=1}^2 \epsilon_i 
{\hat I}_4^i - {b \over 2 \pi} d x_{11} \wedge \sum_{i=1}^2
Q_3^i \} \ , \label{m08}
\ee  
where we used the following definitions ${\hat I}_4^i = (1/2) {\rm tr}
R^2- {\rm tr} F_i^2$, $Q_3^i = (1/2) \omega_{3L}-\omega_{3Y}^i$
($\omega_{3L}$ and $\omega_{3Y}^i$ are Lorentz and gauge Chern-Simons
forms, repectively), $\delta_1 \equiv \delta (x_{11}) dx^{11}$, etc. It is
useful to remember that these definitions are such that 
${\hat I}_4^i = d Q_3^i$. The parameter $b$
can be fixed by a global argument \cite{bde}
\be
\int_{C_5} d G = \int_{C_4 (x_{11}^{(1)})} G - \int_{C_4 (x_{11}^{(2)})} G 
\ , \label{m09}
\ee  
where the 5-cycle $C_5$ has the boundary $\partial C_5 =C_4
(x_{11}^{(2)})+ C_4 (x_{11}^{(1)})$, $- \pi R_{11} < x_{11}^{(1)} <
0$ and  $0 < x_{11}^{(2)} <
\pi R_{11}$. If the standard embedding condition $m_1=p_1=p_2$, $m_2=0$
is not satisfied, then an explicit evaluation of (\ref{m09}) using
(\ref{m5}) and (\ref{m08}) forces upon $b=1$. The gauge and
gravitational anomalies are concentrated on the boundaries and
are given by the standard 10d expressions. Surprisingly enough,
the Green-Schwarz term taking care of their compensation is the 11d
topological Chern-Simons term in (\ref{m1}), since $C$ is not Yang-Mills and
Lorentz invariant. 
In the gauge variation of the three-form $C$ there is actually an additional
arbitrariness \cite{dm}. Making for simplicity the gauge choice
$\delta C_{ABC}=0$, the cancellation between the 10d one-loop
anomaly and the tree-level gauge variation of the Chern-Simons term fixes the relation between
the Yang-Mills couplings and the 11d gravitational coupling to be \cite{bde}
\be
{k_{11}^4 \over \lambda^6} = {12 \over (4 \pi)^5} \ . \label{m010}
\ee  
Compactifications of M-theory can be defined deforming
\cite{witten1} around a space of the form $S^1/Z_2 \times X^6$,
with $X^6$ a Calabi-Yau space of Hodge numbers $(h_{(1,1)}, h_{(2,1)})$,
in a perturbative expansion in $k_{11}^{2/3}$. We denote in the
following by $i,j=1,2,3$ the complex Calabi-Yau indices and by $\mu,\nu,\rho$
the 4d spacetime indices. The resulting 5d bulk theory 
contains as bosonic fields the gravitational multiplet, the universal 
hypermultiplet $({\rm det} g_{i \bar j},C_{\mu \nu \rho},
C_{ijk} \equiv \epsilon_{ijk} a)$, with $a$ a complex scalar,
$h_{(2,1)}$ additional hypermultiplets $(g_{ij},
C_{ij{\bar k}})$ and $h_{(1,1)}-1$ vector multiplets 
$(C_{\mu i {\bar j}}, g_{i {\bar j}})$, with the determinant ${\rm det}
g_{i \bar j}$ of the metric removed here and included in the universal
hypermultiplet. The effect of the nontrivial
Bianchi identity is to produce potential terms for moduli fields such
that the 5d theory becomes a gauged SUGRA \cite{ovrut}. The spectrum on
the two boundaries depends on the solution chosen for (\ref{m7}). For
example, the standard embedding solution $m_1=p_1=p_2$ and $m_2=0$ gives a
gauge group $E_6$ on one boundary with $h_{(1,1)}$ chiral multiplets in the
fundamental representation ${\bf 27}$ of $E_6$ and  $h_{(2,1)}$ chiral
multiplets in the ${\overline {\bf 27}}$, while the other boundary hosts a
super Yang-Mills theory with gauge group $E_8$. Nonstandard embeddings 
and nonperturbative vacua containing fivebranes were also considered 
\cite{nonstandard}. 
   
Some orbifold compactifications of M-theory of the type $T^n/Z_2
\times X^{7-n}$ were also considered in the literature \cite{gm}.
On the other hand, compactifications on particular compact spaces 
$S^1/Z_2 \times S^1 \times X^5$ can be simply related
to Type I compactifications. In order to see this, it is enough to study
the compactification to 9d.   
The compactification of M-theory on $S^1 \times S^1/Z_2$ (with radii
$R_{10}$ and $R_{11}$, respectively) admits two different interpretations \cite{hw}:
\begin{itemize}
\item[1. ]as M-theory on $S^1/Z_2\times S^1$, that according to \cite{hw}  describes
the $E_8 \times E_8$ heterotic string of coupling $\lambda_{E_8}
\!=\!(R_{11}M_{11})^{3/2}$, compactified on a circle $ S^1$ of radius
$R_{E_8}= R_{10}(R_{11}M_{11})^{1/2}$. In this case, a Wilson line must be added, and the
theory is in a vacuum with an unbroken $SO(16) \times SO(16)$ gauge
group. By making a standard T-duality transformation $R_H=1/R_{E_8}M_H^2$, 
$\lambda_H=\lambda_{E_8}/R_{E_8}M_H$, we can relate it to the
$SO(32)$ heterotic string in the vacuum state with gauge group $SO(16) \times SO(16)$,
of coupling $\lambda_H=R_{11}/R_{10}$ and radius
$R_H=1/(R_{10}(R_{11}M_{11})^{1/2})M_H^2$.
\item[2. ]as M-theory on $S^1 \times S^1/Z_2$, that according to
\cite{witten2} describes the IIA theory of coupling $\lambda_{IIA} = (R_{10}M_{11})^{3/2}$,
compactified further on the
$S^1/Z_2$ orientifold of radius $R_{11}(R_{10}M_{11})^{1/2}$. The result is the
Type-I$^\prime$ theory,  T-dual (with respect to the eleventh coordinate) to the Type I
theory (in its $SO(16) \times SO(16)$ vacuum), with coupling
$\lambda_I = R_{10}/R_{11}$, compactified on a circle of radius $1/(R_{11}
(R_{10}M_{11})^{1/2})M_I^2$. In the M-theory regime ($R_{11} >>
R_{10}$), the Type I and Type
I$^\prime$ theories can both be weakly coupled, and can consequently be 
treated as perturbative strings.
\end{itemize}

It is interesting to notice that the above duality relations are in
agreement with the $SO(32)$ heterotic-Type I duality conjecture
$\lambda_H= 1/\lambda_I$, $R_H=R_I/\lambda_I^{1/2}$, which can therefore
be regarded as a prediction in this framework.
  
A further check of these duality chains is found translating in Type
I or heterotic language
the masses of the BPS states of M-theory \cite{hw}. 
Consider first the Kaluza-Klein states of the supergravity multiplet on
$T^2=S^1/Z_2 \times S^1$, together with the wrapping modes of the M2 membrane around the 
torus. Their masses are 
\ba {\cal M}^2 = {l^2 \over R_{11}^2} + {m^2 \over R_{10}^2} +  n^2 R_{10}^2 R_{11}^2
M_{11}^6 \quad , \label{N1}
\ea 
where $(l,m,n)$ is a triplet of integers labelling the corresponding
charges. These masses must have a clear physical interpretation on the Type
I and heterotic sides.  
In Type-I and Type-I$^\prime$ units, the masses of the states (\ref{N1}) are
\ba 
{\cal M}_I^2 = l^2 R_I^2 M_I^4+ {m^2 R_I^2 M_I^4 \over {\lambda_I}^2}  + {n^2 \over
R_I^2} \ ,
\qquad {\cal M}_{I^\prime}^2 = {l^2 \over R_{I^\prime}^2} +  {m^2 M_I^2 \over
{\lambda_{I^\prime}}^2} + n^2 R_{I^\prime}^2 M_I^4  \ . \label{N2}
\ea 
In a similar fashion, in 
$E_8 \times E_8$ and $SO(32)$ heterotic units the states (\ref{N1}) have masses
\ba 
{\cal M}_{E_8}^2 = {l^2 M_H^2 \over {\lambda_{E_8}}^2} + {m^2 \over R_{E_8}^2} + n^2
R_{E_8}^2 M_H^4 \ , \quad {\cal M}_H^2 = l^2 {R_H^2 M_H^4 \over {\lambda_H}^2} + m^2
R_H^2 M_H^4 +  {n^2 \over R_H^2} \quad .
\label{N3}
\ea 
Notice that KK modes $l$ along the eleventh dimension are, according to (\ref{N3}),
nonperturbative in heterotic units but are perturbative states in Type I
and Type I' units (\ref{N2}). In particular, a Scherk-Schwarz type
breaking $l \rightarrow l + \omega$, with some fractional number $\omega$,
describes nonperturbative heterotic physics \cite{dg},\cite{aq} but can
be perturbatively described in Type I strings \cite{ads1}, as we will
show in detail in Section 7.
 
There are also twisted M-theory states associated to the fixed points of 
$S^1/Z_2$, that are charged under the gauge group. 
These include ordinary momentum excitations in the tenth direction and
membrane wrappings in the full internal space, for which 
\be {\cal M}^2={{\tilde m}^2 \over R_{10}^2} + {\tilde n}^2 R_{10}^2 R_{11}^2  M_{11}^6
\quad . 
\label{N4}
\ee 
In type I and Type I$^\prime$ units, their masses become
\ba 
{\cal M}_I^2 = {{\tilde m}^2 R_I^2 M_I^4 \over {\lambda_I}^2} +  {{\tilde n}^2 \over
R_I^2} \ ,
\qquad {\cal M}_{I^\prime}^2 =  {{\tilde m}^2 M_I^2 \over {\lambda_{I^\prime}^2}} +
{\tilde n}^2 R_{I^\prime}^2 M_I^4 \quad , \label{N5}
\ea 
and the wrapping modes are thus perturbative open string states. In $E_8 \times E_8$
and $SO(32)$ heterotic units, the masses of the charged states are
\ba 
{\cal M}_{E_8}^2 = {{\tilde m}^2 \over R_{E_8}^2} + {{\tilde n}^2 R_{E_8}^2}M_H^4 \ ,
\qquad {\cal M}_H^2 = {\tilde m}^2 R_H^2 M_H^4 + {{\tilde n}^2 \over R_H^2} \ . 
\label{N6}
\ea 
The perturbative states labeled by $n$ and $\tilde n$ have counterparts in the Type
I theory that reflect the perturbative heterotic-Type I duality 
(see eqs. (\ref{N2}) and (\ref{N5})). This is effective if the string coupling
$\lambda_I$ is small and $R_I$ is large.
 
The supersymmetric $S^1/Z_2$ compactification is
not the only possibility compatible with the $Z_2$ orbifold structure.
Indeed, there is the possibility of a nontrivial, Scherk-Schwarz type
11d $\rightarrow $ 10d compactification, obtained giving a nontrivial $y_{11}$
dependence to the zero modes in the Kaluza-Klein expansion
\cite{aq,dg}. This is consistent if the 11d theory has an appropriate
discrete symmetry, that in this case is the fermion number. Then the
11d gravitino field $\Psi = (\Psi_1, {\bar \Psi}_2)^T$, where
$\Psi_1, \Psi_2$ are the two Majorana-Weyl spinors, can have the 
nontrivial KK decomposition 
\ba 
\left( 
\begin{array}{c} 
\Psi_1 \\ 
\Psi_2 
\end{array} 
\right)  
&=& U 
\left(
\begin{array}{c}
\sum_{m=0}^{\infty} \cos {m y_{11} \over R_{11}} \Psi_1^{(m)} \\
\sum_{m=1}^{\infty} \sin {m y_{11} \over R_{11}} \Psi_2^{(m)}
\end{array}
\right) \ , \label{N7}
\ea
where $U \equiv exp (M y_{11})$ and $M$ is an antisymmetric matrix. 
Compatibility of the truncation (\ref{N7}) with the orbifold symmetry
$Z_2$ requires $\{ Z_2 , M \} = 0$, 
which fixes $M$ to be the off-diagonal antisymmetric matrix \cite{dg}
$M= i \omega \sigma_2 M_{11}$, where $\sigma_2$ is the Pauli matrix and
$\omega =1/2$ is fixed by the requirement that $U(y_{11}=2 \pi R_{11})
= -I$. The Scherk-Schwarz decomposition in this case reads explicitly
\ba 
\left( 
\begin{array}{c} 
\Psi_1 \\ 
\Psi_2 
\end{array} 
\right)  
&=& 
\left(
\begin{array}{cc} 
\cos {y_{11} \over 2 R_{11}} & \sin  {y_{11} \over 2 R_{11}} \\ 
-\sin  {y_{11} \over 2 R_{11}} & \cos  {y_{11} \over 2 R_{11}}
\end{array}
\right)
\left(
\begin{array}{c}
\sum_{m=0}^{\infty} \cos {m y_{11} \over R_{11}} \Psi_1^{(m)} \\
\sum_{m=1}^{\infty} \sin {m y_{11} \over R_{11}} \Psi_2^{(m)}
\end{array}
\right) \ ,\label{N9}
\ea
and indeed breaks supersymmetry in the eleventh dimension. Notice that 
the surviving gravitini on the two
boundaries $y_{11}=0$ and  $y_{11}=\pi R_{11}$ have opposite chirality 
\cite{dm3}.
The same result holds for the supersymmetric spinor transformation
parameter. Therefore, in order to compensate the gauge and the
gravitational anomalies on the two boundaries we must introduce as usual
the $E_8 \times E_8'$ gauge group, but the chiralities of the gauginos in the two
gauge factors are different. Each boundary preserves one-half of
the original 11d supersymmetry, but the configuration containing both
of them breaks supersymmetry completely \footnote{This argument is equivalent 
to the one recently presented in \cite{fh}.}. In Section 7 we will
present, by compactifying down to 9d, 
a Type I string description of this phenomenon \cite{ads1} and in
Section 10 a 4d compactified description at the field theory level. It will be shown
in Section 7 that the chirality flip means that one boundary contains branes 
and the other boundary antibranes, mutually interacting.  
Interestingly, this field theoretic argument proves that in even
spacetime dimensions, where we can define Weyl fermions, the breaking 
of supersymetry by compactification in one direction $Y$ perpendicular to
the branes is consistent only if the zero mode $Y$-variation of bulk 
fermions gives precisely Weyl fermions at the position of the
branes $Y_i$, while for arbitrary bulk positions $Y$ the zero modes have
no definite chirality. 
  
\section{Type I supersymmetric compactifications to four-dimensions}
  
A particularly simple way of reducing the number of supersymmetries and
of producing fermion chirality is to compactify on orbifolds \cite{dhvw}.
A d-dimensional orbifold $O^d$ can be constructed starting with the
d-dimensional euclidean space $R^d$ or the d-dimensional torus $T^d$ and
identifying points as
\be
O^d \ = \ R^d / S \ = T^d / P \quad , \label{3.01}  
\ee
where the {\it space group} S contains rotations $\theta$ and
translations v and the {\it point group} P is the discrete group of
rotations obtained from the space group ignoring the translations.
A typical element of S acts on coordinates as $X \rightarrow \theta X +$
v and is usually denoted $(\theta,{\rm v})$. The subgroup of S formed by pure
translations $(1,{\rm v})$ is called the {\it lattice} $\Gamma$ of
S. The identification of points of $R^d$ under $\Gamma$ defines the torus $T^d$.
Points of $T^d$ can then be further identified under P to form the
orbifold $O^d$. This is clearly consistent only if P consists of rotations
which are automorphisms of the lattice $\Gamma$.

In most of the following sections we will be interested in 4d ${\cal N}=1$ orientifolds
obtained by orbifolding the six real (three complex) internal coordinates by the twist
$\theta = (e^{2 i\pi v_1},e^{2 i\pi v_2} ,e^{2 i\pi v_3} )$, where ${\bf v} \equiv
(v_1,v_2,v_3)$ is called the twist vector and where for a $Z_N$ orbifold 
$\theta^N = 1$. If $v_1+v_2+v_3=0$ with all $v_i \not=0$, the
orientifold has generically  ${\cal N}=1$ supersymmetry (the ${\cal
N}=2$ of the parent Type IIB model broken to half of it by the
orientifold projection $\Omega$) while if, for example, $v_3=0$ and
$v_1+v_2=0$, the corresponding orientifold has ${\cal N}=2$
supersymmetry. The group structure of the orientifolds we use in the
following\footnote{The group structure is however not unique, see for
example \cite{gj}.} is $(1, \Omega , \theta^k , \Omega \theta^k \equiv \Omega_k)$. The
independent models were classified long time ago \cite{dhvw} and in 4d
the ${\cal N}=1$ orientifolds
are $Z_3$, $Z_4$, $Z_6$, $Z_6'$, $Z_7$,$Z_8$, $Z_8'$, $Z_{12}$, $Z_{12}'$
and $Z_N \times Z_M$ for some integers $N$ and $M$. All of them contain
a set of 32 D9 branes. In addition, the ones containing
$Z_2$-type elements ($Z_4$, $Z_6$, $Z_6'$,$Z_8$, $Z_{12}$, $Z_{12}'$)
have sets of $32$ D5
branes, needed here for the perturbative consistency of the compactified
theory. The presence of the D5 branes can be understood as follows. The
orientifold group element $\Omega \theta^{N/2}$ (and sometimes other
elements, too) has fixed (hyper)planes
called O$5_{+}$ planes, negatively charged under the (twisted) RR fields. By flux
conservation, they ask for a corresponding set of D5 branes with
opposite RR charge. The actual position of the D5 branes is not
completely fixed. They can naturally sit at the orbifold fixed points or
they can live ``in the bulk'' in sets of 2N branes in a $Z_N$ orbifold.
This brane displacement \cite{gp,gj,adds} can be understood as a Higgs
phenomenon breaking the open string gauge group and the sets of 2N bulk branes
can be regarded as one brane and its various images through the orbifold
and orientifold operations.

The three new (in addition to the torus ) Type I one-loop amplitudes for
a $Z_N$ orientifold can be written generically as
\ba
{\cal K} \!\!&\!\!=\!\!&\!\! {1 \over 2N} \sum_{k=0}^{N-1} \int_0^{\infty} {dt
\over t} \ \int {d^4 p \over (2 \pi)^4} Str_{closed} 
\ \Omega \ \theta^k \ q^{\alpha' (p^B p_B+m^2)} \equiv  
\int {dt \over t} (4\pi^2 \alpha' t)^{-2} K \ , \nonumber \\
{\cal A} \!\!&\!\!=\!\!&\!\! {1 \over 2N} \sum_{i,j=1}^{32}
\sum_{k=0}^{N-1} \int_0^{\infty}  {dt \over t} \int {d^4 p \over (2 \pi)^4} 
Str_{(i,j)} \ \theta^k \ q^{\alpha' (p^B p_B+m^2)} \equiv  
\int {dt \over t} (8\pi^2 \alpha' t)^{-2} A \ , \nonumber \\
{\cal M} \!\!&\!\!=\!\!&\!\! {1 \over 2N} \sum_{i=1}^{32}  
\sum_{k=0}^{N-1} \int_0^{\infty} 
{dt \over t}   \int {d^4 p \over (2 \pi)^4} Str_{(i,i)}
\ \Omega \ \theta^k \ q^{\alpha' (p^B p_B+m^2)} \equiv 
\int {dt \over t} (8\pi^2 \alpha' t)^{-2} M \ , \label{3.1}
\ea
where the modular parameters for the three one-loop surfaces are defined
in (\ref{2.5}),(\ref{2.8}) and (\ref{2.12}).    
$\Omega$ acts on the open string oscillators as $\Omega \alpha_m =
\pm e^{i \pi m} \alpha_m$, with the upper plus sign for the NN open
sector and the lower minus sign for the DD open sector. The supertrace
takes into account, as usual, the different statistics of bosons and fermions 
$Str \sim \sum_{bos} - \sum_{ferm}$ and the 4d momentum integrals give rise
to the factors $(8\pi^2 \alpha' t)^{-2}$ in ${\cal A}$ and ${\cal M}$
(and to the corresponding one in ${\cal K}$). The projection operator ${\cal P}$
introduced in Section 3, eqs. (\ref{2.3}), (\ref{2.4}), (\ref{2.9}) and
(\ref{2.13}) for an $Z_N$ orbifold reads
 \be
{\cal P} = {1 \over N} (1+ \theta + \cdots + \theta^{N-1}) \ , \label{3.02}
\ee
and therefore projects into orbifold invariant states ${\cal P} |phys>=|phys>$. 
The untwisted massless closed spectrum is found by first displaying the
right (and left) massless states
\ba
&Sector& \quad \quad State  \qquad \qquad \qquad \qquad \theta^k \qquad \qquad
\qquad {\rm helicity}
\nonumber \\
&NS& \ : \quad \quad \Psi_{-1/2}^{\mu} |0> \quad \quad \quad 1 \quad
\qquad \qquad \pm 1 \nonumber \\
&NS& \ : \quad \quad \Psi_{-1/2}^{j,\pm} |0> \quad \quad 
e^{\pm 2 \pi i k v_j} \qquad \qquad  6 \times 0 \nonumber \\
&R& \ : \quad \quad |s_0s_1s_2s_3> \quad e^{ 2 \pi i k
(s_1v_1+s_2v_2+s_3v_3)} \quad 4 \times (\pm {1 \over 2}) \quad , \label{3.3}
\ea
where $s_i=\pm 1/2$, $s_0+s_1+s_2+s_3=0$ (mod $2$) is the GSO projection in the R
sector and  $j=1,2,3$ denote (complex) compact indices. 
The physical closed string spectrum is obtained taking left-right tensor
products $|L> \otimes |R>$ invariant under the orbifold and orientifold
involution.
Typically the NS-NS spectrum of the orientifold is symmetrized by
$\Omega$, while the RR spectrum is antisymmetrized, but other choices
are possible.
 
The action of a twist element $\theta^k$ in the open N and D sectors 
can be described by $32 \times 32$
matrices $\g_{\theta^k} \equiv \g^k = (\g_{\theta})^k$ acting on the Chan-Paton degrees of 
freedom\footnote{In the case of $B_{\mu \nu}$ backgrounds \cite{bps} and other
discrete backgrounds \cite{carlo} there is a reduction
of the rank of the gauge group and for models with branes and
antibranes the rank of the matrix can be arbitrary, as we shall see later on.}
$\lambda^{(0)}$ for gauge bosons and $\lambda^{(i)}$ ($i=1,2,3$) for
matter scalars. Imposing that vertex operators for the corresponding
physical states be invariant under the orbifold projection defines this
action to be 
\be
\theta^k \quad : \quad \lambda^{(0)} \ \rightarrow \ \g^k \ \lambda \ 
(\g^k)^{-1} \ , \  \lambda^{(i)} \ \rightarrow \ e^{2 \pi i kv_i} \g^k \ \lambda \ 
(\g^k)^{-1} \ . \label{3.6}
\ee
Since $\theta^N=1$, it follows from (\ref{3.6}) that $\g^N = \pm 1$. For
$\g^N = 1$ the gauge groups in the D9 and D5 brane sectors are subgroups
of $SO(32)$, while for $\g^N =- 1$ the D9,D5 gauge groups are subgroups
of $U(16)$. The two choices correspond, in the notation of the previous
section, to ``real'' charges $n$ and to pairs of complex charges
$(m,{\bar m})$.
The corresponding contribution to the one-loop annulus amplitudes
(\ref{3.1}) is multipled by a Chan-Paton multiplicity $ (Tr \gamma^k)^2$.
Similarly, for every element $\Omega \theta^k \equiv \Omega_k $ there is an associated matrix
acting on the CP indices $\gamma_{\Omega_k}$ as
\be
\Omega_k \quad : \quad \lambda^{(0)} \ \rightarrow \ - \g_{\Omega_k} \ 
(\lambda^{(0)})^T \ (\g_{\Omega_k})^{-1} \ , \
 \lambda^{(i)} \ \rightarrow \ - \g_{\Omega_k} \ 
(\lambda^{(i)})^T \ (\g_{\Omega_k})^{-1} \ . 
 \label{3.06}
\ee
Since $\Omega^2=1$ it follows also that $\g_{\Omega}= \pm \g_{\Omega}^T
$. The corresponding
M{\"o}bius amplitudes are multiplied by the multiplicity factor $Tr
(\g_{\Omega_k}^{-1} \g_{\Omega_k}^T)$.
Without loss of generality the matrices $\gamma^k$ can be chosen to be diagonal.
The tadpole consistency conditions fix the Chan-Paton 
matrices $\g$ analogously to (\ref{2.170}), which in turn determine the
gauge group and the charged matter content of the corresponding 4d
orientifold. A generic supersymmetric model contains in the closed and the open
spectrum states having a 10d origin, having a compactification lattice depending
on all six compact coordinates, called the ${\cal N}=4$ sector. There could
also exist states having a 6d origin, with a compactification lattice depending on
two compact coordinates, called ${\cal N}=2$ sectors. Finally, there are
states without any excitations in the compact coordinates, forming the
${\cal N}=1$ sectors.
    
While the structure of the tadpole conditions cannot be described in full
generality, some generic results should however be mentioned.
In all cases, the tadpole conditions corresponding to untwisted forms are
proportional to
\ba
&D9&  : \quad \{ {1 \over 32} (Tr \g_9^0)^2- 2 Tr ( \g_{\Omega,9}^{-1}
\g_{\Omega,9}^T) + 32 \} V_1V_2V_3 \ , \nonumber \\
&D5&  : \quad \{ {1 \over 32} (Tr \g_5^0)^2- 2 Tr ( \g_{\Omega_{N/2},5}^{-1}
\g_{\Omega_{N/2},5}^T) + 32 \} {V_1 \over V_2V_3} \ , \label{3.7}
\ea
where $V_1,V_2,V_3$ are the volumes of the compact torii and we
considered a D5 brane parallel to the first torus $T^1$ and orthogonal
to $T^2$,$T^3$.
The solution to these equations is $\g_9^0=\g_5^0=I_{32}$,
$\g_{\Omega,9}= \g_{\Omega,9}^T$ and $\g_{\Omega_{N/2},5}= \g_{\Omega_{N/2},5}^T$,
asking therefore for one set of D9 branes and, for $N={\rm even}$, of
one set of D5 branes. 
It can also be shown that one can choose conventions  
such that
\be
Tr (\g_{\Omega_k,9}^{-1} \g_{\Omega_k,9}^T)=  Tr(\g_9^{2k}) \quad ,
\quad 
Tr (\g_{\Omega_k,5}^{-1} \g_{\Omega_k,5}^T)= - Tr(\g_5^{2k}) \ . \label{3.8}
\ee
For $Z_N$ orientifolds with N an odd integer, the
twisted tadpole conditions can be easily worked out, too. Indeed, by
using explicit expressions of the partition function on the three
relevant one-loop surfaces \cite{afiv}, one finds the tadpole conditions
\be
\sum_k \{ 32 \prod_{i=1}^3 \sin{2 \pi k v_i} + 2 \prod_{i=1}^3 \sin{
\pi k v_i} (Tr \g_9^k)^2 - 16 \prod_{i=1}^3 \sin{\pi k v_i} (Tr
\g_9^{2k}) \}=0 \ . \label{3.9}
\ee 
For odd N summing over twisted sectors k or over twisted sectors 2k is
however equivalent. We use this in order to rewrite all contributions
in (\ref{3.9}) in terms of $Tr \g_9^k$. We also define the number
of fixed points $N_k=64 (\prod_{i=1}^3 \sin{\pi k v_i})^2$ in an
orbifold. Then in odd orbifolds $N_k=N_{2k}$, implying 
$64 (\prod_{i=1}^3 \cos{\pi k v_i})^2=1$. By combining
these results, we can rewrite the solution of (\ref{3.9}) in the form
\be
Tr \g^{2k} = 32 \prod_{i=1}^3 \cos{\pi k v_i} \ . \label{3.10}
\ee
Some of the models,
$Z_4$,$Z_8$, $Z_{12}$,$Z_{12}'$  ($Z_2 \times Z_2$ models with
discrete torsion) have additional tadpoles from the Klein bottle
\cite{afiv} proportional to $1/V_3$ ($V_1V_2/V_3$), which
cannot be cancelled by adding sets of D5 branes. Surprisingly, these models seem
therefore inconsistent, but it will be shown later on that, at least
some of them allow consistent perturbative realisations with D9 branes and
D${\bar 5}$ (anti)branes, with supersymmetry broken on the antibranes. 

Let us exemplify these results refering to the first 4d Type I chiral model \cite{abpss},
the $Z_3$ orientifold with twist vector ${\bf v}=(1/3,1/3,-2/3)$. The
model has 32 D9 branes and the twisted tadpole condition (\ref{3.10})
reads $Tr \g^{2k}=-4$, for $k=1,2$. The solution of (\ref{3.10}) is $\g =
(\omega I_{12}, \omega^2 I_{12},I_8)$, with $\omega = exp(2 \pi
i/3)$. The untwisted closed spectrum consists of the dilaton, the NS-NS
scalar fields $g_{i {\bar j}}$, $i, {\bar j} =1,2,3$ and the RR axions
$B_{\mu \nu}$ , $B_{i {\bar j}}$. 
The twisted closed spectrum consists of 27 linear multiplets, one per fixed point.

The annulus amplitude in (\ref{3.1}) for the $Z_3$ orientifold can be written
\be
{\cal A} = {\cal A}_{{\cal N}=4}- \frac{1}{6} \sum_{k=1}^{2}
\int_0^\infty \, \frac{dt}t \,
 \, {\cal A}^{(k)} (q) \ , \label{3.010} 
\ee
where ${\cal A}_{{\cal N}=4}$ is the contribution of the ${\cal N}=4$
supersymmetric open spectrum, and  ${\cal A}^{(k)}$ is the contribution of the 
${\g}^k \equiv ({\g})^k$ sectors given by
\be
{\cal A}^{(k)} = {1 \over 8 \pi^4 t^2} \sum_{\a,\b=0,1/2}  \eta_{\a,\b} \
\frac{\vartheta[{\a \atop \b}]}{\eta^3} \ \prod_{i=1}^3 (-2\sin \pi k v_i)
\  \frac{\vartheta[{\a \atop {\b + kv_i}}]}{\vartheta[{1/2 \atop {1/2 +kv_i}}]} \,  \
( {\rm tr} {\g}^{k})^2 \ , \label{3.011}
\ee
by using the definitions (\ref{a2}) in the Appendix.
The M{\"o}bius amplitude can be similarly written as in (\ref{3.010}) by substituting
${\cal A} \rightarrow {\cal M}$, with
\be
{\cal M}^{(k)} = -  {1 \over 8 \pi^4 t^2} \sum_{\a,\b=0,1/2}  \eta_{\a,\b} \
\frac{\vartheta[{\a \atop \b}]}{\eta^3} \ \prod_{i=1}^3 (-2\sin \pi k v_i)
\  \frac{\vartheta[{\a \atop {\b + kv_i}}]}{\vartheta[{1/2 \atop {1/2 +kv_i}}]} \
({\rm tr} {\g}^{2k}) \ . \label{3.012}
\ee
Because of supersymmetry, the amplitudes (\ref{3.011}), (\ref{3.012})
vanish identically using modular identities.

The gauge group and the massless spectrum can be exhibited after expressing
the partition functions (\ref{3.011}),(\ref{3.012}) in terms of conformal
characters \cite{abpss}. To this end, the 10d $SO(8)$ Lorentz characters
are decomposed with respect to the $SO(2) \times SU(3) \times U(1)$
subgroup, where the $SO(2)$ factor corresponds to the (light-cone) spacetime
modes, so that the orbifold action is
\ba
V_8-S_8 &=& C_0+  C_{-}+ C_{+} \quad , \quad
\theta \ (V_8-S_8) = C_0+  \omega C_{-}+ \omega^2 C_{+} \quad , 
\nonumber \\
\theta^2 \ (V_8-S_8) &=& C_0+  \omega^2 C_{-}+ \omega C_{+}
\quad , \label{3.11}
\ea
where $C_0$ are modular functions describing in 4d a chiral multiplet and
$C_+,C_-$ are functions describing 3 chiral multiplets each\footnote{The
functions $C_{0}$, $C_{\pm}$ are defined \cite{abpss} starting from 
$SU(3)$ level-one
characters, from the four $SO(2)$ characters and from the 12 characters of 
the ${\cal N}=2$ superconformal model with central charge $c=1$.}. Then
the amplitudes (\ref{3.011}) and (\ref{3.012}) read
\ba
&&A ={(N+M+{\bar M})^2 \over 6}  (C_0 +
C_{-}+ C_{+}) \ \sum_{m_i} P_{m_i}^{(6)} + \nonumber \\
&&{(N+ \omega  M+ {\bar \omega}  
{\bar M})^2 \over 6}  (C_0 + \omega  C_{-} + {\bar \omega}  
C_{+}) +{(N+{\bar \omega}  M+ \omega  {\bar M})^2 \over 6}
(C_0  +  {\bar
\omega}  C_{-}+ \omega  C_{+})  \ , \nonumber \\
&&M = -{(N+M+{\bar M}) \over 6}  (C_0 +
C_{-} + C_{+}) \ \sum_{m_i} P_{m_i}^{(6)} - \label{3.12} \\
&&{(N+ \omega  M+ {\bar \omega}  {\bar M}) \over 6} (
C_0+  \omega  C_{-} + {\bar \omega} C_{+}) 
- {(N+{\bar \omega} M+ \omega  {\bar M}) \over 6}  (
C_0 +  {\bar
\omega}  C_{-}+ \omega C_{+}) \ , \nonumber 
\ea
where $N,M$ are Chan-Paton factors and $P_{m_i}^{(6)}$ is the momentum
(Kaluza-Klein) compactification lattice. The massless spectrum reads from
(\ref{3.12}) 
\ba
A_0+M_0 = && \bigl[ M {\bar M}+{N(N-1) \over 2} \bigr] (C_0)_0 + 
\bigl[ N {\bar M}+{M(M-1) \over 2} \bigr] (C_{-})_0 \nonumber
\\ +
&& \bigl[ N M+{{\bar M}({\bar M}-1) \over 2} \bigr] (C_{+})_0 \ ,
\label{3.13}
\ea
where the subscript 0 denotes the massless part of the characters. The
Chan-Paton factors are fixed by the tadpole conditions
(\ref{3.7}), (\ref{3.10}) 
\be
N+M+{\bar M}=32 \quad , \quad N-{1 \over 2}(M+{\bar M})=-4 \ , \label{3.14}
\ee
with the solution $M=12$, $N=8$. Therefore (\ref{3.13}) describes an
${\cal N}=1$ chiral model with gauge group   $U(12) \times SO(8)$ and
chiral multiplets in the representations  $3 ({\bf 12,8})_1+ 3 ({\bf 
\overline{66},1})_{-2}$, where the subscripts denote the charges of the
(anomalous) $U(1)_X$ factor contained in $U(12)$.
 
Alternatively, in the formalism of \cite{gp} the gauge group and the
massless matter spectrum can be found from the equations
\ba
&&\lambda^{(0)} = \gamma \ \lambda^{(0)} \ \gamma^{-1}
\quad , \quad \lambda^{(0)} = - \gamma_{\Omega} \ (\lambda^{(0)})^T \ 
\gamma_{\Omega}^{-1} \ , \nonumber \\
&&\lambda^{(i)} = e^{2 \pi iv_i} \gamma \ \lambda^{(i)} \ \gamma^{-1}
\quad , \quad \lambda^{(i)} = - \gamma_{\Omega} \ (\lambda^{(i)})^T \ 
\gamma_{\Omega}^{-1} \ . \label{3.15}
\ea
The matrix $\gamma_{\Omega}= (I_{12} \otimes \sigma_1,I_8)$ (where $\sigma_1$ is
the first, off-diagonal and symmetric Pauli matrix) interchanges the
roots of $\g$ with their complex conjugates. Solving (\ref{3.15}) we
find that the gauge fields $\lambda^{(0)}$ are described by a  general
$12 \times 12$ matrix and by an $8 \times 8$ antisymmetric matrix,
giving indeed the gauge group $U(12) \times SO(8)$. Each of the matter fields
$\lambda^{(i)}$, on the other hand, are described by two $12 \times 8$ matrices
and by one $12 \times 12$ antisymmetric matrix, describing, as before,  chiral multiplets  
in the representations $3 ({\bf 12,8})_1+ 3 ({\bf \overline{66},1})_{-2}$.
\section{Effective action and quantum corrections in Type I strings}

The effective field theory Lagrangian and the quantum corrections in Type I
orbifold compactifications have some distinctive features compared to
the corresponding heterotic compactifications, which will be briefly
reviewed in this section. First of all, it is important to realize that
some of the closed string (twisted and untwisted) axion-type fields are
components of antisymmetric tensors from the RR
sector. Together with the NS-NS scalars and the corresponding NS-R
fermions, these are naturally described (in an ${\cal N}=1$ language) by
linear multiplets. On the other hand, in the heterotic string only the dilaton
superfield was described by a linear multiplet, while all the other
moduli fields fitted into chiral multiplets. 

\vskip 2mm
- {\large \bf Generalized Green-Schwarz mechanism} 
\vskip 2mm

Let us start by defining the Type I compactification moduli, obtained by
a straightforward reduction of the Lagrangian (\ref{1.9}). By defining
complex coordinates $i=1,2,3$ and the associated components of the metric,
$G_i^{\a \b}$, $\a,\b=1,2$ (with the dimension of a squared radius), the dilaton $S$ and
the geometric moduli $T_i,U_i$  for the three complex planes are \cite{afiv}
\be
S=a^{RR}+i {\sqrt{G_1G_2G_3} M_I^6 \over \lambda_I} \ , \ 
U_i = {G_i^{12}+i {\sqrt G_i} \over G_i^{22}} \ , \ T_i = b_i^{RR} + i {{\sqrt G_i} M_I^2 
\over \lambda_I} \ , \label{4.1} 
\ee 
where $G_i \equiv \det G_i^{\a \b}$ and $a^{RR}, b_i^{RR}$ are axionic
fields from the RR sector. Our first goal here is to compute
the tree-level and the one-loop threshold corrections to the gauge
couplings of the Chan-Paton gauge groups. We expect here surprises
compared to the heterotic models, where the tree-level gauge couplings
are universal, $1/g_a^2 = Re \ f_a = k_a Re \ S$ and the numbers $k_a$
denote the Kac-Moody levels. For example, in the $Z_3$ model
described in the previous section, the abelian $U(1)_X$ factor is
anomalous and the mixed $U(1)_X G_a^2$ anomalies  $(C_{SU(12)},C_{SO(8)},C_{U(1)})$
$=$ $(1/4 \pi^2) (-18,36,-432)$ are incompatible with the standard
4d version of the Green-Schwarz mechanism \cite{green}. The solution to this puzzle was
proposed in \cite{iru}, in analogy with the generalized  Green-Schwarz mechanism 
found in 6d by Sagnotti \cite{sagnotti2}. It was conjectured in \cite{iru}
that the gauge fields in $Z_3$ do couple at tree-level to a linear
symmetric combination $M$ of the 27 closed twisted moduli  
\be
f_a = S + s_a M \ . \label{4.2}
\ee 
Under a $U(1)_X$ gauge
transformation with (superfield) parameter $\Lambda$, there are
cubic gauge anomalies. The generalized Green-Schwarz mechanism requires a shift
of the combination $M$ of twisted moduli \cite{lln,abd}
\be
V_X \rightarrow V_X + {i \over 2} (\Lambda-{\bar \Lambda}) \ , \ M
\rightarrow M + {1 \over 2} \ \epsilon \ \Lambda \ , \label{4.3}
\ee
such that the gauge-invariant combination appearing in the K{\"a}hler
potential is $i(M-{\bar M})-\epsilon V_X$. The mixed anomalies are
cancelled provided the following condition holds
\be
{\epsilon \over 4 \pi^2} = {C_{SU(12)} \over s_{SU(12)}}= {C_{SO(8)} \over s_{SO(8)}}=
{C_{U(1)_X} \over s_{U(1)_X}} \ . \label{4.4}
\ee
By supersymmetry arguments, one can also write the D-terms which
encode the induced Fayet-Iliopoulos term 
\be
V_D = {g_X^2 \over 2} (\sum_A X_A K_A \Phi^A + \epsilon 
{\partial K \over \partial M} M_P^2 )^2 \ , \label{4.5}
\ee
where $\Phi^A$ denotes the set of charged chiral fields of $U(1)_X$
charge $X_A$ and $K_A=\partial K / \partial \Phi^A$. 
It was shown in \cite{iru} that actually the mixed anomalies
$C_a$ are proportional to $tr (Q_X \g) tr (Q_a^2 \g)$, where $Q_X$,$Q_a$
are gauge group generators of $U(1)_X$ and of the gauge group factor $G_a$,
respectively. By an explicit check they showed that indeed this
proportionality is valid, and therefore the fields playing a role in
cancelling gauge anomalies are the twisted (linear combination of) fields $M$.
Surprisingly, the dilaton $S$ plays no role in anomaly
cancellation, since, as $tr Q_X=0$, it does not mix with the gauge fields.  
The actual computation of the coefficients $s_a$ and $\e$ (and therefore
the check of the overall normalisation in (\ref{4.4})) was performed in
\cite{abd}, coupling the theory to a background spacetime magnetic
field $B$. In this case, the relevant information is encoded in the
vacuum energy, that is expanded in powers of the magnetic field
\ba
\Lambda (B) &=& - {\cal T}- {1 \over 2} \Bigl( {\cal K} +  {\cal A}(B) +  
 {\cal M}(B) \Bigr) \nonumber \\
&\equiv & \ \Lambda_0 + {1 \over 2} \left({B\over 2\pi}\right)^2
 \Lambda_2 + {1 \over 24}\left({B\over 2\pi}\right)^4 \Lambda_4 + 
\cdots \ .  \label{4.6}
\ea
Computing the divergent piece of the vacuum energy quartic in the magnetic
field it was found, for the slightly more general case of odd $Z_N$ orientifolds,
that
\be
\Lambda_4 =  - {24 \pi^4 \over N } \sum_{k=1}^{N-1} 
( {\rm tr} Q^2 \g^k)^2 \prod_{i=1}^3 |\sin \pi k v_i|
\ \int dl \ , \label{4.7}
\ee
where the terms $({\rm tr} Q^4 \g^k)$ cancel exactly between the annulus and
the M{\"o}bius. The result (\ref{4.7}) can be easily generalized to
arbitrary orientifold vacua.
The interpretation of this term of the type $(tr F^2)^2$ is that twisted NS-NS 
fields $m_k=Im \ M_k$ (the blowing-up modes
of the orbifold) appear at tree-level in the gauge kinetic function of the gauge group 
and generate at one-loop (tree-level in the transverse, closed string picture) a
tadpole.  Notice that the
closed-string propagator for a canonically normalized scalar of mass
$M_c^2$ is
\be
\Delta_{closed} = {\pi \over 2} \int_0^{\infty} \ dl \ e^{-{\pi l \over 2}
(p^{\mu}p_{\mu} + M_c^2)} \ , \label{4.8}
\ee
with $l$ the  modulus of the cylinder. The divergence of an
on-shell propagator can thus be written formally as ${\pi\over
2}\int^\infty dl$. By using this in (\ref{4.7}), one can identify
the additional tree-level contribution to the gauge couplings. The full
tree-level expression is finally
\ba
{4 \pi^2 \over g_{a,0}^2} &=& {1\over {\ell}} +
\sum_{k=1}^{[{N-1 \over 2}]}s_{ak} m_k \label{u12}\\ 
&=& {1 \over \ell} +   \sum_{k=1}^{[{N-1 \over 2}]}
{8 \pi^2 \over \sqrt{2 \pi N}}
( {\rm tr} Q_a^2 \g^k) |\prod_{i=1}^3 \sin \pi k v_i |^{1/2} m_k \ ,
\label{4.9}
\ea 
where $\ell$ is the Hodge dual of the axion $Re S$ in (\ref{4.1}).
Analogously, the coefficient $\e$ in (\ref{4.4}), (\ref{4.5}) can be
found from the mixing between the gauge
fields and the twisted RR antisymmetric tensors, which can be computed
introducing a background magnetic field $B'$ for the abelian gauge factor $U(1)_X$.
Indeed, the quadratic term has a UV divergent part
\be
 {B'^2 \over 4N \pi^2} \sum_{k=1}^{N-1} \prod_{i=1}^3 | \sin \pi k v_i|
({\rm tr} Q_X \g^k)^2 \int \ dl \ . \label{4.10}
\ee
By using the (gauge-fixed) propagator
\be
\Delta^{\mu \nu, \rho \sigma} (k^2) \equiv <C^{\mu \nu} C^{\rho \sigma}> = 
(g^{\mu \rho} g^{\nu \sigma}- g^{\mu \sigma} g^{\nu \rho}){i \over k^2}
\ , \label{4.010}
\ee
for the (RR) antisymmetric-tensor moduli $C^{\mu \nu}$, by factorization
of (\ref{4.10}) we find at the orbifold point $m_k=0$ the coupling
\be
- {1 \over 2 \sqrt{2N \pi^3}} \sum_{k=1}^{[{N-1 \over 2}]} 
\prod_{i=1}^3 | \sin \pi k v_i|^{1 \over 2}
(-i {\rm tr} Q_X \g^k) \ \e_{\mu \nu \rho \sigma} 
C_{\mu \nu}^k F^{\rho \sigma}_X \ , \label{4.11}
\ee
confirming therefore that the dilaton, which would correspond to the $k=0$
(untwisted) contribution in (\ref{4.11}) does not mix with the anomalous gauge field.
The $U(1)_X$ gauge boson thus becomes massive, breaking spontaneously the
symmetry, even for zero VEV's of the twisted fields $m_k$. However, the
corresponding global symmetry $U(1)_X$ remains unbroken\footnote{This
can protect proton decay in phenomenological models with low-string scale.}, since the
Fayet-Iliopoulos terms vanish in the orbifold limit $m_k=0$
\cite{poppitz}. {}From (\ref{4.11}) we find 
\be
\e =  \sqrt{2 \over N \pi^3} \sum_k \prod_{i=1}^3 | \sin \pi k v_i|^{1 \over 2}
(-i {\rm tr} Q_X \g^k) \ . \label{4.12}
\ee 
The above discussion generalizes easily to other models, with additional
anomalous $U(1)_\a$ (${\a}=1 \cdots N_X)$ and corresponding linear combinations of twisted
moduli fields $M_k$ coupling to the gauge fields. In this case the gauge
kinetic function becomes
\be
f_a = S + \sum_k s_{ak} M_k \ , \label{4.13}
\ee
and (\ref{4.3}) generalizes to 
\be
V_\a \rightarrow V_\a + {i \over 2} (\Lambda_\a -{\bar \Lambda}_\a) \ , \ M_k
\rightarrow M_k + {1 \over 2} \ \epsilon_{k\a} \ \Lambda_\a \ , \label{4.14}
\ee
in an obvious notation. The cancellation of the gauge anomalies ${\rm tr} X_\a Q_a^2$
described by the coefficients $C_{{\a}a}$ asks for the generalized Green-Schwarz conditions
\be
C_{{\a}a} = {1 \over 4 \pi^2} \sum_k s_{ak} \epsilon_{k\a} \ , \label{4.15}
\ee
valid for each ${\a},a$. The gauge-invariant field combination appearing in the
K{\"a}hler  potential is $ i(M_k-{\bar M}_k)- \sum_\a \epsilon_{k\a} V_\a$ and
generates, by supersymmetry, the D-terms
\be
V_D = \sum_\a {g_{\a}^2 \over 2} (\sum_A X_A^\a K_A \Phi^A + \sum_k
\epsilon_{k\a} 
{\partial K \over \partial M_k} M_P^2 )^2 \ . \label{4.16}
\ee
A similar analysis for gravitational anomalies in orientifold models
can be found in \cite{scrucca}. 
The K{\"a}hler potential for the twisted moduli $M_k$ has not yet been computed
in orientifolds. It is however known that, close
to the orientifold point $m_k=0$, it starts with a quadratic term
$K=\sum_k M_k^{\dagger} M_k$. It is certainly important to work out the
full K{\"a}hler potential
for twisted moduli in the various known orbifold examples and to study
the consequences of (\ref{4.16}), especially for
phenomenological problems like fermion masses and mixings and
for supersymmetry breaking in models with anomalous $U(1)$ symmetries.   

\vskip 2mm
- {\large \bf Threshold corrections}
\vskip 2mm

The one-loop threshold corrections to gauge couplings can also be computed
by the same method \cite{bf,abd} and are related to the quadratic term
in (\ref{4.6}).  
The general structure of the corrections is
\be 
{4\pi^2 \over g_a^2 (\mu) }\ 
 = \ {4\pi^2 \over g_{a}^2 (\mu_0)}  + \Lambda_{2,a} \ \equiv
\ {4\pi^2 \over g_{a}^2 (\mu_0)}  \  + \
  \int_{1 / \mu_0^2}^{1 / \mu^2} {dt \over 4t} 
{\cal B}_a (t) \ \ ,  \label{4.17}
\ee
with the upper and lower limits corresponding, respectively, to the IR
and the UV regions in the open channel.  It is more convenient
technically to implement the infrared cutoff with the help of
a function $F_{\mu}(t)$. In the transverse channel,
for example, one possible choice is
\be
F_{\mu}(l)=1-e^{-{l / \mu^2}} \ , \label{4.18}
\ee
with the same cutoff for the two relevant diagrams, the annulus and the
M{\"o}bius. As explained  in \cite{bf,bachas2,bk},
the  integral must converge  in the UV   
if all the tadpoles have been  canceled globally, and if the
background field has no
component along an  anomalous $U(1)$ factor. The potential IR
divergences, on the other hand, are due to massless charged particles
circulating in the loop, so that 
\be
 {\rm lim}_{t\to\infty}\ {\cal B}_a (t) = b_a \  \label{4.19}
\ee
is the $\beta$-function coefficient of the effective field theory
at energies much lower than  the first massive threshold. 

The threshold corrections encoded in the function ${\cal B}_a (t)$ can
be computed in a generic ${\cal N}=1$ model containing D9 and D5
branes \cite{abd}. The various contributions can, in analogy with the
case of heterotic models \cite{dkl}, be grouped into two parts. The first comes from
the ${\cal N}=1$ sectors, i.e. sectors in which the orbifold operation
acts in a nontrivial way on all three complex planes. This sector gets
contributions both from the compactification lattice and from the string oscillator 
states. The second comes from the ${\cal N}=2$ sectors, i.e. sectors
in which the orbifold operation leaves one compact torus fixed and
rotates the two others. This sector gets contributions only from the
compactification lattice, more precisely from the compact torus left
fixed by the orbifold operation. The result can be understood \cite{bf,bk}
noticing that the oscillator states are non-BPS and only BPS
states can contribute to the threshold corrections from ${\cal N}=2$ sectors.
The same is true for ${\cal N}=4$ sectors, which however give
vanishing contributions. The corresponding
contribution can be computed in a closed form
and the result, obtained sending the UV cutoff to infinity, turns out to be  
\ba
\Lambda_{2,a} = {1 \over 12} \sum_i b_{ai}^{({\cal N}=2)} \int_0^{\infty} {dt \over
t} F_{\mu}(t) \sum_{(m_i^1,m_i^2)} \left[
4 \ e^{-{\pi t \over {\sqrt G_i}{\rm Im}U_i} |m_i^1+U_i m_i^2|^2} - \ 
e^{-{\pi t \over {\sqrt
G_i}{\rm Im}U_i} |m_i^1+U_im_i^2|^2} \right] \nonumber \\
=  {1 \over 3 \pi } {\sqrt G_i} \sum_i b_{ai}^{({\cal N}=2)}
\int_0^{\infty} \ dl
(1-e^{-{l \over \mu^2}}) \sum_{(n_i^1,n_i^2)} \left[
 e^{-{ {\sqrt G_i} \over \pi \ {\rm Im}U_i} |n_i^2+U_i n_i^1|^2 l} - 
\ e^{- { {\sqrt G_i} \over \pi \ {\rm Im}U_i} |n_i^2+U_i n_i^1|^2 l}
\right] ,  \label{4.20}
\ea
where $b_{ai}^{({\cal N}=2)}$
is the effective theory beta function coefficient of the corresponding ${\cal N}=2$
sector\footnote{Notice that our definition of $b_{ai}^{({\cal N}=2)}$ differs
from the definition of ref. \cite{dkl}. Our definition represents the 
contribution of the ith ${\cal N}=2$ sector to the total beta function, and therefore
equals $b_{ai}^{({\cal N}=2)}/ind$ in their notation.} and $m_i^1,m_i^2$
are Kaluza-Klein momenta of the compact torus $T^i$. 
By explicitly computing (\ref{4.20}), we find the result 
\ba
&&\Lambda_{2,a} = -{1 \over 4} \sum_i b_{ai}^{({\cal N}=2)} 
 \ln ( {\sqrt G_i} |\eta(U_i)|^4 {\rm Im}U_i \mu^2) = \nonumber \\
&& -{1 \over 4} \sum_i b_{ai}^{({\cal N}=2)} 
 \ln \left[ \left({{\rm Im}S \ {\rm Im}T_i \over {\rm Im} T_j \ {\rm Im} 
T_k}\right)^{1/2} \ |\eta(U_i)|^4 {\rm Im}U_i 
{\mu^2 \over M_I^2} \right] 
\ , \label{4.21}
\ea 
with $j \not= k \not= i$. 

The corrections (\ref{4.21}) are similar to the 
heterotic ones \cite{dkl} in the ${\rm Im T}_i \rightarrow \infty$ limit, taking
into account that on the heterotic side the complex
structure moduli have the same definition (\ref{4.1}), while
\be
S = a + i {\sqrt{G_1G_2G_3} M_H^6 \over \lambda_H^2} \ , \ T_i=b_i + i
{\sqrt G_i} M_H^2 \ . \label{4.22}
\ee
Taking the infrared limit of the threshold function ${\cal B}_a(t)$, by 
using (\ref{4.19}) one can compute the beta function of the effective
field theory in a generic $Z_N$ ${\cal N}=1$ orientifold. The result is
\ba
&b_a& = {4 \over N} \sum_{k \not=N/2} [({\rm tr} Q_a^2 \g_9^k)({\rm tr}
\g_9^k)- 2({\rm tr} Q_a^2 \g_9^{2k})]
( \prod_{i=1}^3 \sin{\pi kv_i}) \sum_{j=1}^3 {\cos{\pi kv_j} \over \sin{\pi kv_j}}
\nonumber \\
&+& {1 \over N} \sum_{i, k \not= N/2} 
({\rm tr} Q_a^2 \g_9^k)({\rm tr} \g_{5_i}^k) \cos{\pi kv_i}
+{24 \over N} {\rm tr} Q_a^2 \ , \label{4.23}
\ea
where the last contribution 
in the right-hand side of (\ref{4.23}) comes from the D9-D5 part of the
cylinder ${\cal A}_{95}^{(0)}$
and from the M{\"o}bius with the insertion of a $\theta^{N/2}$ twist
${\cal M}_{99}^{(N/2)}$. The second line in (\ref{4.23}) exists only
for even $N$ orientifolds. It can be checked case by case that
(\ref{4.23}) indeed agrees with the field-theoretical definition of the
beta-function
\be
b_a =  - 3 T_a(G) + \sum_r T_a(r) \ , \label{4.24}
\ee
where $T_a(r)$ denotes, as usual, the Dynkin index of the representation
$r$. 

The results presented above from \cite{abd} were derived for the D9
branes. They apply however, with minimal modifications, also to D5
branes in the appropriate models. For example, for a D5 brane parallel
to the third complex plane, the tree-level gauge couplings, analogous to
(\ref{u12}), read
\be
{4 \pi^2 \over g_{a,0}^2} = {\rm Im} T_3 +
\sum_{k=1}^{[{N-1 \over 2}]}s'_{ak} m_k \ . \label{4.25}
\ee
If the D5 brane is stuck to a fixed plane of the orbifold , the
coefficients $s'_{ak}$ are different from zero only for twisted fields
living at that particular fixed point (hyperplane). If the D5 brane is
moved to the bulk, the coefficients  $s'_{ak}$ are all
vanishing, since the corresponding gauge fields cannot couple to the twisted
fields that are confined to the fixed points.
 
Some more phenomenological aspects of 4d orientifolds can be found, for
example, in \cite{penn}.

\section{Type I string mechanisms for breaking supersymmetry.}

Without D-branes, in heterotic and Type II models the only known 
perturbative mechanism to spontaneously break
supersymmetry\footnote{Supersymmetry is also broken by orbifolding
the internal space. However, the resulting breaking is not soft, in the
sense that there is typically no trace of the original supersymmetry in the
resulting spectrum.} in superstrings  \cite{kp} is the string generalization of the
Scherk-Schwarz mechanism \cite{ss}. In this case, there
are tree-level gaugino masses $m_{1/2} = \omega/R$, where $R$ is the
compact radius used for the breaking and $\omega$ a parameter that is
arbitrary in field theory but
quantized in string theory. The reason for this is that the gauge fields
live in the full (bulk) 10d space and directly feel \cite{ablt} supersymmetry breaking.
Phenomenological reasons ask therefore for radii of the TeV size, a rather 
unnatural possibility in heterotic models \cite{antoniadis}, since it asks
for a string coupling of the order of $10^{32}$. 

On the other hand, the presence of D-branes in Type I models, with gauge 
fields and matter confined on them, offers 
new possibilities for breaking supersymmetry compared to the 
heterotic constructions. They can generically be classified into
three classes:

(i) Breaking by compactification.

Here there are two subclasses. In the first, the D brane under
consideration is parallel to the direction of breaking and the massless
D brane spectrum feels at tree-level supersymmetry breaking.
This situation was called ``Scherk-Schwarz'' breaking in \cite{ads1}, since
it is the analog of the heterotic constructions \cite{kp} and the
spectrum is a discrete deformation of a supersymmetric model. The
corresponding spectra have heterotic duals.
In the second class, the D brane under
consideration is perpendicular to the direction of the breaking and the massless
D brane spectrum is supersymmetric at tree-level. This was called 
``M-theory breaking'' in \cite{ads1} (also called ``Brane Supersymmetry'' in
\cite{tye}, which proposed to extend the phenomenon to the whole massive
spectrum, a situation then realized in \cite{bg}), 
since it describes in particular 
supersymmetry breaking in M-theory along the eleventh dimension
\cite{dg,aq}, as shown in Section 4. These models ask also for
the presence of antibranes (and antiorientifold planes) in the spectrum,
interacting with the branes. Supersymmetry breaking 
is transmitted by radiative corrections from the brane massive 
states or from the gravitational sector to the massless modes.

All RR and NS-NS tadpoles can be set to zero in both subclasses of these
models and we shall confine our attention to this choice.
Moreover, in these models the closed (gravitational) sector
has a softly broken supersymmetry. 

(ii) Models containing brane-antibrane systems: Brane supersymmetry
breaking. 

In these constructions \cite{ads2,au,aadds}, the closed (bulk) 
sector is exactly
supersymmetric to lowest order. We can also distinguish here between two
subclasses of models. In the first subclass, tadpole conditions, and
therefore the consistency of the theory, require the introduction of
antibranes in the system. The closed sector is supersymmetric but is
different from the standard supersymmetric one. These models contain
D9-D${\bar 5}$ tachyon-free brane configurations. In the second subclass,
the closed sector is the standard supersymmetric one. The RR tadpole
conditions ask therefore for a minimal number of D-branes and the whole
spectrum can thus be supersymmetric. However, one can consistently introduce
additional brane-antibrane pairs of the same type that break
supersymmetry. These configurations
interact and are tachyonic, but if the branes and the antibranes are
suitable separated, the tachyons can be lifted in mass. 

(iii) Breaking by internal magnetic fields

Internal background magnetic fields $H_i$ in a compact torus $T^i$ (of
radii $R_1^{(i)}$, $R_2^{(i)}$) can couple to the open
string endpoints \cite{ft}, carrying charges $q_L^{(i)}$,$q_R^{(i)}$ under $H_i$. 
Particles of different spin couple differently to the magnetic field and
acquire different masses, breaking supersymmetry
\cite{bachas1}. Defining $\pi \epsilon_i = arctan (\pi q_L^{(i)} H_i ) + 
arctan (\pi q_R^{(i)} H_i)$, the mass splittings of all string states
can be summarized by the formula   
\be  
\delta m^2 = (2 n +1) |\epsilon_i | + 2 \Sigma_i \epsilon_i \ ,
\label{7.01}
\ee
where $n$ are the Landau levels of the charged particles in the magnetic
field and $\Sigma_i$ are internal helicities. Possible values of the
magnetic fields satisfy a Dirac quantization condition 
$H_i \sim k / (R_1^{(i)} R_2^{(i)})$. For weak fields, $\epsilon_i \simeq
(q_L^{(i)}+q_R^{(i)}) H_i$ and the resulting mass splittings are
inversely proportional to the area of the magnetized torus 
$m_{SUSY}^2 \sim k / (R_1^{(i)} R_2^{(i)})$ \cite{bachas1}. The spectrum
generically contains charged tachyons coming from scalars having internal helicities
$\Sigma_i = -1$ ($\Sigma_i = 1$) for positive (negative) magnetic
field, which can however be avoided in special models. The mechanism can 
easily accomodate several magnetic fields pointing out in several
compact torii and can also be implemented in orbifold models.

Models of type (ii) and (iii) are characterized by the fact that all RR 
tadpoles
cancel, while some NS-NS tadpoles are left uncanceled. As discussed in
Section 3, the proper interpretation of the NS-NS tadpoles is that scalar
potentials are generated for appropriate NS-NS moduli fields. 
 
We now turn to a more detailed presentation of the mechanisms (i) and (ii).
For simplicity of notation, throughout this section {\it we leave implicit
the contribution of transverse bossons}, $1/\eta^8$ in the 9d and 10d
models and $1/\eta^4$ in the 6d model discussed in the third paragraph.
 
\vskip 2mm
- {\large \bf Breaking by compactification I: direction parallel 
to the brane (Scherk-Schwarz breaking)}
\vskip 2mm

These models are constructed performing a Scherk-Schwarz deformation
in the closed sector on the Kaluza-Klein momentum states $m \rightarrow m +
\omega$. The brane under consideration is parallel to the breaking
direction, i.e. it has associated momentum (KK) modes which feel a similar
breaking. All these models contain the geometrical objects present
in supersymmetric models, in particular 32 D9 branes and 32 O$9_{+}$
planes, and are discrete deformations of supersymmetric models. 
 
The simplest such example is provided by a 9d model which, in
the closed sector, can be described as a Type OB/g orbifold, the
orbifold operation being $g=-(-1)^{G_L} (-1)^n$, where $G_L$ is the (left) world-sheet 
fermion number. Alternatively, after a redefinition of the radius $R
\rightarrow 2R$, the model can be described as the orbifold
$IIB/g=(-1)^F I$, where $F=F_L+F_R$ is the spacetime fermion number and
$I$ is the shift $I: X_9 \rightarrow X_9 + \pi R$, acting on the states
as $(-1)^m$. The closed spectrum is tachyon free for $R \ge M_I^{-1}$
and supersymmetry is restored in the $R \rightarrow \infty$ limit
\footnote{Some of the models in this paragraph were studied also in 
\cite{blum}.}

The relevant amplitudes to consider, in the notations of (\ref{3.1}), are  
\ba
K_1 &=& \frac{1}{2}  \ (V_8 - S_8) \ \sum_m\ P_m \ , \nonumber \\
A_1 &=& \frac{n_1^2 + n_2^2}{2} \sum_m ( V_8 P_{m} - S_8 P_{m +
  1/2} ) + n_1 n_2 \sum_m (V_8
P_{m + 1/2} - S_8 P_{m} ) \ , \nonumber \\ 
M_1 &=& - \frac{ n_1 + n_2 }{2} \sum_m ( {\hat
V}_8 P_{m} - {\hat S}_8 P_{m + 1/2} ) \ , \label{7.1}
\ea
where, as usual, $n_1,n_2$ denote Chan-Paton charges and
$P_m$($P_{m+1/2}$) denote integer (half-integer) momentum states. The
spectrum corresponds to a
family of gauge groups $SO(n_1) \times SO(n_2)$, with
$n_1+n_2=32$ fixed by the tadpole conditions. For integer momentum levels, the
spectrum consists of  vectors\footnote{In 9d the 10d vector actually comprises a
vector and a scalar.} in the representations 
$({\bf n_1(n_1\!\!-\!\! 1)/2},{\bf 1})$ + $({\bf 1},{\bf n_2(n_2\!\! -\!\! 1)/2})$ and 
fermions in the representation $({\bf n_1},{\bf n_2})$. On
the other hand, for half-integer levels, the spectrum consists of fermions in the
$({\bf n_1(n_1\!\!-\!\! 1)/2},{\bf 1})$ + $({\bf 1},{\bf n_2(n_2\!\!-\!\!1)/2})$ 
and vectors in the $({\bf n_1},{\bf n_2})$.

This model can easily be understood as a discrete deformation by the
fermion number $(-1)^F$ of the supersymmetric model described by
\ba 
K_1 &=& {1 \over 2} (V_8-S_8) \sum_m P_m \ , \nonumber \\  
A_1 &=& (V_8-S_8) \sum_m \left (
\frac{n_1^2 + n_2^2}{2} P_{m} + n_1n_2 P_{m+1/2}
\right ) \ , \nonumber \\ 
M_1 &=& - \frac{ n_1 + n_2 }{2} ( V_8 - S_8) \sum_m P_{m} \ ,
\label{7.2} 
\ea 
obtained breaking the compactified $SO(32)$ Type I model with the
Wilson line $W=(I_{n_1},-I_{n_2})$. The fact that (\ref{7.1}) is a
discrete deformation of (\ref{7.2}) reflects the spontaneous
character of the breaking, which disappears in the decompactification limit
$R \rightarrow \infty$. Moreover, in this model a scalar potential is
induced  that for large radius behaves as $1/R^9$, and thus
dynamically tends to decompactify the theory to 10d and to restore supersymmetry.

A large class of models can be constructed
compactifying on orbifolds \cite{ads1,adds}, with different patterns of
supersymmetry breaking: ${\cal N}=4 \rightarrow {\cal N}=0$, 
 ${\cal N}=4 \rightarrow {\cal N}=2$,   ${\cal N}=4 \rightarrow {\cal
N}=1$,  ${\cal N}=2 \rightarrow {\cal N}=0$ and
${\cal N}=2 \rightarrow {\cal N}=1$.

\vskip 2mm
- {\large \bf Breaking by compactification II: direction orthogonal 
to the brane (M-theory breaking or Brane Supersymmetry) }
\vskip 2mm

The starting point in constructing these models is a shift $n
\rightarrow n + \omega$ in the winding modes of the closed sector of the
parent Type IIB superstring. The brane under consideration is
perpendicular to the direction of the breaking. This is easy to
vizualize in the T-dual picture, where windings shifts become standard
momentum shifts, but the compact direction becomes perpendicular to the
brane. A simple prototype is again provided by a 9d example,
that in the parent IIB theory is simply obtained by interchanging
the KK momenta with the windings in the
(Scherk-Schwarz) breaking. The open sector, on the other hand, is
completely different, due to the momentum/winding asymmetry in the open
sector resulting from the standard $\Omega$ projection.

The relevant amplitudes are in this case 
\ba
K_2 &=& \frac{1}{2}  \ (V_8 - S_8) \ \sum_m P_{2m} + \frac{1}{2} \ (O_8 - C_8) \
\sum_m P_{2m+1} \ , \nonumber \\  
A_2 &=&  
\frac{N_1^2 + N_2^2}{2} ( V_8 - S_8 ) \sum_m P_{m} + N_1 N_2 (O_8 -
C_8 ) \sum_m P_{m+1/2} 
\ , \nonumber
\\ M_2 &=& - \frac{ N_1 + N_2 }{2} {\hat V}_8 \sum_m P_{m} +
\frac{ N_1 + N_2 }{2} {\hat S}_8 \sum_m (-1)^m P_{m} \  \label{7.3}
\ea 
and the tadpole conditions are $N_1=N_2=16$, and are satisfied also
in the $R \rightarrow 0$ limit. Notice that the massless open spectrum
is supersymmetric, since
\be 
A_2 + M_2 =  
\frac{N_1^2 + N_2^2}{2} ( V_8 - S_8 ) -
\frac{ N_1 + N_2 }{2} ( {\hat V}_8 - {\hat S}_8 ) + {\rm massive} \ . 
\label{7.4}
\ee 
The open spectrum is actually supersymmetric for all even momenta and describes a
vector and a spinor in the adjoint of $SO(16) \times SO(16)$. On the other hand, for 
odd momenta the vector is again in the adjoint, while the spinor is in the symmetric
representations $({\bf 135},{\bf 1})$$+$$2({\bf 1},{\bf 1})$$+$$({\bf 1},{\bf 135})$. 
Finally, there are scalars and spinors in the $({\bf 16},{\bf 16})$ representation
with half-integer momenta.

The duality arguments of Section 4 associate the closed sector of this Type-I model,
after a T-duality, to a Scherk-Schwarz deformation affecting the momenta of the Type-IIA
string. In the corresponding Type-I$^\prime$ representation, however, the open strings
end on D8-branes perpendicular to the direction responsible for the breaking of
supersymmetry. Therefore, as we have just seen, all open string modes with even
windings, and in particular the massless ones, are unaffected. The soft
nature of this breaking is less evident than in the previous example, but
the very soft nature of the radiative corrections was shown in
\cite{ads1} and is related to the local tadpole cancellation properties
of the model.
Resorting again to the duality arguments of Section 4, it is clear that
this breaking corresponds to a non-perturbative phenomenon on the heterotic side and
realizes the Scherk-Schwarz deformation along the eleventh coordinate of M theory.
An additional nice argument concerning the relation to M-theory is that,
in addition to global tadpole conditions, the model satisfies {\rm
local} tadpole conditions in the breaking direction and therefore 
has a consistent $R \rightarrow 0$ limit, equivalent in M-theory
language to $R_{11} \rightarrow \infty$. 

It is interesting to notice from (\ref{7.3}) that the model actually
contains 16 D9 branes and 16 D${\bar 9}$ antibranes. In the T-dual
picture ($R_{\perp}=1/RM_I^2$) the 16 D8 branes are at the origin
$y= 0$ and the 16 D${\bar 8}$ branes are at the orientifold fixed point
$y=\pi R_{\perp}$. To be more precise, there are 16 D8 branes
and 16 O$8_{+}$ planes at $y= 0$ and  16 D${\bar 8}$ antibranes
and 16 O${\bar 8}_{+}$ antiplanes at $y=\pi R_{\perp}$, such that
supersymmetry is still preserved in the vicinity of each fixed points
$y= 0, \pi R_{\perp}$ where local tadpole conditions are satisfied
as well. In order to substantiate this picture and to better define the
notion of antibranes and antiorientifold planes, let us describe in more
detail the effective lagrangian describing the interactions of SUGRA fields
with branes/orientifold planes in this model. This can be easily done
writing the amplitudes (\ref{7.3}) in the tree-level (closed) channel
\ba
{\tilde K}_2 &=& \frac{2^5}{2}  \ (V_8 \sum_n W_{2n} - S_8 \sum_n W_{2n+1})
 \ , \nonumber \\  
{\tilde A}_2 &=& \frac{1}{2^6} \sum_n  
\{ [N_1 + (-1)^n N_2]^2 V_8 - [N_1 - (-1)^n N_2]^2 S_8 \} W_n 
\ , \nonumber
\\ {\tilde M}_2 &=& -  (N_1 + N_2)  ( {\hat V}_8 \sum_n W_{2n} -
 {\hat S}_8 \sum_n W_{2n+1} ) \ .  \  \label{7.05}
\ea 

The effective lagrangian can be found by writing the vacuum energy
(without the torus contribution)
\be
{\tilde K}_2 + {\tilde A}_2 + {\tilde M}_2 = {1 \over 64} \{
[N_1-16+(-1)^n (N_2-16)]^2 V_8 - [N_1-16-(-1)^n (N_2-16)]^2 S_8 \} \sum_n W_n
\ . \label{7.06}
\ee 
By factorization the effective bulk-D8/O8 action is easily found and reads
\be
S \!=\! \int \ d^{10} x \{ \sqrt{G} \ {\cal L}_{SUGRA} - (N_1\!-\!16) T_8 
(\sqrt {G} e^{-\Phi} \!+\! A_9) \delta (y)- \nonumber \\
(N_2\!-\!16) T_8  (\sqrt {G} e^{-\Phi} \!-\! A_9) \delta (y\!-\! \pi R) \}
\ , \label{7.07}
\ee
where
\be
{\cal L}_{SUGRA} \!=\! {1 \over 2 k_{10}^2} \{ e^{-2 \Phi}
[R + 4 (\partial \Phi)^2 ] - {1 \over 2 \times 10 ! } F_{10}^2 \} \label{7.08}
\ee   
is the bosonic Type I supergravity action. Notice that the
corresponding fermionic fields are massive, due to the breaking of supersymmetry
by compactification.

In (\ref{7.07})- (\ref{7.08}),
$A_9$ is the RR 9-form coupling to the D8 (D${\bar 8}$) and  O$8_{+}$ (O${\bar 8}_{+}$) 
systems and $F_{10}$ is its corresponding field strength, $k_{10}$ defines
the 10d Planck mass, $T_8$ is the D8 brane tension and $G$ is the 
10d metric. The change in sign in the RR part of
the last term in (\ref{7.07}) confirms precisely that the model contains
16 D8/O8 at $y=0$ and 16 D${\bar 8}$/O${\bar 8}$ at $y=\pi R$.

Branes and antibranes attract each
other and for sufficiently small (large) distances $R_{\perp}$ ($R$), a tachyon appears
\cite{bas}, as is easily seen also in (\ref{7.3}). The configuration
is therefore in principle unstable, although a more detailed analysis of
the potential for the radius is needed in order to settle this question.
The full vacuum energy for large $R_{\perp}$
can be estimated to be (after a rescaling $l \rightarrow l/R^2$) \cite{ads1}
\be
\Lambda \sim - {1 \over R_{\perp}^9} - M_I^9 R_{\perp} \int_0^{\infty} dl \
e^{-\pi R_{\perp}^2M_I^2 l} \sum_n (-1)^n e^{-\pi l n^2 \over 4}
\nonumber \\
\simeq \ - {1 \over R_{\perp}^9} - {\pi \over 8}M_I^9 e^{- 2 \pi
R_{\perp}M_I} \ ,
\label{7.04}
\ee
where numerical coefficients were set to one in (\ref{7.04}). In
(\ref{7.04}), the first term
is the one-loop torus contribution piece proportional to $1/R_{\perp}^9$
and the second attractive term comes from (anti)branes-orientifold
contributions proportional to
$exp(-R_{\perp}M_I)$, where the exponential supression appears due to
the local tadpole property of the model. There the induced potential for the
radius does not seem to have
a minimum. However, this conclusion cannot be reliably drawn from the
approximate expression  (\ref{7.04}), and new effects can certainly
appear, as for instance tachyon condensation, which could render the configuration stable.

To the best of our knowledge,
this is the first construction of a perturbative Type I model containing 
simultaneously branes and antibranes\footnote{Notice, however, that models
with branes and antibranes were constructed previously 
in orientifolds of Type O models, \cite{bs,augusto}.}. 
Other models with different patterns of supersymmetry breaking can be
constructed compactifying on orbifolds \cite{ads1,adds} and some duality relations
with heterotic constructions were investigated in \cite{gregori}.
  
\vskip 2mm
- {\large {\bf Brane Supersymmetry Breaking I: non-BPS stable configurations}}
\vskip 2mm

As already mentioned, the main idea of Brane Supersymmetry Breaking
is to put together branes and antibranes. In general, such 
systems contain (open string) tachyons stretched between branes and 
antibranes, that reflect the attractive force between them. However, in
some models the consistency of the theory (the RR tadpole conditions) asks for
D9-D${\bar 5}$ stable, tachyon-free non-BPS systems.  
The simplest model of this type we are aware of is the $T^4/Z_2$
orbifold model with orientifold projection in the twisted sector with a
flipped sign \cite{ads2,au}.
 
Following \cite{bs}, let us introduce the convenient combinations of 
$SO(4)$ characters
\ba 
Q_o = V_4O_4-C_4C_4 \ , \qquad Q_v = O_4V_4-S_4S_4 \ , \nonumber\\ 
Q_s = O_4C_4-S_4O_4 \ , \qquad Q_c = V_4S_4-C_4V_4 \ , \label{7.5} 
\ea 
which describe in 6d at massless level the propagation of a vector multiplet ($Q_o$),
of a hypermultiplet ($Q_v$), of half of a hypermultiplet ($Q_s$) and
where $Q_c$ contains only massive particles.
With these notations, there are two consistent inequivalent
choices for the Klein bottle, described by 
\be 
K_3 = \frac{1}{4} \biggl\{ ( Q_o + Q_v ) ( \sum_{m_i} P_{m_i}^{(4)} + 
\sum_{n_i} W_{n_i}^{(4)} ) + 2 \epsilon
\times 16 ( Q_s + Q_c ){\biggl(\frac{\eta}{\theta_4}\biggr)}^2 \biggr\}
\ , \label{7.6}
\ee  
where $P^{(4)}$ ($W^{(4)}$) denotes the momentum (winding) lattice sum and
$\epsilon = \pm 1$. For both choices of
$\epsilon$, the closed string spectrum has $(1,0)$ supersymmetry, but
the two resulting projections are quite different.  The usual choice
($\epsilon=1$) \cite{bs,gp} leaves 1 gravitational multiplet, 1
tensor multiplet and 20 hypermultiplets, while $\epsilon=-1$ leaves 1 
gravitational multiplet, 17 tensor  multiplets and 4 hypermultiplets. 
The case  $\epsilon=-1$ is our primary focus here, since in this case
the flipped $\Omega$ action in the closed twisted sector is equivalent
with the replacement O$5_{+} \rightarrow O5_{-}$. However, O$5_{-}$ RR
charge is positive and the RR tadpole conditions require for consistency
the presence of 32 D${\bar 5}$ (anti)branes in the spectrum.

The annulus amplitude is simpler to understand in the transverse channel
\ba
{\tilde A}_3 &=& \frac{2^{-5}}{4} \biggl\{ (Q_o + Q_v) \biggl( N^2 v
\sum_{n_i} W_{n_i}^{(4)}  +
\frac{D^2 \sum_{m_i} P_{m_i}^{(4)}}{v} \biggr) + 2 N D (Q'_o - Q'_v) 
{\biggl(\frac{2 \eta}{\theta_2}\biggr)}^2  \label{7.7} \\ &+&  
16 (Q_s + Q_c) \biggl( R_N^2 + R_D^2 \biggr){\biggl(\frac{
\eta}{\theta_2}\biggr)}^2 + 8 R_N R_D ( V_4 S_4 - O_4 C_4 - S_4 O_4 +
C_4 V_4 ){\biggl(\frac{
\eta}{\theta_3}\biggr)}^2
\biggr\} \ , \nonumber
\ea  
where we introduced the Chan-Paton multiplicities $N,R_N,D,R_D$ and the primed
characters \cite{ads1} are related by a
chirality change $S_4 \leftrightarrow C_4$ to the unprimed ones
defined in eq. (\ref{7.5}). For the spacetime part, this simply means
a change of fermion chirality, as shown in more detail below.
Notice that (\ref{7.7}) is identical in
structure to the corresponding amplitude for the supersymmetric
$T^4/Z_2$ Type I orbifold \cite{bs},
\cite{gp}, except that in the D9-D${\bar 5}$ sector the signs of the RR
terms are reversed, in order to correctly take into account the
(negative) charge of D${\bar 5}$ (anti)branes. 
The direct-channel annulus is obtained by an
S-transformation, and reads
\ba 
A_3 &=& \frac{1}{4} \biggl\{ (Q_o + Q_v) ( N^2 \sum_{m_i}
P_{m_i}^{(4)} + D^2 \sum_{n_i} W_{n_i}^{(4)} ) + 
2 N D (Q'_s + Q'_c) {\biggl(\frac{\eta}{\theta_4}\biggr)}^2 
\label{7.8}
\\ &+& (R_N^2 + R_D^2) (Q_o - Q_v) {\biggl(\frac{2
\eta}{\theta_2}\biggr)}^2 + 2 R_N R_D ( - O_4 S_4 - C_4 O_4 + V_4 C_4 +
S_4 V_4 ){\biggl(\frac{
\eta}{\theta_3}\biggr)}^2 \biggr\} \ . \nonumber
\ea  
In (\ref{7.8}), the 9${\bar 5}$ ($ND$) term is similar to the corresponding one
in the supersymmetric $T^4/Z_2$ orientifold, but with fermions of
flipped 6d chirality, the 9${\bar 5}$ term with orbifold insertion $R_N R_D$ is
nonsupersymmetric (but does not contribute to the vacuum energy since 
twisted tadpoles ask for $R_N = R_D=0$, see below), while the other
terms in (\ref{7.8}) are precisely the supersymmetric ones.
 
Finally, the M{\"o}bius amplitude describes the
propagation between branes and  orientifold planes (holes and
crosscaps). All D9-O$9_{+}$
terms, the D${\bar 5}$-O$5_{-}$ terms in the R-R sector and the  D${\bar
5}$-O$9_{+}$ terms in the NS-NS sector are as in the standard $T^4/Z_2$
orientifold, while the signs of all D9-O$5_{-}$ terms, of the  D${\bar
5}$-O$5_{-}$  terms in the NS-NS sector and of the  D${\bar 5}$-O$9_{+}$ terms
in the R-R sector are inverted. In particular, this implies that the
M{\"o}bius amplitude  breaks supersymmetry at tree level in the
D${\bar 5}$ sector, an effect felt by all open-strings ending on the
D${\bar 5}$ branes. The direct (open string) M{\"o}bius amplitude is
\ba  
&\!\!\!\!M_3\!\!\!\!& = - \! \frac{1}{4} \biggl\{ N \sum_{m_i}
\! P_{m_i}^{(4)} ( \hat{O}_4 \hat{V}_4 \!+\! \hat{V}_4 \hat{O}_4 \!-\! 
\hat{S}_4 \hat{S}_4 \!-\! \hat{C}_4
\hat{C}_4 ) \!-\!  D \! \sum_{n_i} W_{n_i}^{(4)} ( \hat{O}_4
\hat{V}_4  \!+\! \hat{V}_4 \hat{O}_4 \!+\! \hat{S}_4 \hat{S}_4 \!+\! \hat{C}_4
\hat{C}_4 ) \nonumber \\ &-&\!\!\!\!\! N( 
\hat{O}_4 \hat{V}_4 \!-\! \hat{V}_4 \hat{O}_4 \!-\! \hat{S}_4 \hat{S}_4
\!+\! \hat{C}_4 \hat{C}_4 )\left(
{2{\hat{\eta}}\over{\hat{\theta}}_2}\right)^2  \!\!+\! D( \hat{O}_4
\hat{V}_4 \!-\! \hat{V}_4 \hat{O}_4 \!+\! \hat{S}_4 \hat{S}_4
\!-\! \hat{C}_4 \hat{C}_4)\left(
{2{\hat{\eta}}\over{\hat{\theta}}_2}\right)^2  \biggr\} \ ,
\label{7.9} 
\ea  
and parametrizing the Chan-Paton charges as
$N=n_1+ n_2$, $D=d_1+ d_2$, $R_N=n_1- n_2$, $R_D=d_1- d_2$,
the RR tadpole conditions $N=D=32,R_N=R_D=0$
determine the gauge group $[ SO(16) \times
SO(16) ]_9 \times  [ USp(16) \times USp(16) ]_{\bar 5}$.

The $99$ spectrum is supersymmetric, and comprises the (1,0)
vector multiplets for the $SO(16) \times SO(16)$ gauge group and a
hypermultiplet in the representations ${\bf\! (16,16,1,1)}$ of the
gauge group. On the other hand, the ${\bar 5} {\bar 5}$ DD spectrum is
not supersymmetric, and contains, aside from the gauge vectors of $[
USp(16) \times USp(16) ]$, quartets of scalars in the ${\bf (1,1,16,16)}$,
right-handed Weyl fermions in the $ {\bf (1,1,120,1)}$ and in the ${\bf
(1,1,1,120)}$, and left-handed Weyl fermions in the
${\bf (1,1,16,16)}$. Finally, the ND sector is also non supersymmetric,
and comprises doublets of scalars in the ${\bf (16,1,1,16)}$ and in
the 
${\bf (1,16,16,1)}$, together with additional (symplectic)
Majorana-Weyl fermions in the ${\bf (16,1,16,1)}$ and ${\bf
(1,16,1,16)}$. These Majorana-Weyl fermions are a peculiar feature of
six-dimensional spacetime,  where the fundamental Weyl fermion, a
pseudoreal spinor of
$SU^*(4)$, can be subjected to an additional Majorana condition,  if
this is supplemented by a conjugation in a pseudoreal representation
\cite{wsi}. In this case, this is indeed possible, since the ND
fermions are valued in the fundamental representation of $USp(16)$.

{}From the D9 brane point of view, 
the diagonal combination of the two $USp(16)_{\bar 5}$ gauge groups
acts as a global symmetry. This corresponds
to having complex scalars and symplectic Majorana-Weyl fermions in the
representations $16\times[{\bf (16,1)+(1,16)}]$ of the D9 gauge group.
As a result, the bose-fermi degenerate ND spectrum looks effectively 
supersymmetric, and indeed all $9 \bar{5}$ terms do not
contribute to the vacuum energy.  However, as in the 
6D temperature breaking discussed in \cite{ads1},
the chirality of the fermions in $Q'_s$ is not the one
required by 6D supersymmetry.   This chirality flip 
is a peculiar feature of models with branes and antibranes. 

Brane-antibrane interactions have been discussed recently in the
literature in the context of stable non-BPS systems
\cite{sen}. Our results for the D9-D${\bar 5}$ system, restricted to
the open sector, provide particular examples of Type-I vacua
including non-BPS stable configurations of BPS branes with vanishing interaction energy for
all radii, as can be seen from the vanishing of the ND
annulus amplitude.

The breaking of supersymmetry gives rise to a vacuum energy localized
on the D$\bar{5}$ branes, and thus to a tree-level potential for the 
NS moduli, that can be extracted from the corresponding uncancelled NS
tadpoles. A simple
inspection shows that the only non-vanishing ones correspond to the NS
characters
$V_4O_4$ and $O_4V_4$ associated to the 6D dilaton $\phi_6$ and to the
internal volume $v$:
\be  {2^{-5}\over 4}\left\{ \left( (N-32){\sqrt v}+{D+32 \over{\sqrt
v}}\right)^2 V_4O_4 +\left( (N-32){\sqrt v}-{D+32 \over{\sqrt
v}}\right)^2 O_4V_4
\right\} \, . \label{7.10}
\ee  
Using factorization and the values $N=D=32$ needed to cancel the
RR tadpoles, the potential (in the string frame) is:
\be  V_{\rm eff}=c{e^{-\phi_6}\over{\sqrt v}}=ce^{-\phi_{10}} ={c\over
g_{\rm YM}^2}\, , \label{7.11}
\ee  
where $\phi_{10}$ is the 10D dilaton, that determines the
Yang-Mills coupling
$g_{\rm YM}$ on the D${\bar 5}$ branes, and $c$ is some {\it positive}
numerical  constant. The potential (\ref{7.11}) is clearly localized on
the D${\bar 5}$ branes, and is positive. This can be understood by noticing that
the O$9_{+}$  plane contribution to vacuum energy is negative  and exactly 
cancels for $N=32$. This fixes
the D${\bar 5}$ brane contribution to the vacuum energy, that is thus 
positive, consistently with the interpretation of this mechanism as
global supersymmetry breaking. The potential (\ref{7.11}) has the usual
runaway behavior in the dilaton field, as expected by general arguments.

\vskip 2mm
- {\large {\bf Brane Supersymmetry Breaking II: brane-antibrane pairs }}
\vskip 2mm

In the previous example, the breaking of supersymmetry on the antibranes 
is directly enforced by the consistency of the model, which contains D9
branes and D${\bar 5}$ antibranes, a (non-BPS ) stable configuration
without tachyons. 
Somewhat different scenarios have been recently proposed in 
\cite{sugimoto,au,aadds}.
In the resulting models, a supersymmetric open sector is deformed 
allowing for the simultaneous 
presence of branes and antibranes of the same type. Whereas tadpole
conditions only fix the total RR charge, the option of saturating it by 
a single type of D-brane, whenever available, stands out as the only one 
compatible with space-time supersymmetry.
However, if one relaxes this last condition, there are no 
evident obstructions to
considering vacuum configurations where branes and antibranes with
a fixed total RR charge are simultaneously present. 

Branes and antibranes of the same type are mutually interacting systems.
The brane-antibrane vacuum energy in 10d, for concreteness, can be 
summarized by comparing
the corresponding annulus amplitude with the usual brane-brane one
\ba
\ \ \ \ \ \ \ \ \ \ \ \ &{\rm open \ channel}& \ \ \ \ 
{\rm closed \ channel} \  \nonumber \\
{\rm brane-brane} : &V_8-S_8&  \ \ \ \ \ V_8-S_8 \ , \nonumber \\
{\rm brane-antibrane} : &O_8-C_8& \ \ \ \ \ V_8+S_8 \ . \label{7.012}
\ea
In the closed channel, the sign change in the RR ($S_8$) term simply
reflects the flipped (positive) RR charge of antibranes corresponding
to branes. In the open channel, this reflects into the propagation
of a charged open string tachyon ($O_8$) and of a fermion ($C_8$) of
opposite chirality compared to the brane-brane spectrum.
  
The rules for 
constructing this wider class of models can be simply presented referring 
to a ten-dimensional example \cite{sugimoto,aadds} that requires an open sector
with a {\it net} number
of 32 (anti)branes in order to cancel the resulting RR tadpole.  The closed
part in these models is the usual supersymmetric one. The open
amplitudes, on the other hand, are 
\ba
A_4 = \frac{N_{+}^{2} + N_{-}^{2}}{2}\, (V_8 - S_8) + N_+ N_- \,
(O_8 - C_8) \ , \nonumber \\
M_4 = \pm \frac{1}{2} (N_+ + N_-) \, \hat V _8 + \frac{1}{2} (N_+ -
N_-) \, \hat S _8 \ , \label{7.12}
\ea
where $N_+$ and $N_-$ count the total numbers of D9 and 
D${\bar 9}$ branes. The strings streched between D9-D${\bar 9}$ branes,
with Chan-Paton factor $ N_+ N_{-}$ in the annulus, reflect the opposite
GSO projections for open
strings stretched between two D-branes of the same type (99 or
${\bar 9}{\bar 9}$) and of different types ($9{\bar 9}
$ or ${\bar 9} 9$) \cite{bas,sen}. While the former yields
the supersymmetric Type I spectrum, the latter eliminates the vector and 
its spinorial superpartners, and retains the
tachyon and the spinor of opposite chirality. As a
result, supersymmetry is broken and an instability, signaled
by the presence of the tachyonic ground state, emerges.
The M{\"o}bius amplitude now involves naturally an undetermined sign for $V_8$, whose
tadpole is generally incompatible with the one of $S_8$, and
is to be relaxed. Together with $A$, the two signs lead to
symplectic or orthogonal gauge groups with $S_8$ fermions in (anti)symmetric 
representations and tachyons and $C_8$ fermions in bi-fundamentals. The
minus sign corresponds to O$9_{+}$ planes and the plus sign to O$9_{-}$
planes. In this last case, however, $N_{+}$ describes antibranes
while $N_{-}$ describes branes.

In these ten-dimensional models, the only way to eliminate the tachyon
consists in introducing only D9-branes (or only D${\bar 9}$ branes). Depending on
the signs in the M{\"o}bius amplitude, one thus recovers either
the SO(32) superstring or the USp(32) model of \cite{sugimoto}. 

On the other hand, more can be done if one
compactifies the theory on some internal manifold. In this case, one
can introduce Wilson lines (or, equivalently, separate the branes) 
in such a way that in the open strings 
stretched between separated D9 and D$\bar 9$ branes the tachyon 
becomes massive. 
It is instructive to analyze in detail the simple case of circle 
compactification.
A Wilson line $W=(I_{N_{+}}, -I_{N_{-}})$ affects the annulus amplitude, that in the
direct-channel now reads
\be
A_4 = \frac{N_{+}^{2} + N_{-}^{2}}{2} \, (V_8 - S_8)\, \sum_m P_m +
N_+ N_- \,(O_8 - C_8)\, \sum_m P_{m+1/2} \ , \label{7.13}
\ee
where $P_{m+1/2}$ denotes a sum over $\frac{1}{2}$-shifted momentum states.
In this case both the tachyon and the $C_8$ spinor are lifted in mass.
The open sector is completed by the M{\"o}bius amplitude 
\be
M_4 = \frac{1}{2} \left[ \pm (N_+ + N_- )\, \hat V_8 + (N_+ -
N_-) \, \hat S_8 \right]  \sum_m P_m \ \label{7.14}
\ee
and at the massless level comprises gauge bosons in the adjoint of
${\rm SO}(N_+) \times {\rm SO}(N_-)$ (or  
${\rm USp}(N_+) \times {\rm USp}(N_-)$, depending on the sign of
$\hat V _8$ in $M$) and $S$ spinors in (anti)symmetric representations.
We display here  the effective action of branes/O-planes with
supergravity fields in the case of a minus sign in (\ref{7.14})
\ba
&&S = \int \ d^{10} x \{ \sqrt{G} \ {\cal L}_{SUGRA} - 
(N_{+}-16) T_8 (\sqrt {G} e^{-\Phi} + A_9) \delta (y) 
\nonumber \\
&&- T_8 [ \sqrt {G} (N_{-}-16) e^{-\Phi} - (N_{-}+16) A_9 ] \delta (y-\pi
R) \} \ , 
\label{7.15}
\ea
where ${\cal L}_{SUGRA}$ is the 10d supergravity lagrangian. Notice in
(\ref{7.15}) the peculiar interaction of D${\bar 8}$ antibranes with 
O-planes.

Interestingly enough, it was recently realized \cite{aadds} that the sign flip of the
RR charge of the O5 plane in some 4d orientifold models with no
supersymmetric solution ($Z_2 \times Z_2$ with discrete torsion or $Z_4$)
defines consistent models with supersymmetry on branes and broken
supersymmetry on antibranes. Indeed, for example in the $Z_4$ model the
supersymmetric construction leads to a tadpole in the Klein bottle (of the type
$1/V_3$) which cannot be cancelled by the existing (D9 and D5) branes in
the model. By changing the charge of the O5 plane (or, equivalently, the
$\Omega$ projection in the closed string $Z_2$ twisted sector), this
tadpole becomes massive and the open spectrum can be consistently
constructed without further obstructions.
Another interesting feature of brane-antibrane systems is the presence
of mutual forces. It was suggested in \cite{au} and was explicitly shown
in \cite{aadds}, that by adding D9-D${\bar 9}$ and D5-D${\bar 5}$ pairs,
scalar potentials are generated by the NS-NS tadpoles such that some or
all radii of the compact space are stabilized. These models present some 
phenomenological interest, as will be seen later in this review. A generic feature of 
all models with supersymmetry broken on a collection of (anti)branes, however, is that
there is a dilaton tadpole, which means that the correct background is not the Minkowski
one with maximal symmetry. Identifying the correct background is
therefore an important step
in unravelling the properties of these models. A step forward was
recently made in \cite{dm4}, where
a background (with $SO(9)$ symmetry) of the 10d Type I model with gauge
group $USp(32)$ was found. 
The tenth coordinate turns out to be spontaneously compactified, so
that the length of the tenth
dimension is finite. The geometry of the background is $R^9 \times
S^1/Z_2$, with the zero-mode  
of the graviton localized near one of the boundaries of the interval.

All the models with broken supersymmetry discussed in this Section face
the problem of the cosmological constant \cite{weinberg}. It seems difficult to find
models with naturally zero (or very small) vacuum energy. There exist 
however explicit perturbative Type II examples and also Type I
descendants for the class of models with supersymmetry breaking by
compactification \cite{ks}. To date, there are no similar 
models exhibiting the phenomenon of brane supersymmetry breaking.  
    
\section{Millimeter and TeV$^{-1}$ large extra dimensions}

The presence of branes in Type I, Type II, Type O strings and M-theory opens new
perspectives for particle physics phenomenology. Indeed, we already saw
in (\ref{1.12}) that in Type I strings the string scale is not
necessarily tied to the Planck scale. In view of the new D-brane
picture, let us take a closer look at the simplest example of compactified Type I
string, with only D9 branes present. We found in (\ref{1.12}) that the
string scale can be in the TeV range if the string coupling is
extremely small, $\lambda_I \sim 10^{-32}$. Then, from the second
relation (\ref{1.11}) one can see that in this case the compact volume is
very small $V M_I^6 \sim 10^{-32}$. Let us split the compact volume into two
parts, $V=V^{(1)} V^{(2)}$, where $V^{(1)}$, of dimension
$6-n$, is of order one in string units and $V^{(2)}$, of dimension
$n$, is very small. The Kaluza-Klein states of the brane fields
along $V^{(2)}$ are much heavier than the string scale and therefore
are difficult to excite.
The physics is then better captured in this case performing
T-dualities along $V^{(2)}$, which read
\be
\lambda_I^{'} = {\lambda_I \over V^{(2)}M_I^n} \quad , \quad
V_{\perp} = {1 \over  V^{(2)} M_I^{2n}} \ . \label{8.1}
\ee
In the T-dual picture, neglecting
numerical factors, the relations (\ref{8.1}) become
\be
M_P^2 \sim {1 \over \alpha_{GUT} \lambda_I^{'}} V_{\perp} M_I^{2+n} \quad ,
\quad 
{1 \over  \alpha_{GUT}} \sim {V_{||} M_I^{6-n} \over \lambda_I^{'}} 
\ , \label{8.2} 
\ee
where for transparency of notation we redefined $V^{(1)} \equiv  V_{||}$.
After the $n$ T-dualities, the D9 brane becomes a D(9-n) brane, since
the T-dual winding modes of the bulk (orthogonal) compact space are very heavy
and therefore the brane fields cannot propagate in the bulk. As seen
from (\ref{8.2}), for a very large bulk volume the string
scale can be very low $M_I << M_P$. The geometric picture here is that
we have a D-brane
with some compact radii parallel to it, of order $M_I^{-1}$, and
some very large, orthogonal compact radii.
In particular, if the full compact space is
orthogonal to the brane ($n=6$), from (\ref{8.2}) the T-dual string
coupling is fixed by the unified coupling $\lambda_I^{'} \sim \alpha_{GUT}$,    
and therefore we find \cite{aahdd}
\be
M_P^2 \sim {1 \over \alpha_{GUT}^2} V_{\perp} M_I^{2+n} \ , \label{8.3}
\ee 
a relation similar to that proposed in the field-theoretical scenario of 
\cite{add}. 
   
Let us now imagine a ``brane-world'' picture\footnote{For earlier
proposals of such a ``brane-world'' picture, see \cite{braneworld}.}, in
which the Standard Model gauge group and charged fields are confined to
the D-brane under consideration. We can then ask a very important question:
what are the present experimental limits on parallel $R_{||}$ and perpendicular
$R_{\perp}$ type radii ? The Standard Model fields have light KK states
in the parallel directions  $R_{||}$. Their possible effects in
accelerators were studied in detail \cite{antoniadis}, and the
present limits are $R_{||}^{-1} \ge 4-5$ TeV. On the other hand,
Standard Model excitations related to  $R_{\perp}$ are very heavy
and are thus irrelevant at low energy. The main constraints on  $R_{\perp}$
come from the presence of very light winding (KK after T-dualities)
gravitational excitations, which can therefore generate deviations from
the Newton law of gravitational attraction. The actual experimental
limits on such deviations are limited to the cm range and experiments in
the near future are planned to improve them \cite{long}. For $M_I \sim
$ TeV in (\ref{8.3}), the case of only one extra dimension is clearly
excluded, since it asks for $R_{\perp}^{-1} \sim 10^8$ Km. However, for two
extra dimensions, we find  $R_{\perp}^{-1} \sim$ 1mm, not yet
excluded by the present experimental data. On the other hand, if all
compact dimensions are
perpendicular and large, one finds  $R_{\perp}^{-1} \sim$ fm,
distance scale completely inaccessible for Newton law measurements.   
Such a physical picture with $M_I \sim$ TeV provides  in
principle a
new solution to the gauge hierarchy problem, i.e. of why
the Higgs mass $M_h$ is much lower than the 4d Planck mass $M_P$,   
provided the physical cutoff $M_I$ has similar values $M_I \simeq M_h$.

In Type I strings, the brane we considered can be a D9 or a D5 brane,
up to T-dualities. Our brane world can live on any of the branes; let us
choose for concreteness that our Standard Model gauge group be on a D9 brane.
Notice that, while D9 branes fill (before
T-dualities) the full 10d space, D5 branes fill only six
dimensions. The D5 degrees of freedom can of course propagate in what we
called previously bulk space, and can change slightly our previous picture. 
The relation between the corresponding D9 and D5 gauge couplings is
\be
{g_9^2 / g_5^2} = V_{\perp} \quad , \label{8.4}
\ee
where $V_{\perp}$ denotes here (before T-dualities) the compact volume
perpendicular to the D5 brane. If  $V_{\perp} >> 1$ in string units,
then D5 branes live in (at least part of) the bulk and, by (\ref{8.4}) their
gauge coupling is very suppressed compared to our (D9) gauge
coupling. In particular, if $V_{\perp}$ in (\ref{8.4}) is as
in (\ref{8.3}), the D5 gauge couplings are of gravitational strength. 
The fields in mixed 95 representations are charged under both gauge
groups. Then, due to their very small gauge couplings, the D5 gauge
groups manifest themselves as global symmetries on our D-brane, and could be used for
protecting baryon and lepton number nonconservation processes. Indeed,
global symmetries are presumably 
violated by nonrenormalizable operators suppressed by the
fundamental scale $M_I$ and, since $M_I$  can be very low, we need
suppression of many higher-dimensional operators. 
    
There are clearly many challenging questions that such a scenario must
answer \cite{bdn} in order to be seriously considered as an alternative to the
conventional ``desert picture'' of supersymmetric unification at
energies of the order of $10^{16}$ GeV. The gauge hierarchy problem
still has a counterpart here, understanding the possible mm size of 
the compact dimensions (perpendicular to our brane) in a theory with
a fundamental length (energy) in the $10^{-16}$ mm (TeV) range.
There are several ideas concerning this issue in the literature 
\cite{ddgr}, which however need further studies in realistic models in
order to prove their viability. A serious theoretical question 
concerns gauge coupling unification, that in this case, if it exists, must
be completely different from the conventional MSSM (Minimal
Supersymmetric Standard Model) one. Moreover, there
is more and more convincing evidence for neutrino masses and mixings,
and the conventional picture provides an elegant explanation of their
pattern via the
seesaw mechanism \cite{seesaw} with a mass scale of the order of the
$10^{12}-10^{15}$ GeV, surprisingly close to the usual GUT scale. From this
viewpoint, neutrino masses can be considered as the first
experimental manifestation of physics beyond the Standard Model (see
for example \cite{ramond}).     
The new scenario described above should therefore provide at least a
qualitative picture for neutrino masses and mixings. Cosmology,
astrophysics \cite{add,sacha}, accelerator physics
\cite{antoniadis,grw} and flavor physics \cite{ddg,flavor} put
additional strong constraints on the low-scale string scenario.
   
\section{Gauge coupling unification}

Models with gauge-coupling unification at low energy triggered by
Kaluza-Klein states were independently proposed in
\cite{ddg}, at the same time as brane-world models with a low-string
scale. Both provide possible solutions to
the gauge hierarchy problem. It is transparent, however, that low-scale
string models are the
natural framework for this fast-driven unification. In this chapter we
separate the discussion into two steps: we begin with the
field-theoretic picture originally
proposed in \cite{ddg}, and then move to the Type I string approach 
developed in
\cite{bachas2,abd} which brings some new, interesting features.

\vskip 2mm
- {\large \bf Field theory approach}
\vskip 2mm
 
The essential ingredient in this approach are the KK excitations of the
Standard Model gauge bosons and matter multiplets and their contribution
to the energy evolution of the physical gauge couplings. In
the early paper \cite{tv}, Taylor and Veneziano pointed out that the KK
excitations give power-law corrections that   
at low energy can be interpreted as threshold corrections. Actually, as shown in 
\cite{ddg}, if
the energy is higher than the KK compactification scale $1/R$, these
corrections should be interpreted as power-law accelerated evolutions of gauge
couplings that, under some reasonable assumptions, can bring these
couplings together at low energies. 

The one-loop evolution of gauge couplings in 4d between energy scales $\mu_0$ and
$\mu$ can be computed with standard methods, and the final result can be
cast into the form
\be 
{1 \over \alpha_a (\mu)} ={1 \over \alpha_a (\mu_0)} + {1 \over 2 \pi }
\sum_r {\rm Str} \int_{1 / \mu^2}^{1 / \mu_0^2} {dt \over t} Q_{a,r}^2 
({1 \over 12} - \chi_r^2) e^{-t m_r^2} \ , \label{8.5} 
\ee 
where $Q_{a,r}$ is the gauge group generator in the representation $r$
of the gauge group, $m_r^2$ is the mass operator and
$\chi_r$ is the helicity of various charged particles
contributing in the
loop. In 4d (\ref{8.5}) can be readily integrated as usual in order to obtain,
for example
\be
{1 \over \alpha_a (\mu)}={1 \over \alpha_a (M_Z)} - {b_a \over 2 \pi}
\ln {\mu \over M_Z} \ , \label{8.6}
\ee 
where $b_a$ are the beta-function coefficients defined as in
(\ref{4.24}) for a supersymmetric theory.

Let us start, for reasons to be explained later on, with the
MSSM in 4d and try to extend it to 5d,
where the fifth dimension is a circle of radius $R_{||}$, in the notation 
introduced in the previous section. In this case (\ref{8.5}) generalizes to 
\be 
{1 \over \alpha_a (\mu)} ={1 \over \alpha_a (\mu_0)} + {1 \over 2 \pi }
\sum_r {\rm Str} \int_{1 / \mu^2}^{1 / \mu_0^2} {dt \over t}  Q_{a,r}^2 ({1 \over
12} - \chi_r^2) (\sum_n e^{-t m_{n,r}^2 (R_{||})} +  e^{-t m_r^2}) \ , \label{8.7} 
\ee 
where we separated the mass operator into a part containing fields
with KK modes and a part containing fields without
KK modes. Indeed, consider again for concreteness gauge couplings
of a D9 brane and consider $\delta$ large compact dimensions $R_{||}M_I
>> 1$ parallel to D9 and orthogonal to
D5. Then the 99 states will have associated KK states, but 95 states will not.
Evaluating (\ref{8.7}) with $\mu_0 = M_Z$, one finds
\ba
{1 \over \alpha_a (\mu)}&=&{1 \over \alpha_a (M_Z)} - {b_a \over 2 \pi}
\ln {\mu \over M_Z} - {{\tilde b}_a \over 2 \pi} 
\int_{1 / \mu^2}^{1 / M_Z^2} {dt \over t} \theta_3^{\delta} ({i t
\over \pi R_{||}^2}) \nonumber \\
&\simeq& {1 \over \alpha_a (M_Z)} - {b_a \over 2 \pi}
\ln {\mu \over M_Z} + {{\tilde b}_a \over 2 \pi}
\ln ({\mu R_{||}}) -  {{\tilde b}_a \over 2 \pi} [ ({\mu
R_{||}})^{\delta}-1] \ . \label{8.8}
\ea
The coefficients ${\tilde b}_a$ in (\ref{8.8}) denote one-loop beta-function      
coefficients of the massive KK modes, to be computed in each specific model. The 
important term contained in (\ref{8.8}) is the power-like term  
$({\mu R_{||}})^{\delta} >>1$, which overtakes the logarithmic terms
in the higher-dimensional regime and governs the eventual unification
pattern. 

The power-like term is proportional to the coefficients ${\tilde b}_a$, 
that {\it are not} the usual 4d MSSM ones which successfully predict
unification. Therefore, from this point of view the MSSM
unification would just be an accident, and this fact is disappointing. Let us
however go on and find the {\it minimal} possible embedding of the MSSM
in a 5d spacetime. Before doing it, notice that compactifying
on a circle a supersymmetric theory in 5d gives a 4d theory with at least 
${\cal N}=2$ supersymmetries. The simplest way to avoid this is to
compactify on an {\it orbifold}, a singular space defined in Section 5.
We consider as example the case of a $Z_2$ orbifold which breaks
supersymmetry down to ${\cal N}=1$. 5d fields can be even or odd under
this operation, in particular 5d Dirac fermions in 4d truncate into
one even Weyl fermion containing a zero mode and its KK tower and one
odd Weyl fermion,
with no associated zero mode, and its KK tower. It is easy to realize
that a 4d chiral multiplet $(\psi_1, \phi_1)$ can arise from a 5d 
hypermultiplet containing KK modes  
$(\psi_1^{(n)},\psi_2^{(n)},\phi_1^{(n)},\phi_2^{(n)} )$ or from a 5d
vector multiplet.
Similarly, a 4d massless vector multiplet $(\lambda, A_{\mu})$ arises from a 5d
vector multiplet containing the KK modes  $(\lambda^{(n)},\psi_3^{(n)},
A_{\mu}^{(n)},a^{(n)} )$, where $\psi_i^{(n)}$, $i=1,2,3$ are 4d Weyl 
fermions and $\phi_i^{(n)}$, $a^{(n)}$ are complex scalars.  
The massive KK representations are clearly nonchiral, while chirality is
generated at the level of zero modes. 
 
\begin{figure}
\centerline{ \epsfxsize 3.25 truein \epsfbox {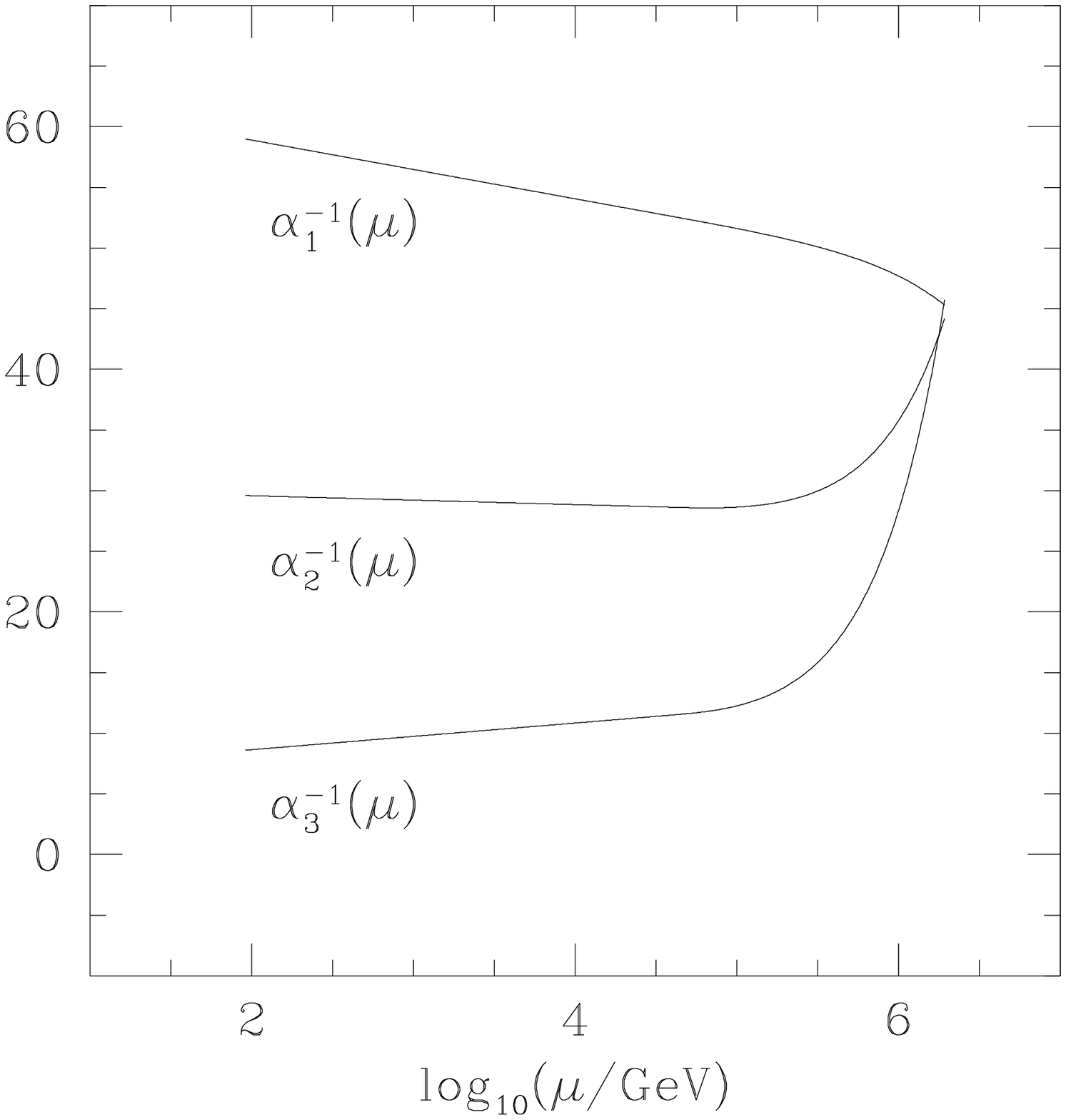}
             \epsfxsize 3.25 truein \epsfbox {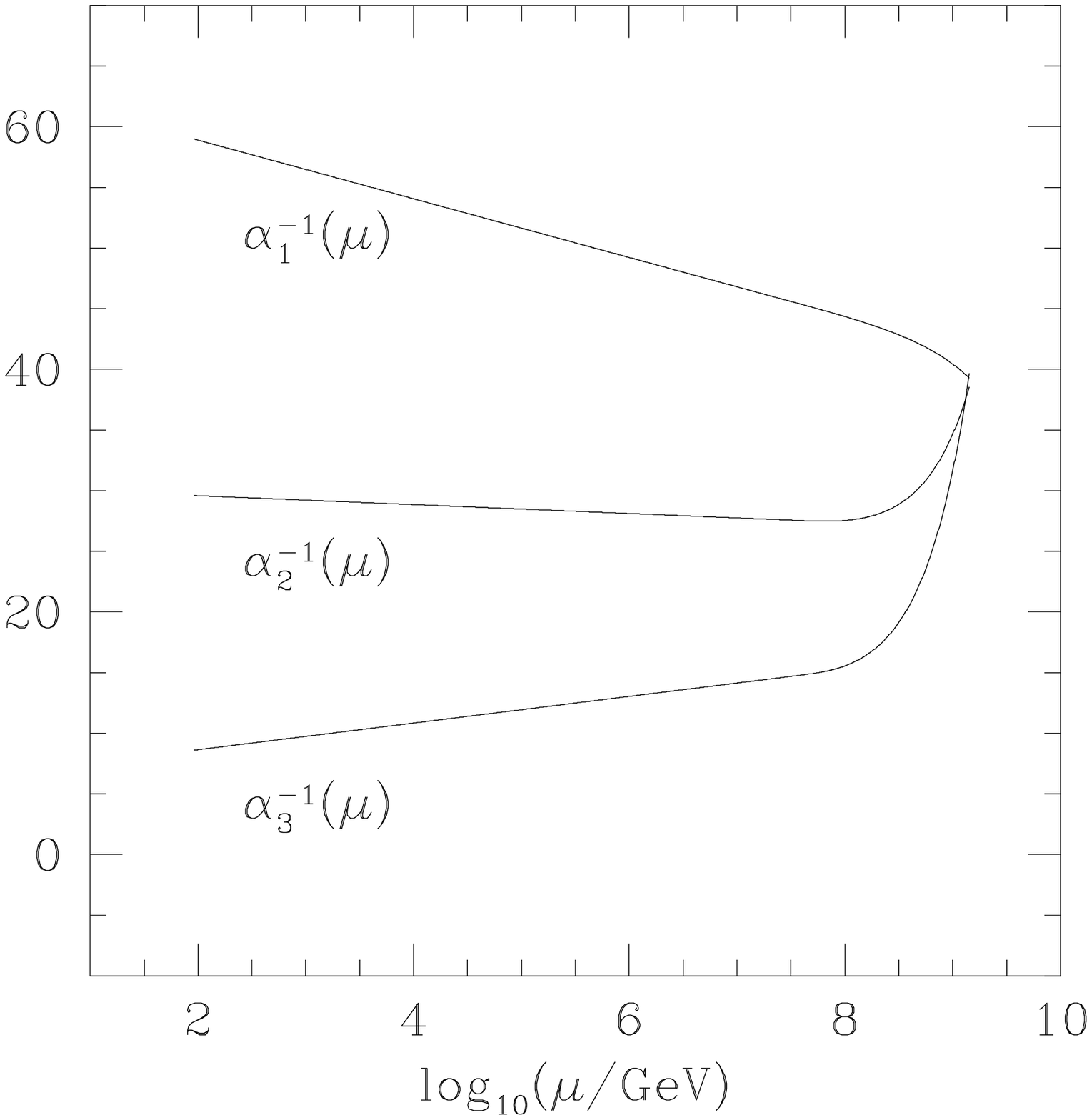}}
\caption{Unification of gauge couplings in the presence of
     extra spacetime dimensions.
     We consider two representative cases:
          $R^{-1} =  10^{5}$ GeV (left),
          $R^{-1} =  10^{8}$ GeV (right).
      In both cases we have taken $\delta=1$ and $\eta=0$.  }
\label{unifII}
\end{figure}

The simplest embedding of the MSSM in 5d is the following 
\cite{ddg}. The gauge bosons and the two Higgs multiplets of MSSM are 
already in
real representations of the gauge group and naturally extend to
KK representations  $(\lambda^{(n)},\psi_3^{(n)},
A_{\mu}^{(n)},a^{(n)} )$ and  $(\psi_1^{(n)},\psi_2^{(n)},H_1^{(n)},
H_2^{(n)})$, respectively\footnote{As one of the two Higgses in
a hypermultiplet is odd under $Z_2$, the simplest
extension actually has one KK Higgs hypermultiplet and one Higgs
without KK excitations.}. The matter fermions of MSSM, being chiral, can
either contain only zero modes or, alternatively, can have
associated mirror fermions and KK excitations for $\eta=0,1,2,3$ families.   
The unification pattern does not depend on $\eta$ (the value of the
unified coupling, on the other hand, does), since each family forms
a complete $SU(5)$ representation. The massive beta-function
coefficients for this simple 5d extension of the MSSM are
\be
({\tilde b}_1,{\tilde b}_2, {\tilde b}_3)= ({3 \over 5}, -3,-6) + \eta
(4,4,4) \ , \label{8.9}
\ee   
where, as usual, we use the $SU(5)$ embedding ${\tilde b}_1 \equiv (3/5)
{\tilde b}_Y$. These coefficients in the case $\eta=3$ are not the same as
the MSSM ones $(b_1,b_2, b_3)= (33/5,1,-3)$. However,
interestingly enough, as seen from Figure \ref{unifII}, the couplings unify
with a surprisingly good precision, for any compact radius 
$10^{3}$ GeV $ \le R_{||}^{-1} \le
10^{15}$ GeV, at a energy scale roughly a factor of 20 above the
compactification scale $R_{||}^{-1}$. The algebraic reason for this is that, in order to have
MSSM unification, the conditions that must be fulfilled are
\be
{B_{12} \over B_{13}} ={B_{13} \over B_{23}}=1 \quad , {\rm where} 
\quad B_{ac} \equiv { {\tilde b}_a-{\tilde b}_c \over  b_a- b_c } \
. \label{8.10}
\ee
Although these relations are not satisfied exactly in our case, they are
nonetheless approximately satisfied
\be
{B_{12} \over B_{13}} = {72 \over 77} \simeq 0.94 \quad ,
\quad {B_{13} \over B_{23}} = {11 \over 12} \simeq 0.92 \ . \label{8.11}
\ee

\begin{figure}[ht]
\centerline{ \epsfxsize 4.0 truein \epsfbox {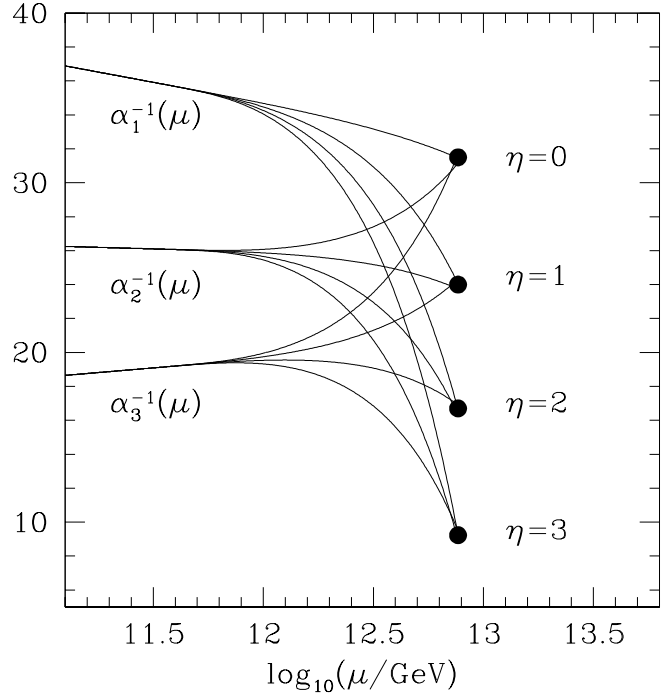}}
\caption{Unification of gauge couplings in the presence of
     extra spacetime dimensions.
     Here we fix $R^{-1}= 10^{12}$ GeV, $\delta=1$,
      and we vary $\eta$.
      For this value of $R^{-1}$, we see that the unification
      remains perturbative for all $\eta$.}
\label{unifnew}
\end{figure}

This fast unification with KK states is another numerical miracle,
similar to the MSSM unification and may be regarded as one serious hint 
pointing into the possible relevance of extra dimensions in
our world. There are clearly a lot of questions that this scenario can raise,
which were discussed in detail in the literature \cite{ddg,gr,kt}, the
most important ones being:

- The perturbative nature. 
 
Indeed, even if unified coupling in Figure \ref{unifII} is about 
$\alpha_{GUT}
\simeq 1/50$ for $M_{GUT} \sim 10$ TeV, the parameter controlling the
loop expansion is $N_{KK} \alpha_{GUT}$, where $N_{KK}$ is here the number (of order
20) of KK states with masses ligher then the unification scale
$M_{GUT}$. This parameter is large (of order $2/5$), and therefore
perturbativity seems to be lost. However, things are slightly better
than expected. Indeed, massive modes come into ${\cal N}=2$
multiplets. In ${\cal N}=2$ theories beta functions get contributions
only at one loop. Therefore the higher-order loops must contain
zero-mode propagators, which have reduced ${\cal N}=1$ supersymmetry
but have no KK modes. For example, the (two loop)/(one-loop) effects are
naively of order  $N_{KK} \alpha_{GUT}$ but, due of the 
argument above, they are actually only of order
$\alpha_{GUT}$. Two-loop contributions induced by Yukawa coupling
corrections could in principle be larger, but actually they are still
under control (see, for example, M. Masip in \cite{gr}).
Notice that the most perturbative case is $\eta =0$, and increasing
$\eta$ from 0 to 3 renders the model less and less perturbative (see
Figure \ref{unifnew}). 

- The sensitivity to high-energy thresholds 
 
The result of the computation is more sensitive to 
high-energy thresholds than the MSSM result is, if we consider the 
unification in the sense of a Grand Unified Theory. A related question
is the supression of baryon number violating operators, in low-scale
strings, which are induced by states of the GUT theory.
If we consider the unification in the string sense, as will be shown in
the next paragraph string thresholds affect only the ${\cal N}=1$
sectors of the theory, while in ${\cal N}=2$ sectors, responsible for the
power-law evolution, the string states decouple \cite{bf,bachas2} and the 
corrections come only from KK massive states. The running stops at a
higher winding scale, without the need of a GUT gauge group and new
thresholds there.

- The need for supersymmetry 

We considered above a higher-dimensional extension of MSSM. We could in
principle try a similar extension of the Standard Model, without
invoking supersymmetry. In this case the unification is still possible
\cite{ddg}, at the price of introducing additional gauge group
representations. The extension is consequently not minimal by any
means. Therefore, it is amusing that even in case of large extra
dimensions, supersymmetry seems to play a role in the unification
of gauge couplings. 
 
In order to have a physical interpretation of the unification scale
discussed above, we now turn to the string approach, using results derived in Section 6. 

\vskip 2mm
- {\large \bf String theory approach}
\vskip 2mm
 
In a superstring model, the threshold corrections to gauge couplings
(\ref{4.17}) can be generically written as
\be
{\cal B}_a = {\cal B}_a^{({\cal N}=4)} + {\cal B}_a^{({\cal N}=2)} +  
{\cal B}_a^{({\cal N}=1)} \ , \label{8.12}
\ee 
where the different terms in the rhs of (\ref{8.12}) denote
contributions from ${\cal N}=4$, ${\cal N}=2$ and ${\cal N}=1$ sectors, respectively.  
The ${\cal N}=4$ sectors, containing the full $\Gamma^{(6)}$ lattice in
the notation of Section 3, have a 10d origin and give no contribution to
threshold corrections. The  ${\cal N}=2$ sectors contain the lattice of
one compact torus  $\Gamma^{(2)}$. In these sectors only BPS KK
states contribute to threshold corrections and string oscillators
decouple \cite{bf, bachas2}. Therefore, their contribution to the evolution of
gauge couplings does not stop at the string scale $M_I$, but
rather, as we will see, at a heavy KK scale.    
The  ${\cal N}=1$ sectors have no KK
excitations and give a moduli-independent contribution to
threshold-corrections, interpreted as the ${\cal N}=1$ contribution to 
gauge couplings, running up to $M_I$.  

The string one-loop threshold corrections coming from ${\cal N}=2$
sectors were computed in (\ref{4.21}). We also saw in Section 6
that, in addition to the dilaton ${\rm Im} S \sim 1/l$, there are
tree-level (disk) contributions from
couplings of gauge fields to the twisted moduli $m_k$, displayed in 
(\ref{u12}). Putting all the terms together, we find the complete
one-loop gauge couplings 
\be
{4\pi^2 \over g_a^2 (\mu)} = {1 \over l} + \sum_k s_{ak} m_k
+{1 \over 4} b_a^{({\cal N}=1)} \ln {M_I^2 \over \mu^2} - {1 \over 4}
\sum_{i=1}^3  b_{ai}^{({\cal N}=2)} \ln ({\sqrt{G_i} \mu^2 |\eta(U_i)|^4
{\rm Im} U_i}) \ , \label{8.13}
\ee  
where for a rectangular torus of radii $R_1,R_2$, we have $\sqrt{G_i}=R_1R_2$
and ${\rm Im} \ U= R_1/R_2$. In (\ref{8.13}), $b_{ai}^{({\cal N}=2)}$ 
denote beta function coefficients from ${\cal N}=2$ sectors having KK excitations in the
compact torus $T^i$. The total beta function (\ref{4.24}) of the model
is, in these notations,
\be
b_a =  b_a^{({\cal N}=1)} + \sum_{i=1}^3  b_{ai}^{({\cal N}=2)} \
. \label{8.14}
\ee

Let us now consider the field-theory limit of the corrections given by
an ${\cal N}=2$ sector, depending on a torus of radii $R_{1,2}$.
In the limit $R_1 \rightarrow \infty$ and $R_2$ fixed, the corrections
are linearly divergent as $\Lambda_2 \sim R_1/R_2$. These power-law
corrections can be used to lower the unification
scale \cite{ddg} in models with a low value of the string scale $M_I$.
Notice that, in all the above computations, $\mu$ denoted an infrared
energy scale, smaller than any KK mass scales. 
Actually, for energies $\mu >> R_1^{-1}$, relevant for the $R_1
\rightarrow \infty$ limit, it can be seen that the previous factor $R_1/R_2$ 
really becomes $R_1 \mu$, thus reproducing the field-theory
derivation (\ref{8.8}) with $\delta =1$. In this case, to get unification
one needs $10^{3}$ GeV $ \le R_1^{-1} \le 10^{15}$ GeV. On the other hand, it is
at first sight surprising that in the opposite limit of very heavy KK 
states (windings after
T-duality) $R_1 \rightarrow 0$, there is a divergent contribution 
$\Lambda_2 \sim R_2/R_1$. In particular, this applies to mm perpendicular 
dimensions and can therefore spoil the solution to the hierarchy problem
\cite{bachas2, ab}. These corrections can however be avoided in a class
of Type I models that satisfy local tadpole cancellation in the
corresponding direction \cite{pw}, \cite{ads1}. 
 
Another interesting and unexpected feature is that in the limit $R_1,R_2 \rightarrow \infty$
with $R_1/R_2$ fixed, $\Lambda_2 \sim \ln (R_1R_2 \mu^2)$, instead of
the quadratic divergence ($\delta=2$ in (\ref{8.8})) expected in the
field theory approach. The same result holds in the $R_1,R_2 \rightarrow
0$ limit. This result can be understood by the
following argument \cite{ab}. After T-duality, the two directions are
very large and perpendicular to the brane under consideration. One-loop
threshold corrections can also be understood as tree-level coupling of gauge
fields to closed sector fields, which have a bulk variation reproducing
the threshold dependence on the compact space. The bulk variation can be
computed in a supergravity approximation, solving classical field 
equations for closed fields coupled to
various sources subject to global neutrality (or global tadpole cancellation)
in the compact space. As the Green function in two dimensions has a
logarithmic behaviour, this explains the logarithmic term $\ln (R_1R_2
\mu^2)$. The same argument in one compact dimension explains the
linearly divergent term previously discussed.
    
The logarithmic evolution $\Lambda_2 \sim \ln (R_1R_2 \mu^2)$ can also
be used to achieve unification at 
a high energy scale, even if the fundamental string scale has much lower 
values \cite{bachas2}, by ``running'' beyond the string scale.

Notice that both the power-law and the logarithmic evolution of gauge
couplings use ${\cal N}=2$ beta-functions. As shown in the field-theory
approach, a simple higher-dimensional extension succeeds in producing
unification with ${\cal N}=2$ sectors. In this case however, MSSM unification
would be just a miraculous accident. It would be useful to see if one 
can obtain
the usual MSSM unification in models with a low string scale. One possibility  
recently proposed in \cite{ibanez2} takes advantage of the couplings to
twisted fields present in (\ref{8.13}) in models without ${\cal N}=2$ sectors. 
Let us assume that in some models $s_{ak} =c_k b_a$ and, in addition,
that the
twisted fields have some vevs $<m_k>$. In terms of
$ \sum_k c_k m_k \equiv c \ m $, the one-loop relation (\ref{8.13}) becomes
\be
{4\pi^2 \over g_a^2 (\mu)} = {4\pi^2 \over g_a^2}|_{\rm tree} + c \ b_a <m>
+ {b_a \over 2} \ln {M_I \over \mu} = {4\pi^2 \over g_a^2}|_{\rm tree}
+  {b_a \over 2} \ln {M_I \ e^{2c <m>} \over \mu} \ . \label{8.15}
\ee
The real unification scale is therefore $M_{GUT} = exp({2c <m>}) \ M_I$,
which, depending on the sign in the exponential, can be much larger than
$M_I$. There are some Type I models where indeed  the proportionality
relation $s_{ak} =c_k b_a$ holds \cite{imr, ibanez2}. In these
models, unfortunately,
$<m>=0$ and this ``mirage'' unification does not occur \cite{abd}. It is still
reasonable to hope that in some other models all conditions are
fulfilled and that the mechanism can be implemented\footnote{A proposal
trying to combine \cite{bachas2} and \cite{ibanez2} was also recently
studied in \cite{adm}.}. A possible scenario is the
following. Suppose that gaugino condensation takes place in a gauge 
group factor $G_a$, coupling to $m$, giving rise to a nonperturbative
superpotential for the complex chiral superfields $S,M$ of the form
\be
W(S,M) \sim e^{- \alpha (S+ s_a M )} \ , \label{8.16}
\ee
where $\alpha$ is a numerical factor depending on the gauge factor $G_a$
and its matter content. For a large class of K{\"a}hler potentials for
the moduli $M$ and if the dilaton $S$ is stabilized, a nonzero value 
$<M> \sim M_P$ is easily generated and can provide the (mirage)
unification discussed above.
 
\section{Supersymmetry breaking}  

- {\large \bf Breaking through compactification in field theory: the
Scherk-Schwarz mechanism }

The Scherk-Schwarz mechanism for breaking supersymmetry takes advantage
of the presence of compact spaces in compactifications of
higher-dimensional supersymmetric field theories or of superstrings. 
The main idea is to use symmetries ${\cal S}$ of the higher-dimensional theory
which do not commute with supersymmetry, typically R-symmetries or the fermion
number $(-1)^F$. Then, after being transported around the compact space (a circle of
radius $R$, for concreteness and coordinate $0 \le y \le 2 \pi R$), 
bosonic and fermionic fields $\Phi_i$ return to the initial value 
(at $y=0$) only up to a symmetry operation
\be
\Phi_i (2 \pi R, x ) = U_{ij} (\omega) \Phi_j (0, x ) \ , \label{9.01}   
\ee
where the matrix $U \in {\cal S}$ is different for bosons and fermions
and $x$ are noncompact coordinates. At the field theory level, (\ref{9.01})
implies that the Kaluza-Klein decomposition on the circle is changed 
so that zero modes acquire a nontrivial dependence on the $y$
coordinate, according to
\be
\Phi_i (y,x) = U_{ij} (\omega,y/R) \sum_m e^{i m y \over R} 
\Phi_j^{(m)} (x) \ , \label{9.02}
\ee 
where $\omega$ is a number, quantized in String Theory.
The matrix $U$ satisfies some additional constraints in supergravity
in order for the generated scalar potential to be positive definite.
The ansatz considered by Scherk and Schwarz is $U = \exp (M y)$, where
$M$ is an antihermitian matrix. Then kinetic terms in the $y$ direction
generate mass terms and break supersymmetry, the resulting fermion-boson 
splittings being equal to $\omega / R$. This twisting procedure is very similar
to  the breaking of supersymmetry at {\it finite temperature} and, because of
this, the terms breaking supersymmetry are UV finite, even at the field 
theory level.

The mechanism can be applied in globally supersymmetric models,
in supergravity models and in superstrings. The breaking is induced
by the different boundary conditions for bosons and fermions and
is therefore an explicit breaking. However, at the supergravity level,
it appears to be {\it spontaneous}. In order to clarify this point,  
let us consider a simple global model and a local (supergravity) one.  
\vskip .2cm
i) a globally supersymmetric model
 
The model has one hypermultiplet in 5d, containing one Dirac
fermion $\Psi$ and two complex scalars $\phi_1,\phi_2$, described by the 
free lagrangian
\be
{\cal L} = - {i \over 2} ( {\bar \Psi} \gamma^M \partial_M \Psi -
 \partial_M {\bar \Psi} \gamma^M \Psi ) -|\partial_M \phi_1|^2
- \partial_M \phi_1|^2 \ . \label{9.03}
\ee  
The model (\ref{9.03}) has several symmetries. Let us choose the
R-symmetry\footnote{P. Fayet \cite{ss} was the first to use 
R-symmetries in order to produce phenomenologically interesting soft
masses in global supersymmetric
models. He proposed compact radii in the TeV range for this purpose. }
that leave the fermion invariant and rotates the two complex scalars
between themselves. The Scherk-Schwarz matrix reads $M=i (\omega /R) \sigma_2$,
where $\sigma_2$ is the second Pauli matrix. After the modified KK reduction
(\ref{9.02}), one finds the resulting 4d lagrangian
\ba
&{\cal L}& =  - \sum_m \{ {i \over 2} ( \Psi_1^{(m)} \sigma^{\mu} 
\darr{\partial}_{\mu} {\bar \Psi}_1^{(m)} +
\Psi_2 \sigma^{\mu} \darr{\partial}_{\mu} {\bar \Psi}_2^{(m)} )
+ |\partial_{\mu} \phi_1^{(m)}|^2+|\partial_{\mu} \phi_1^{(m)}|^2
\nonumber \\
&+& {m \over R} ( \Psi_1^{(m)} \Psi_2^{(m)} + h.c.) + {m^2 + \omega^2
\over R^2} 
(| \phi_1^{(m)}|^2 + | \phi_2^{(m)}|^2 ) \} \ , \label{9.04}
\ea
where we defined the Weyl fermions $\Psi^{(m),T} =
(\Psi_1^{(m)},{\bar \Psi}_2^{(m)})^T$. 
The lagrangian (\ref{9.04})  describes the free propagation
of massive Weyl spinors $\Psi_1^{(m)} , \Psi_2^{(m)}$, of mass $m/R$
and  the free propagation
of massive complex scalars $\phi_1^{(m)}, \phi_2^{(m)}$, of mass squared
$(m^2+\omega^2) / R^2$. Supersymmetry is explicitly broken by the {\it soft}
mass term $\omega^2/R^2$.  

In orbifolds, there is a compatibility condition between the orbifold
action $\theta$ and the Scherk-Schwarz twisting matrix $[\theta ,
U]=0$. In particular, if the coordinate $y$ in question is orbifold
invariant $\theta y=y$, this implies $[\theta , M]=0$, while if it is $Z_2$
twisted $\theta y=-y$, this implies $\{\theta , M \}=0$. 

ii) a local model: supersymmetry breaking in M-theory

The example we discuss now is the compactified version of
supersymmetry breaking in M-theory, discussed in general terms in
Section 4 and realized also in Type I strings in Section 7.

Consider the simplest truncation of 11d supergravity down to 5d, keeping
only the breathing mode of the compact space $g_{i \bar j}= \delta{i
\bar j} \exp(\sigma)$, and
concentrate for simplicity on zero modes only. In this case, the only matter
multiplet in 5d (in addition to the 5d gravitational multiplet with bosonic
fields ($g_{MN},C_M$), where $C_{M i \bar j}= (1/6) A_M \delta_{i \bar j}$
is a vector field originating from the 3-from of 11d supergravity) is the
universal hypermultiplet of bosonic fields ($\sigma, C_{MNP}, a$), with
$C_{ijk} = (1/6) \epsilon_{ijk} a$, whose scalar fields parametrize
the coset ${{SU(2,1)}/ {SU(2) \times U(1)}}$ \cite{ferrara}.
The bosonic 5d supergravity lagrangian is\footnote{There are also terms
coming from the modified Bianchi identity (\ref{m5}) correcting the
lagrangian (\ref{9.05}). We neglect them here for simplicity, but they
can be found in \cite{ovrut}, for example.} 
\ba
&\!\!{\cal S}_5\!\!& \!=\! {1 \over 2k_5^2} \int d^5 x \sqrt{g} \{ R
\!-\! {9 \over 2} (\partial_M
\sigma)^2 \!-\! {1 \over 24} e^{6 \sigma} G_{MNPQ}  G^{MNPQ} -{3 \over 2}
F_{MN} F^{MN} \!-\! 2 \ e^{-6 \sigma} |\partial_M a|^2 \} \nonumber \\
&-& {1 \over k_5^2} \int d^5 x \epsilon^{MNPQR} \{ {i \over \sqrt{2}} 
C_{MNP} \partial_Q a \partial_R a^{\dagger} + {1 \over 2 \sqrt{2}}
A_{M} F_{NP} F_{QR} \} \
, \label{9.05}
\ea 
where $F_{MN} = \partial_M A_N- \partial_N A_M$.  
The compactification from 5d to 4d is on the orbifold $S^1/Z_2^{HW}$, 
with orbifold
action $Z_2^{HW}$ defined in (\ref{m3}) of Section 4. 
The lagrangian of the universal hypermultiplet can be derived from the
4d K{\"a}hler potential \cite{ferrara}
\be
{\cal K} =- \ln \ (S+S^{\dagger} - 2 a^{\dagger} a) \ , \label{eq:s1}
\ee
lifted back to 5d, where $S=\exp (3\sigma ) + a^{\dagger} a + i a_1$  
and the axion $a_1$ is defined by the Hodge duality
$\sqrt{2} \exp (6 \sigma ) G_{MNPQ} =\epsilon_{MNPQR} 
( \partial^R a_1 + i a^{\dagger} \darr{\partial}^R a) $. The 
lagrangian (\ref{9.05}) has a global 
$SU(2)_R$ symmetry, acting linearly on the redefined hypermultiplet fields
\be
z_1 = {1-S \over 1+S} \quad , \quad z_2 = {2 a \over 1+S} \ , \label{eq:s2}
\ee
which form a doublet $(z_1, z_2)$. In the gravitational multiplet,
the 5d Dirac gravitino is equivalent to two 4d Majorana gravitinos, 
transforming as an $SU(2)_R$ doublet. One of the
gravitini is even under $Z_2^{HW}$ and has a zero mode (before the
Scherk-Schwarz twisting), while the other is odd and has only massive KK 
excitations. The $Z_2^{HW}$ projection acts on the hypermultiplet as
$Z_2^{HW} S =S$, $Z_2^{HW} a = -a$, which translates on the $SU(2)$ doublet
in the obvious way
\ba 
Z_2^{HW}
\left( 
\begin{array}{c} 
z_1 \\ 
z_2 
\end{array} 
\right)  
&=& 
\left(
\begin{array}{cc} 
1 & 0 \\ 
0 & -1
\end{array}
\right)
\left(
\begin{array}{c}
z_1 \\
z_2
\end{array}
\right) \ . \label{eq:s3}
\ea

The Horava-Witten projection then asks for using the $U(1)_R$ subgroup
of $SU(2)_R$ and the corresponding Scherk-Schwarz decomposition reads \cite{dg}
\ba 
\left( 
\begin{array}{c} 
{\hat z}_1 \\ 
{\hat z}_2 
\end{array} 
\right)  
&=& 
\left(
\begin{array}{cc} 
\cos M_0 x_5 & \sin M_0 x_5 \\ 
-\sin M_0 x_5 & \cos M_0 x_5
\end{array}
\right)
\left(
\begin{array}{c}
z_1 \\
z_2
\end{array}
\right) \ ,\label{eq:s4}
\ea
corresponding to the matrix $M=i M_0 \sigma_2$. Notice that, 
thanks to the anticommutation relation  $\{ Z_2^{HW} , M \}=0$,
the fields ${\hat z}_i$ have the same $Z_2^{HW}$ parities as the fields $z_i$.
The 4d complex superfields of the model are $S$ (with the zero mode
$a=0$) and $T$, where $T=g_{55} + i \ C_5 $ and the axion $C_5$ is the
fifth component of the vector field in the 5d gravitational multiplet.  
The resulting scalar potential in 4d in the Einstein frame is computed from the 
kinetic terms of the $({\hat z}_1, {\hat z}_2)$ fields derived form (\ref{eq:s1}). 
After putting $z_2=0$ at the zero mode level, the result is
\be
V = \int dx_5 \sqrt{g_{55}} \ g^{55} {\cal K}^{a \bar b} \ \partial_5 z_a 
\partial_5 z_{\bar b} =
{4 M_0^2 \over (T+T^{\dagger})^3} {|1-S|^2 \over S+S^{\dagger}}
\ , \label{eq:s40}
\ee
where $a,b=1,2$ and ${\cal K}^{a \bar b}$ is the inverse of the
K{\"a}hler metric ${\cal K}_{a \bar b}= \partial_a \partial_{\bar b}
{\cal K}$.
This result is interpreted as a superpotential generated for $S$. The 4d
theory is completely described by
\be
{\cal K} =-\ln (S+S^{\dagger}) - 3 \ln (T+T^{\dagger}) \quad ,
\quad W= 2 M_0 (1 + S) \ . \label{eq:s5}
\ee
Notice that the superpotential corresponds to a
non-perturbative effect from the heterotic viewpoint.
The minimum of the scalar potential is $S=1$ and corresponds to a 
spontaneously broken supergravity with zero cosmological constant. 
The order parameter for supersymmetry breaking
is the gravitino mass $m_{3/2}^{2} =e^{{\cal K}} |W|^2 = {2 M_0^2 / 
(T+T^{\dagger})^3}$.
If in 4d supergravity units we define $M_0 = \omega M_P$, then
$m_{3/2}=\omega /R_5$, where $R_5$ is the radius of the fifth
coordinate. This is consistent with the fact that the gravitino mass was
affected by the R-symmetry. The Goldstone fermion is the fifth component
of the ($Z_2^{HW}$ even) 5d gravitino $\Psi_5$.
The important point about (\ref{eq:s5}) or any other
supergravity example is that the breaking of supersymmetry {\`a} la
Scherk-Schwarz appears to be spontaneous, of the F-type, with a zero
cosmological constant. The resulting models are of no-scale type 
\cite{noscale}. 

In heterotic strings the only available perturbative method of breaking
supersymmetry is the Scherk-Schwarz mechanism. In this case, soft
masses $M_{SUSY} \sim R^{-1}$ are generated at tree-level for the gauginos,
so that phenomenologically interesting values require 
$R^{-1} \sim $TeV. In this case however the model cannot be controlled
\cite{antoniadis}, in view of the large value
of the heterotic string scale $M_H \sim 5 \times 10^{17}$ GeV.
The most popular mechanism invoked in this case for breaking
supersymmetry is gaugino condensation in a hidden sector
\cite{nilles} $<\lambda \lambda> = \Lambda^3$, while the transmission 
to the observable sector is mediated by gravitational interactions,
and thus
\be
M_{SUSY} \sim {\Lambda^3 \over M_P^2} \ , \quad {\rm where} \quad
\Lambda \sim M_P \ e^{-1/(2b_0g^2(M_P))} \ . \label{9.1}
\ee
This mechanism singles out intermediate scales $\Lambda \sim 
10^{12}-10^{13}$ GeV, naturally realized by the one-loop
running of the hidden sector gauge coupling and could also be useful 
for purposes like
neutrino masses or PQ axions. Gaugino condensation, however, is a 
nonperturbative field theory phenomenon and there is little hope to
discover a string theory description of it.
A third possibility is to start directly with
nonsupersymmetric strings, possibly interpreted as
models with supersymmetry broken at the string scale $M_{SUSY} \sim
M_H$. As $M_H$ is very large, however, this possibility was completely
ignored since there was no clear way to solve the hierarchy problem  
in this case. 

In models with D-branes there are many different ways in which
supersymmetry can be broken in a phenomenologically interesting way.
This is due to the two main new features of these theories:

- The Standard Model can be confined to a subspace (D-brane) of the full
ten or eleven dimensional space.

- The string scale in these models can be lowered all the way down
to the TeV range.

\vskip 2mm
- {\large \bf Perturbative supersymmetry breaking with branes}
\vskip 2mm
 
The simplest string constructions of this type were presented in Section 
7. Even if several distinct mechanisms are available, they can be
splitted for phenomenological purposes into two classes :

i) Supersymmetry broken in the bulk. 

ii) Supersymmetry broken on some branes. 

The class i) contains the models with breaking through compactification,
in which generically a one-loop cosmological constant is generated
in the closed sector $E_0 \sim R^{-4}$, where $R$ is the radius of the
compact dimension $Y$ breaking
supersymmetry {\`a} la Scherk-Schwarz. If the Standard Model lives on a
brane parallel to $Y$, then $M_{SUSY} \sim R^{-1}$ and the phenomenology
is very close to the analogous heterotic models \cite{antoniadis}.
On the other hand, if the Standard Model lives on a brane perpendicular
to $Y$, then at tree-level the massless brane spectrum is
supersymmetric \cite{ads1} and supersymmetry breaking on the brane is
transmitted via radiative corrections. If the bulk ($Y$ in this case)
contains only gravity, then $M_{SUSY} \sim R^{-2}/M_P$ and we need here
some intermediate radius $R^{-1} \sim 10^{11}$ GeV, natural in
M-theory \cite{aq,dg} or intermediate scale string scenarios \cite{benakli}.
If the bulk contains also some other branes, there is also a Standard Model gauge mediation
coming from states charged under both Standard Model and the bulk gauge
groups \cite{adq,benakli}. If the bulk volume is large, the bulk gauge
coupling is volume suppressed with respect to the Standard Model
couplings. Consequently, in this case also $M_{SUSY} \ll R^{-1}$ and the
gauge transmission has all the known advantages concerning the
flavor-blind structure of the resulting soft breaking terms (for a
review and extensive references, see \cite{giudice}). 
 
Class ii) contains the ``Brane supersymmetry breaking'' models, with branes D
and antibranes ${\bar D}$ and a tree-level supersymmetric bulk spectrum. 
Supersymmetry is broken on the antibranes at the string
scale $M_I$, while the
tree-level spectrum of the branes is supersymmetric. If the Standard
Model lives on the antibranes, the string scale $M_I$ should be in the TeV range. 
Supersymmetry breaking on the branes can be transmitted, as before, by gravitational
interactions, in which case $M_{SUSY} \sim M_I^2/M_P$ and, if the Standard
Model lives on the branes, phenomenology asks for $M_I \sim 10^{11}$
GeV. Alternatively, if the bulk volume is sufficiently small, the
transmission can be gauge mediated and proceed through the massive
brane-antibrane excitations.   
As discussed at the end of Section 3 and in Section 7, these models have
uncancelled NS-NS tadpoles that translate, in physical terms, into scalar potentials
for the dilaton and the moduli fields describing the compact space. 
In order to understand qualitatively some of their features, we briefly discuss
a 6d model based on a toroidal compactification with D9 and D${\bar 9}$
branes with Chan-Paton factors $N_{+},N_{-}$ and D5 and
D${\bar 5}$ branes with Chan-Paton factors $D_{+},D_{-}$, worked out in \cite{aadds}. 
The RR tadpole conditions read
\be
N_{+}-N_{-} = 32 \quad , \quad  D_{+}-D_{-} = 0 \ , \label{9.2}
\ee
and the scalar potential induced by the NS-NS tadpoles is
\be
V_{eff} \sim e^{-\phi_6} \left[ (N_{+}+N_{-}-32) \sqrt{v} +
{(D_{+}+D_{-}) \over \sqrt{v}} \right]
\ee
in string units, where $\Phi_6$ is the 6d dilaton. The potential has 
a minimum and stabilizes the internal
space at the value $v_0 = (D_{+}+D_{-})/(N_{+}+N_{-}-32)$. We see
that, in order to have a very large (very small) compact space, we need 
a very large number of D9+D${\bar 9}$ (D5+D${\bar 5}$) branes,
compatible with (\ref{9.2}). This is in principle possible, but of
course asks for a dynamical explanation of the very large number of branes
and antibranes in the model\footnote{The possible existence of a 
large number of
(anti)branes in these models is a nontrivial and interesting
possibility, since in supersymmetric compactifications  the number of
D9 or D5 branes of a given type is always equal or less than 32.}.   
The dilaton potential, on the other hand, has a runaway behavior. This
is an unavoidable consequence of the perturbative nature of this
mechanism. This problem is related to the dilaton tadpole, which asks
for a redefinition of the
spacetime background. The class ii) also contains models with
supersymmetry breaking induced by internal magnetic fields.

\vskip 2mm
- {\large \bf Nonperturbative supersymmetry breaking}
\vskip 2mm
 
Despite of the serious progress achieved in the perturbative breaking
of supersymmetry
in Type I strings, we probably need nonperturbative effects for at
least one reason mentioned above, the dilaton stabilization
problem. Indeed, even if the problem can be circumvented searching
for a nontrivial background {\`a} la Fischler-Susskind, the explicit
example worked out in \cite{dm4} suggests that nonperturbative effects
in the string coupling (dilaton) appear in this background.
The present-day technology forces us to rely here on field theoretical
arguments, like holomorphy in supersymmetry. There are here several 
scenarios proposed
in the literature, which take advantage of the various brane 
configurations, each brane different and far away from ours being
a potential hidden sector breaking supersymmetry. A novelty in Type
I is that twisted fields $M_k$ can easily participate to supersymmetry
breaking. A simple example is provided by gaugino condensation
(\ref{8.16}) with a large class of K{\"a}hler potentials for twisted
moduli fields, and in particular the minimal one, $M_k^{\dagger} M_k$.

\section{Bulk physics: Neutrino and axion masses with large extra dimensions}

There is more and more convincing evidence for the existence of neutrino
masses and mixings, in light of the recent SuperKamiokande results
\cite{superk}. Any extension of the Standard Model should therefore
address this question, at least at a qualitative level. The most
elegant mechanism for explaining the smallness of neutrino masses
postulates the existence of right-handed neutrinos with very large
Majorana masses $10^{11}$ GeV $\le M \le 10^{15}$ GeV. Via the
seesaw mechanism \cite{seesaw} very small neutrino masses, of the
order of $m_{\nu} \sim v^2/M$, are generated, where $v \simeq 246$ GeV
is the vev of
the Higgs field. This suggests the presence of a large (intermediate or
GUT) scale in the theory, related to new physics. On the other
hand, low-scale string models do not have such a large scale and
therefore superficially have problems to acommodate neutrino masses.
Similarly, the strong CP problem in the Standard Model finds its most
natural explanation by postulating a global continuous $U(1)_{PQ}$ symmetry
with $U(1)_{PQ}[SU(3)]^2$ anomalies. In this case, the $\theta$
parameter of QCD
becomes a dynamical field $\theta \rightarrow \theta + (1/f) a$, where
$a$ is called the Peccei-Quinn {\it axion} \cite{pq}. The symmetry
$U(1)_{PQ}$ is spontaneously
broken at a large scale $f$ and, by instanton effects, an axion
potential is generated such that $\theta + (1/f) <a>=0$, dynamically solving
the strong CP problem. The experimentally allowed window for the axion
is considered to be   $10^{8}$ GeV $\le f \le  10^{12}$ GeV. 
These arguments were used in \cite{benakli} for arguing that the string
scale is likely to be at some intermediate value $M_I \sim 10^{11}$ GeV. 
  
In this Section it will be argued that there is actually a natural way
to find very small neutrino and invisible axion masses, taking
advantage of the fact that right-handed neutrinos and axions, that are
Standard Model gauge singlets, can be placed in the bulk space. 
These scenarios have interesting new features
compared to the standard 4d mechanisms due to the higher-dimensional
nature of the gauge singlets. 
 
\vskip 2mm
- {\large \bf Neutrino masses}
\vskip 2mm
 
The scenario we present here is based on the observation that
right-handed neutrinos can be put in the bulk of a very
large (mm size) compact space \cite{ddg2,addm,smirnov}, perpendicular to the
brane where we live. Consider for
simplicity the case of one family of neutrinos. The model consists of
our brane with the left-handed neutrino $\nu_L$ and Higgs field confined
to it
and one bulk Dirac neutrino,  $\Psi = (\psi_1,\bar\psi_2)^T$ in Weyl
notation, invading a space with (again for simplicity) one compact
perpendicular direction $y$. The compact direction is taken here to be an
orbifold $S^1/Z_2$, since as is well known circle compactifications are not
phenomenologically realistic. The $Z_2$ orbifold acts on the spinors as
$Z_2 \Psi (y)=\pm \gamma_5 \Psi (-y)$, so that one of the two-component
Weyl spinors, \eg, $\psi_1$, is even
under the $Z_2$ action $y\to -y$, while the other spinor
$\psi_2$ is odd. If the left-handed neutrino $\nu_L$ is restricted
to a brane located at the orbifold fixed point $y=0$,
$\psi_2$ vanishes at this point and so $\nu_L$ couples only to
$\psi_1$. This then results in a Lagrangian of the form
\ba
{\cal L} &=& - {1 \over 2} \int d^{4} x \,dy ~M_s\, \biggl\lbrace
{\bar\psi} i{\bar\gamma}^M \partial_M \psi
-  \partial_M {\bar\psi} i{\bar\gamma}^M \psi   \biggr\rbrace
                                 \nonumber\\
      && - ~\int d^4 x ~ \biggl\lbrace
         {\bar\nu}_L i{\bar\sigma}^\mu D_\mu \nu_L
         ~+~({\hat m} \nu_L \psi_1|_{y=0} + {\rm h.c.})     \biggr\rbrace~.
\label{10.1}
\ea
Here $M_s$ is the mass scale of the higher-dimensional fundamental
theory (\eg, a reduced Type~I string scale) and the spacetime
index $M$ runs over all five dimensions:  $x^M\equiv(x^\mu,y)$.
The first line describes the kinetic-energy term for
the 5d $\Psi$ field, while the second line describes the kinetic energy
of the 4d two-component neutrino field $\nu_L$,
as well as the coupling between $\nu_L$ and $\psi_1$.
Note that in 5d, a bare Dirac mass term for $\Psi$
would not have been invariant under the action of the $Z_2$ orbifold,
since $ {\bar \Psi}\Psi\sim \psi_1\psi_2 +$ h.c.

Now compactify the Lagrangian (\ref{10.1}) down to 4d,
expanding the 5d $\Psi$ field in Kaluza-Klein
modes.  The orbifold relations $\psi_{1,2}(-y)=\pm \psi_{1,2}(y)$
imply that the Kaluza-Klein decomposition takes the form
\be
     \psi_1(x,y) = {1\over \sqrt{2\pi R}}\,\sum_{n=0}^\infty
         \psi_1^{(n)}(x)\,\cos (ny/R) \ , \ 
     \psi_2(x,y) = {1\over \sqrt{2\pi R}}\,\sum_{n=1}^\infty
         \psi_2^{(n)}(x)\,\sin (ny/R)~ \ .
\label{10.2}
\ee

However, a more general possibility emerges naturally from the
Scherk-Schwarz compactification \cite{ss}.
Recall that our original 5d Dirac spinor field $\Psi$
is decomposed in the Weyl basis as $\Psi=(\psi_1,\bar\psi_2)^T$,
where $\psi_1$ and $\psi_2$ have the
mode expansions given in (\ref{10.2}).
Let us consider performing a local rotation in $(\psi_1,\psi_2)$
space of the form
\be
         \pmatrix{ \hat \psi_1 \cr \hat \psi_2 } ~\equiv~  U
         \pmatrix{ \psi_1 \cr \psi_2 } ~~~~~~~{\rm where}~~~
     U ~\equiv ~ \pmatrix {
               \cos (\omega y/R) & -\sin (\omega y/R) \cr
               \sin (\omega y/R) & \cos (\omega y/R) \cr}~ \ , 
\label{10.12}
\ee
with $\omega$ an (for the moment) arbitrary real number. The effect of 
the matrix $U$ in
(\ref{10.12}) is to twist the fermions after a $2 \pi R$ rotation on
$y$. Such twisted boundary conditions, as we have seen, are  
allowed in field and in string theory if  the higher-dimensional theory
has a suitable $U(1)$ symmetry. The 4d Lagrangian of the component
fields coming from the 5d Lagrangian is found from (\ref{10.1}) by
replacing everywhere $\psi_i \rightarrow {\hat \psi}_i$, and includes
the mass terms  
\ba
&&{\cal L}_{kin} = - {1 \over 2} \int dy ~M_s\, \biggl\lbrace
{\bar {\hat \psi}} i{\gamma}^5 \partial_5 {\hat \psi}
-  \partial_5 {\bar{\hat \psi}} i{\gamma}^5 {\hat \psi} 
\biggl\rbrace = \nonumber \\ 
&&- \sum_{n=0}^{\infty} \biggl\lbrace {n \over R} \psi_1^{(n)} 
\psi_2^{(n)} + {M_0 \over 2} \ ( \psi_1^{(n)} \psi_1^{(n)} + \psi_2^{(n)} 
\psi_2^{(n)}) +h.c \biggl\rbrace \ , \label{10.02} 
\ea
where $M_0= \omega / R$. 
For convenience, let us define the linear combinations
$N^{(n)}\equiv(\psi_1^{(n)}+\psi_2^{(n)})/\sqrt{2}$
and $M^{(n)}\equiv(\psi_1^{(n)}-\psi_2^{(n)})/\sqrt{2}$ for all $n>0$.
 
Inserting (\ref{10.12}), (\ref{10.02}) into (\ref{10.1}) and
integrating over the compactified dimension then yields
\ba
  {\cal L} &=& - \int d^4 x
     ~ \Biggl\lbrace
    {\bar\nu}_L i{\bar\sigma}^\mu D_\mu \nu_L
     + {\bar \psi}_1^{(0)} i{\bar\sigma}^\mu\partial_\mu \psi_1^{(0)}
     + \sum_{n=1}^\infty \left(
     {\bar N}^{(n)} i{\bar\sigma}^\mu\partial_\mu N^{(n)}
     +{\bar M}^{(n)} i{\bar\sigma}^\mu\partial_\mu M^{(n)} \right) \nonumber\\
    && ~~~~+~ \biggl\lbrace {1 \over 2} \, M_0 \,\psi_1^{(0)} \psi_1^{(0)}
        ~+~
         {1 \over 2} \sum_{n=1}^\infty \, \left\lbrack
         \left(M_0 + {n\over R}\right) N^{(n)} N^{(n)}
         + \left(M_0 - {n\over R}\right) M^{(n)} M^{(n)} \right\rbrack
            \nonumber\\
    && ~~~~+~ m \left[ \nu_L \psi_1^{(0)} +
          \nu_L \sum_{n=1}^\infty \left( N^{(n)}
         +   M^{(n)} \right) \right] ~+~ {\rm
h.c.}\biggr\rbrace\Biggr\rbrace~.
\label{10.3}
\ea
Here the first line gives the four-dimensional kinetic-energy terms,
while the second line gives the Kaluza-Klein and Majorana mass terms.
The third line of (\ref{10.3}) describes the coupling
between the 4d neutrino $\nu_L$ and the 5d field $\Psi$.
Note that in obtaining this Lagrangian it is necessary to rescale
the Kaluza-Klein modes  $\psi_1^{(0)}$, $N^{(n)}$, and $M^{(n)}$ 
so that their 4d kinetic-energy terms are canonically normalized.
This then results in a suppression of the Dirac neutrino mass $\hat m$
by the factor $(2\pi M_s R)^{1/2}$.
In the third line, we have therefore defined the effective Dirac
neutrino mass couplings 
\be
            m ~\equiv~  {{\hat m}\over \sqrt{2}\,\sqrt{\pi M_sR}}~ \ .
\label{10.4}
\ee
In the Lagrangian (\ref{10.3}), the Standard-Model neutrino
$\nu_L$ mixes with the entire tower of Kaluza-Klein states
of the higher-dimensional $\Psi$ field.
Indeed, if for simplicity we restrict our attention to the case of only one
extra dimension and define
\be
        {\cal N}^T ~\equiv~ (\nu_L, \psi_1^{(0)},
                  N^{(1)}, M^{(1)},
                  N^{(2)}, M^{(2)}, ...)~ \ ,
\label{10.5}
\ee
we see that the mass terms in the Lagrangian (\ref{10.3})
take the form $(1/2) ({\cal N}^T {\cal M} {\cal N}+{\rm h.c.})$,
where the mass matrix is 
\be
      {\cal M} ~=~ \pmatrix{
         0 &  m   &   m &   m  &   m  &  m & \ldots \cr
         m &  M_0 &   0  &   0  &   0  &  0  & \ldots \cr
         m &  0   &   M_0+1/R  &   0  &   0  &  0  & \ldots \cr
         m &  0   &   0  &   M_0-1/R  &   0  &  0  & \ldots \cr
         m &  0   &   0  &   0  &   M_0+2/R  &  0  & \ldots \cr
         m &  0   &   0  &   0  &   0   &  M_0-2/R  & \ldots \cr
         \vdots  &  \vdots &   \vdots  &   \vdots &   \vdots &  \vdots  &
\ddots \cr}~ \ .
\label{10.6}
\ee

Let us start for simplicity by disregarding the possible bare Majorana
mass term, setting $M_0=0$.
In this case, the characteristic polynomial which determines
the eigenvalues $\lambda$ of the mass matrix (\ref{10.6}) can
be worked out exactly and takes the form
\be
   \left\lbrack \prod_{k=1}^\infty \left({k^2\over R^2}-\lambda^2\right)
         \right\rbrack \,
    \left[\lambda^2-m^2 +2 \lambda^2 m^2 R^2\sum_{k=1}^\infty
    {1\over k^2-\lambda^2 R^2 }\right] ~=~ 0~ \ ,
\label{10.7}
\ee
clearly invariant under $\lambda\to -\lambda$.  From this
we immediately see that all eigenvalues fall into {\it degenerate},
pairs of opposite sign.
In order to solve this eigenvalue equation, it is convenient to note
that $\lambda = k/R$ is never a solution (unless of course $m=0$),
as the cancellation that would occur in the first factor
in (\ref{10.7}) is offset by the divergence of the second factor.
We are therefore free to disregard the first factor entirely, and
focus on solutions for which the second factor vanishes.
The summation in second factor can be performed exactly, resulting
in the transcendental equation
\be
          \lambda R ~=~ \pi (m R)^2 \,\cot ( \pi \lambda R)~ \ .
\label{10.8}
\ee
All the eigenvalues can be determined from this equation, as functions
of the product $mR$. This equation can be analyzed graphically \cite{ddg2}, 
and in the limit $mR\to 0$ (corresponding to $m\to 0$),
the eigenvalues are $k/R$, $k\in Z$, with a double eigenvalue at $k=0$.
Conversely, in the limit $mR\to\infty$, the eigenvalues with $k>0$ shift
smoothly toward $(k+1/2)/R$, while those with $k<0$ shift smoothly
toward $(k-1/2)/R$. Finally, the double zero eigenvalue splits toward the values $\pm 1/(2R)$.
In order to derive general analytical expressions valid
in the limit $mR\ll 1$, we can solve (\ref{10.8}) iteratively by
power-expanding the cotangent function.
To order ${\cal O}(m^5 R^5)$, this gives the solutions
\be
  \lambda_{\pm} ~=~ \pm m\,\left( 1- {\pi^2 \over 6} m^2 R^2 + ...\right)
      ~,~~~~~\lambda_{\pm k} ~=~
        \pm {k\over R}\,
        \left(1 +  {m^2 R^2 \over k^2} - {m^4 R^4\over k^4} +...\right)~,
\label{10.10}
\ee
where $\lambda_{\pm k}$ are the two eigenvalues at each Kaluza-Klein
level $k$ and $\lambda_\pm$ are the ``light'' eigenvalues
at $k=0$.
Finally, it is also straightforward to solve explicitly for the light
mass eigenstates $|\tilde \nu_\pm\rangle$ corresponding to $k=0$.
To leading order in $mR$, we find
\be
       |\tilde \nu_\pm \rangle ~=~ {1\over \sqrt{ 2}} \,
          \left\lbrace
         \left( 1- {\pi^2\over 6} m^2 R^2 \right)
        | \nu_L \rangle  ~\pm~ |\psi^{(0)}_1\rangle
            - mR \,\sum_{k=1}^\infty
         {1\over k}\,\left\lbrack |N^{(k)}\rangle -|M^{(k)}\rangle
                   \right\rbrack \right\rbrace~.
\label{10.11}
\ee
This implies that the overlap between the light mass
eigenstates and the neutrino gauge eigenstate is generically less
than half in this scenario. The important prediction of this scenario is
that the gauge neutrino and the (lightest) sterile neutrino are degenerate in
mass, a possibility that can be experimentally tested.

Let us now return to the more general case $M_0 \not=0$.
To this end, it is useful to define
\be
         k_0 ~\equiv~ \lbrack M_0 R\rbrack ~,~~~~~
         \epsilon ~\equiv~ M_0 - {k_0\over R}~ , 
\label{10.13}
\ee
where $[x]$ denotes here the integer nearest to $x$.
Thus, $\epsilon$ is the smallest diagonal entry in the mass matrix
(\ref{10.6}), corresponding to the excited Kaluza-Klein state
$M^{(k_0)}$.  In other words, $\epsilon \equiv M_0$ (modulo $R^{-1}$)
satisfies $-1/(2R) < \epsilon \leq 1/(2R)$.
The remaining diagonal entries in the mass matrix
can then be expressed as $\epsilon \pm k'/R$, where $k'\in Z^{+}$.
Reordering the rows and columns of our mass matrix,
we can therefore cast it into the form
\be
      {\cal M} ~=~ \pmatrix{
         0 &  m   &   m &   m  &   m  &  m & \ldots \cr
         m &  \epsilon &   0  &   0  &   0  &  0  & \ldots \cr
         m &  0   &   \epsilon+1/R  &   0  &   0  &  0  & \ldots \cr
         m &  0   &   0  &   \epsilon-1/R  &   0  &  0  & \ldots \cr
         m &  0   &   0  &   0  &   \epsilon+2/R  &  0  & \ldots \cr
         m &  0   &   0  &   0  &   0   &  \epsilon-2/R  & \ldots \cr
         \vdots  &  \vdots &   \vdots  &   \vdots &   \vdots &  \vdots  &
\ddots \cr}~ \ .
\label{10.14}
\ee
While this is of course nothing but the original mass matrix (\ref{10.6}),
the important consequence of this rearrangement is that the {\it heavy}\/ mass
scale $M_0$ has been replaced by the {\it light} mass scale $\epsilon$.
Unlike $M_0$, we see that $|\epsilon| \sim {\cal O}(R^{-1})$.
Thus, the heavy Majorana mass scale $M_0$ completely {\it decouples}
from the physics.  Indeed, the value of $M_0$ enters the results only
through its determinations
of $k_0$ and the precise value of $\epsilon$.
Therefore, interestingly enough, the presence of the infinite
tower of regularly-spaced Kaluza-Klein states ensures
that only the value of $M_0$ modulo $R^{-1}$
plays a role.

The easiest way to solve (\ref{10.14}) for
the eigenvalues $\lambda_{\pm}$ is to integrate out the Kaluza-Klein modes.
It turns out that there are two relevant cases to consider,
depending on the value of $\epsilon$.  If $|\epsilon|  \gg m$
(which can arise when $mR\ll 1$), {\it all}\/ of the Kaluza-Klein
modes are extremely massive relative to $m$,
and we can integrate them out to obtain an effective
$\nu_L \nu_L$ mass term of size
\ba
   |\epsilon|  \gg m:~~~~~~~~
      m_\nu &=&  m^2/\epsilon ~+~ m^2 \,\sum_{k'=1}^\infty
         \left( {1\over \epsilon+k'/R} + {1\over \epsilon-k'/R}
\right)\nonumber\\
         &=& \pi m^2 R \cot\left(\pi R\epsilon\right)~.
\label{10.15}
\ea
We shall discuss the special case $\epsilon= 1/2R$ later on.
Alternatively, if $|\epsilon| \not \gg m $, the lightest Kaluza-Klein
mode $M^{(k_0)}$ should not be integrated out, and the end result is an effective
$\nu_L \nu_L$ mass term of size $\mu$, where
\ba
   |\epsilon| \not\gg m:~~~~~~~~
       \mu &\equiv& -m^2 \,\sum_{k'=1}^\infty
         \left( {1\over \epsilon+k'/R} + {1\over \epsilon-k'/R}
\right)\nonumber\\
         &=& {m^2\over \epsilon} - \pi m^2 R \cot\left(\pi R\epsilon\right)~.
\label{10.16}
\ea
Note that $\mu \to 0$ smoothly as $\epsilon\to 0$, with $\mu$ otherwise
of size ${\cal O}(m^2 R)$.
Diagonalizing the final $2\times 2$ mass matrix
mixing $\nu_L$ and $M^{(k_0)}$
in the presence of this mass term then yields
\be
   |\epsilon|  \not\gg m:~~~~~~~~
   \lambda_\pm ~=~ {1 \over 2} \left\lbrack
     (\mu+\epsilon) ~\pm~
          \sqrt{ (\mu-\epsilon)^2 + 4 m^2 }\right\rbrack~ \ . \label{10.17}
\ee
Thus, as $M_0 \rightarrow 0$ (or as $M_0 \rightarrow n/R$ where $n\in Z$),
we see that $\epsilon,\mu \to 0$, and we recover
the eigenvalues given in (\ref{10.10}).

We therefore conclude that, although we may have started with
a bare Majorana mass $M_0 >> R^{-1}$,
in all cases the final neutrino mass remains of order $m^2 R$.
Even though we might have expected
a neutrino mass of order $m^2/M_0$ from
the mixing between $\nu_L$ and the original zero-mode $\psi_1^{(0)}$,
the contribution $m^2/M_0$ from the
zero-mode is completely canceled by the summation over
the Kaluza-Klein tower, while the
seesaw between $\nu_L$ and $M^{(k_0)}$ becomes dominant.
It is this feature that causes the heavy scale $M_0$ to be effectively
replaced by the radius $R^{-1}$, so that once again
our effective seesaw scale
is $M_{\rm eff}\sim {\cal O}(R^{-1})$.

In string theory, however,  there are additional topological
constraints (coming from the preservation of the form of the worldsheet
supercurrent) that permit only {\it discrete}\/ values of $\omega$ \cite{kp}.
In particular, in a compactification from five to four dimensions,
this restriction allows only one non-trivial
possibility, $\omega=1/2$.
Taking $\omega=1/2$ then implies $\psi_{1,2}(2\pi R) = -\psi_{1,2}(0)$, which
shows that lepton number is broken globally (although not locally)
as the spinor is taken around the compactified space. 
In order to obtain the corresponding neutrino mass,
we note that for $\epsilon=1/2R$ the assumption $mR\ll 1$
translates into $\epsilon >> m$, whereupon the result (\ref{10.15})
is valid.  Thus, for $\epsilon= 1/2R$ we find the remarkable
result that $m_\nu =0$ !
In obtaining this result, one might worry that (\ref{10.15}) is
only approximate because it relies on the procedure of integrating
out the Kaluza-Klein states rather than on a full diagonalization of
the corresponding mass matrix.  However, it is straightforward to show
that when $\epsilon=1/2R$, the characteristic eigenvalue
equation $\det ({\cal M}-\lambda I)=0$ for the mass matrix
(\ref{10.6}),(\ref{10.14}) becomes
\be
      \lambda R
       \left\lbrack
     \prod_{k=1}^\infty
            (\lambda^2 R^2 - (k -{1 \over 2})^2) \right\rbrack
        \left\lbrack
       1 - 2 m^2 R^2 \sum_{k=1}^\infty {1\over  \lambda^2 R^2 - (k-1/2)^2 }
       \right\rbrack ~=~ 0~.
\label{10.19}
\ee
This has an exact trivial solution $\lambda=0$, corresponding to an
exactly massless neutrino.
Indeed, the characteristic polynomial for the
mass matrix in this case has the form
\be
       \lambda R ~=~ -\pi (mR)^2 \,\tan\left(\pi \lambda R\right)~ \ . 
\label{10.20}
\ee
It is then clear than the zero eigenvalue is always present,
irrespective of the value of the radius.
In fact, by changing the value of $M_0$, we see that
it is possible to smoothly {\it interpolate}\/ between
the scenario with $M_0=0$ and the scenario we are discussing here \cite{ddg2}.
This also provides another explanation of why only
the value $\epsilon \sim M_0$ (modulo $R^{-1}$) is relevant physically.
The regular, repeating aspect of the infinite towers of Kaluza-Klein
states is now manifested graphically in the periodic nature of
the cotangent function.

We can also solve for the full spectrum of eigenvalues as a function
of $mR$. We find that the
non-zero eigenvalues are identical to those given in (\ref{10.10})
for $k\not=0$, but now $k\to k-1/2$.  
Note that the massless neutrino eigenstate
is {\it primarily}\/ composed of the neutrino gauge
eigenstate $\nu_L$, for $mR\ll 1$.
Although this neutrino mass eigenstate also contains
a small, non-trivial admixture of Kaluza-Klein states, its dominant
component is still the
gauge-eigenstate neutrino $\nu_L$, as required phenomenologically.
It should be stressed that this combined neutrino mass eigenstate is
exactly massless in the limit
that the full, infinite tower of Kaluza-Klein states participates
in the mixing\footnote{Actually, our field theory approach breaks down
for KK masses of the order of the fundamental string scale $M_s$. If
we cut our summation at $k_{max} = RM_s$,
the physical neutrino is not exactly massless anymore, but
aquires a small mass $m_{\nu} \sim m^2/M_s$. For
phenomenologically interesting values $m \sim R^{-1} \simeq 10^{-2} eV$
and $M_s \sim $ TeV, this mass is however negligibly small $m_{\nu} \sim
10^{-15} eV$.}. This result is valid
{\it regardless}\/ of the value of neutrino Yukawa coupling $m$ or of
the scale $R^{-1}$ of the Kaluza-Klein states.

It is also interesting to notice that the desired value of
the Majorana mass $M_0=1/2R$ emerges naturally from a Scherk-Schwarz 
decomposition, for reasons that are {\it topological}\/ and hence do not 
require any fine-tuning. It should however be stressed that in this case
lepton number is not broken if we consider the full tower of KK
states. Indeed, it can be easily shown that for $\omega = 1/2$ KK 
states pair up so that the full lagrangian still preserves lepton number. 

The scenario(s) presented have also other interesting consequences.
The neutrino eigenstate can now oscillate into an infinite tower of
right-handed KK neutrinos with a probability that can be reliably
estimated and experimentally tested. Moreover, even if in the last
scenario presented the physical neutrino is massless, its
probability of oscillation into the tower of KK states is nonvanishing.
In particular, a neutrino mass difference $\Delta m \sim 10^{-2} eV$,
that fits the experimental data, could well be explained by an oscillation of
the massless neutrino into the first KK state, for a radius $R^{-1} \sim
10^{-2} eV$, precisely in the mm region we are interested in !  

\vskip 2mm
- {\large \bf Bulk axion masses}
\vskip 2mm
 
The most elegant explanation of the strong CP problem is provided by the
Peccei-Quinn (PQ) mechanism \cite{pq}, in which the CP violating angle $\bar\Theta$
( $\bar\Theta$ by definition includes the contribution of weak
interactions) is set to zero
dynamically as a result of a global, spontaneously broken $U(1)_{PQ}$ Peccei-Quinn
symmetry.  However, associated with this symmetry there is a new Nambu-Goldstone boson,
the axion \cite{ww}, which essentially replaces
the $\bar\Theta$ parameter in the effective Lagrangian.  
This then results in an effective Lagrangian of the form 
\be
         {\cal L}^{\rm eff}~=~ {\cal L}_{\rm QCD} ~-~ 
         {1 \over 2} \partial_\mu a \partial^\mu a ~+~
             {a \over f_{\rm PQ}}\, \xi\, 
            {g^2\over 32\pi^2} F^{\mu \nu}_a \tilde F_{\mu\nu a}~ \ , \label{10.22}
\ee
where $f_{\rm PQ}$ is the axion decay constant, associated with the scale
of PQ symmetry breaking.  Here $\xi$ is a model-dependent parameter 
describing
the PQ transformation properties of the ordinary fermions, and we have not
exhibited other terms in the Lagrangian that describe axion/fermion couplings.
The mass of the axion is then expected to be of the order
\be
           m_a ~\sim~ {\Lambda_{\rm QCD}^2\over f_{\rm PQ}}~ \ , \label{10.23}
\ee
where $\Lambda_{\rm QCD} \approx 250$ MeV;  likewise, the couplings of axions
to fermions are suppressed by a factor of $1/f_{\rm PQ}$.
Thus, heavier scales for PQ symmetry breaking generally imply lighter axions
that couple more weakly to ordinary matter.

Ordinarily, one might have preferred to link the scale $f_{\rm PQ}$ to
the scale of electroweak symmetry breaking, thus implying an axion mass
$m_a\approx {\cal O}(10^2)$ keV. However, so far all experimental searches for
such axions have been unsuccessful \cite{kim}, and indeed only a narrow
allowed window exists:
\be
           10^{10} \, {\rm GeV} ~\le ~ f_{\rm PQ} ~\le~
           10^{12} \, {\rm GeV} ~ \ , \ 10^{-5} \, {\rm eV} ~\le ~ m_a ~\le~
           10^{-3} \, {\rm eV} ~.
\label{10.24}
\ee
The resulting axion is exceedingly light
and its couplings to ordinary matter are exceedingly suppressed.
These bounds generally result from various combinations
of laboratory, astrophysical, and cosmological 
constraints.  In all cases, however, the crucial ingredient is the
correlation between the {\it mass}\/ of the axion and the strength of its
 {\it couplings}\/ to matter, since both are essentially determined by
the single parameter $f_{\rm PQ}$.

This situation may be drastically altered
in theories with large extra spatial dimensions.
We shall consider the
consequences of placing the PQ axion in the ``bulk'' (\ie, perpendicular
to the brane that contains the Standard Model) so that 
it accrues an infinite tower of Kaluza-Klein excitations \cite{add,cty,ddg3}.  
This is reminiscent of the option of placing the right-handed neutrino in the
bulk discussed above. In order to generalize the PQ mechanism,
we will assume that there exists a complex scalar field $\phi$ in
higher dimensions which transforms under a global $U(1)_{\rm PQ}$ symmetry
$ \phi ~\rightarrow~ e^{i\Lambda} \phi$. 
This symmetry is assumed to be spontaneously broken by the bulk dynamics
so that
$\langle \phi \rangle = f_{\rm PQ}/\sqrt{2}$, where $f_{\rm PQ}$ is
the energy scale associated with the breaking of the PQ symmetry.
We thus write our complex scalar field $\phi$ in the form
\be
       \phi ~\approx~ {f_{\rm PQ} \over \sqrt{2}} \,e^{i a/f_{\rm PQ}} \
, \label{10.27}
\ee
where $a$ is the Nambu-Goldstone boson (axion) field.  
If we concentrate on the case of 5d for concreteness,
the kinetic-energy term for the scalar field takes the form
\be
  {\cal S}_{\rm K.E.} ~=~ - \int d^4x\,dy~ M_s \,\partial_M \phi^\ast
   \partial^M \phi ~=~ - 
         \int d^4x\,dy~ M_s \, {1 \over 2}
    \partial_M a \partial^M a~ \ , \label{10.28}
\ee
where we have neglected the contributions of the radial mode.
Here $x^\mu$ are the 4d coordinates and
$y$ is the coordinate of the fifth dimension.
Note that there is no mass term for the axion, 
as this would not be invariant under the $U(1)_{\rm PQ}$ transformation
$a~\rightarrow~ a + f_{\rm PQ}\Lambda $.
Furthermore, as a result of the chiral anomaly, we will
also assume a bulk/boundary coupling of the form 
\be
     {\cal S}_{\rm coupling}~=~ \int d^4 x \, dy ~{\xi\over f_{\rm PQ}}\,  
   {g^2\over 32\pi^2} \,a \,F^{\mu\nu}_a \tilde F_{\mu\nu a} \,\delta (y)~
 \ , \label{10.29}
\ee
where $F_{\mu\nu a}$ is the (4d) QCD field strength confined to a four-dimensional subspace
(\eg, a D-brane) located at $y=0$. 
Thus, our effective 5d action takes the form
\be
   {\cal S}_{\rm eff} ~=~ \int d^4x~dy~ \left[ \, - {1 \over 2} \,M_s\,\partial_M a
   \partial^M a ~+~ 
     {\xi \over f_{\rm PQ}}\,  
   {g^2\over 32\pi^2} \,a \,F^{\mu\nu}_a \tilde F_{\mu\nu a} \,\delta
(y)\right]~ \ . \label{10.30}
\ee
While we have assumed that the spontaneously broken $U(1)_{\rm PQ}$ is parametrized
by $f_{\rm PQ}$, one still has to address the fact that gravitational effects can 
also break the $U(1)_{\rm PQ}$ symmetry, since gravitational 
interactions generically break global symmetries \cite{kw}.  
We will assume, however, that the gravitational contributions to the axion mass 
are indeed suppressed, and that $U(1)_{\rm PQ}$ remains a valid symmetry even in
the presence of gravitational effects.

In order to obtain an effective 4d theory, our next step is to
compactify the fifth dimension.  For simplicity,
we shall assume that this dimension is compactified on the $Z_2$
orbifold that we considered in the neutrino case. This implies that the axion
field will have a Kaluza-Klein decomposition of the form
\be
    a(x^\mu,y)~= {1 \over \sqrt{2 \pi R}} ~\sum_{n=0}^{\infty} \,a_n(x^\mu) 
\,\cos\left({n y\over R}\right) \ , \label{10.31}
\ee
where $a_n(x^\mu)\in R$ are the Kaluza-Klein modes and 
where we have demanded that the axion field be 
symmetric under the $Z_2$ action (in order to have a zero-mode 
that we can identify with the usual 4d axion).

It is also interesting to note that for the
Kaluza-Klein axion modes $a_n$, the Peccei-Quinn transformation 
takes the form
\be
            \cases{      a_0 \to a_0 + f_{\rm PQ} \Lambda & \cr  
                         a_k \to a_k~ ~~~~~~~~~~~~~{\rm for~all}~k>0~. & \cr}
\ . \label{10.32}
\ee
Thus, only $a_0$ serves as the true axion transforming 
under the PQ transformation, while the excited Kaluza-Klein modes $a_k$ 
remain invariant.

Substituting (\ref{10.31}) into (\ref{10.30}) and integrating over the
fifth dimension, we obtain the effective four-dimensional Lagrangian density
\be
   {\cal L}_{\rm eff} ~=~ {\cal L}_{\rm QCD} ~-~ {1 \over 2} 
   \sum_{n=0}^{\infty} (\partial_\mu a_n)^2
            ~-~ {1 \over 2} \sum_{n=1}^{\infty} {n^2\over R^2} a_n^2 
                ~+~
        {\xi \over \hat f_{\rm PQ}}\, {g^2\over 32\pi^2}
    \left(\sum_{n=0}^{\infty} r_n a_n\right)~
            F^{\mu\nu}_a \tilde F_{\mu\nu a}~ \ , 
\label{10.33}
\ee
where  
\be
        r_n ~\equiv ~\cases{ 1 & if $n=0$\cr
                          \sqrt{2} &  if $n>0$ ~.\cr} 
\label{10.34}
\ee
Note that in order to obtain (\ref{10.33}) and (\ref{10.34}),  we had to 
rescale each of the Kaluza-Klein modes $a_n$ in 
order to ensure that they have canonically normalized kinetic-energy terms. 
We have also defined $\hat f_{\rm PQ}\equiv (V M_s)^{1/2} f_{\rm PQ}$,
where $V$ is the volume of our compactified space.
For $\delta$ extra dimensions, this definition generalizes to
$\hat f_{\rm PQ}\equiv (V M_s^\delta)^{1/2} f_{\rm PQ}$.
Note that while $f_{\rm PQ}$ sets the overall mass scale for
the breaking of the Peccei-Quinn symmetry, it is the volume-renormalized
quantity $\hat f_{\rm PQ}$ that parametrizes the coupling between the axion
and the gluons. In general, since $M_s \gg R^{-1}$,
we find that $\hat f_{\rm PQ} \gg f_{\rm PQ}$.
Therefore, as pointed out in Ref.~\cite{add}, this
volume-renormalization of the brane/bulk coupling can be used
to obtain sufficiently suppressed axion/gauge-field couplings
even if $f_{\rm PQ}$ itself is taken to be relatively small.
Notice that, if we were to take $\delta=n$ for the 
current axion case, (\ref{8.3}) would
imply either that $\hat f_{\rm PQ}\sim M_{\rm Planck}$ (which would presumably
overclose the universe), or $M_s\ll {\cal O}$(TeV) (which would clearly 
violate current experimental bounds).  Therefore, if we assume an 
isotropic compactification with all equal radii, an intermediate
scale $\hat f_{\rm PQ}$ can be generated only if $\delta<n$.
In other words, the
axion must be restricted to a {\it subspace}\/ of the full higher-dimensional bulk.

Let us now proceed to verify that this 
higher-dimensional PQ mechanism still cancels the CP-violating phase,
and use this to calculate the {\it mass}\/ of the axion.
In the one-instanton dilute-gas approximation, it is straightforward
to show that 
\be
      \langle F^{\mu\nu}_a \tilde F_{\mu\nu a}\rangle ~=~
      - \Lambda_{\rm QCD}^4\, \sin\left(
           {\xi\over \hat f_{\rm PQ}} 
        \sum_{n=0}^{\infty} r_n a_n + \bar \Theta \right)~ \ . \label{10.35}
\ee
This gives rise to an effective potential for the axion modes:
\be
    V(a_n)  ~=~ {1 \over 2} \sum_{n=1}^{\infty} {n^2\over R^2} a_n^2 
        ~+~ {g^2\over 32\pi^2}\, \Lambda_{\rm QCD}^4
       \left[ 1 - \cos\left( 
           {\xi\over \hat f_{\rm PQ}} 
        \sum_{n=0}^{\infty} r_n a_n + \bar \Theta \right)\right]~. \label{10.36}
\ee
In order to exhibit the PQ mechanism, we now minimize the axion effective potential,
\be
      {\partial V\over \partial a_n} ~=~
           {n^2\over R^2} a_n ~+~  r_n
              {\xi\over \hat f_{\rm PQ}}\, {g^2\over 32\pi^2}\,
            \Lambda_{\rm QCD}^4\, \sin\left( {\xi\over \hat f_{\rm PQ}} 
           \sum_{n=0}^{\infty}r_n a_n
             + \bar\Theta \right)~=~0~, \label{10.37}
\ee
obtaining the unique solution
\ba
     \langle a_0 \rangle &=&  
         {\hat f_{\rm PQ}\over\xi}(-\bar\Theta + \ell \pi) ~,~~~ \ell \in
2 Z~\nonumber\\
         \langle a_k \rangle &=&  0~~~~       {\rm for~all} ~k>0~.
\label{10.38}
\ea
Note that while any value $\ell\in Z$ provides an extremum of the potential,
only the values $\ell\in 2 Z$ provide the desired {\it minima}.
Thus, this higher-dimensional Peccei-Quinn mechanism still
solves the strong CP problem:  
$a_0$ is the usual Peccei-Quinn axion which solves the strong CP problem 
by itself by cancelling the $\bar\Theta$ angle,
while all of the excited Kaluza-Klein axions $a_k$ for $k>0$ have 
vanishing VEVs.
This makes sense, since only $a_0$ is a true massless Nambu-Goldstone field from
the 4d perspective (see the PQ transformation properties
(\ref{10.32})).  
However, these excited Kaluza-Klein 
axion states nevertheless have a drastic effect on the axion mass matrix.
Indeed, the mass matrix derived from (\ref{10.36}) is
\be
     {\cal M}^2_{nn'} ~\equiv~ {n^2\over R^2} \delta_{nn'} ~+~ 
      \xi^2 {g^2\over 32\pi^2} \, 
     {\Lambda_{\rm QCD}^4 \over \hat f_{\rm PQ}^2} \,r_n r_{n'}\,\cos\left(
           {\xi\over \hat f_{\rm PQ}}
       \sum_{n=0}^\infty r_n a_n
          + \bar\Theta\right)\Biggl |_{\langle a \rangle}~, \label{10.39}
\ee
and in the vicinity of the minimum (\ref{10.38}) becomes
\be
          {\cal M}^2_{nn'} ~=~ {n^2\over R^2} \delta_{nn'} ~+~ 
           \xi^2 {g^2\over 32\pi^2} {\Lambda_{\rm QCD}^4\over \hat f_{\rm PQ}^2} 
              \,r_n r_{n'}~ \ .
\label{10.40}
\ee
Let us now define
\be
     m^2_{\rm PQ} \equiv \xi^2 {g^2\over 32\pi^2} {\Lambda_{\rm QCD}^4 
          \over \hat f^2_{\rm PQ}} \ , \ 
     y \equiv {1\over m_{\rm PQ} R}~ \ ,
\label{10.41}
\ee
so that $m_{\rm PQ}$ is the expected mass 
that the axion would ordinarily have acquired in four dimensions
(depending on $\hat f_{\rm PQ}$ rather than $f_{\rm PQ}$ itself),
and $y$ is the ratio of the scale of the extra dimension and $m_{\rm PQ}$.
Our mass matrix then takes the form
\be
          {\cal M}_{nn'}^2 ~=~ 
            m_{\rm PQ}^2 \, \left( r_n r_{n'} ~+~ y^2 \,n^2 \,\delta_{nn'} \right)~,
\label{10.42}
\ee
or equivalently
\be
      {\cal M}^2 ~=~  m_{\rm PQ}^2 \, \pmatrix{   1 & \sqrt{2} & \sqrt{2} & \sqrt{2} &\ldots \cr
                  \sqrt{2} & 2+y^2 & 2 & 2& \ldots \cr
                  \sqrt{2} & 2 & 2+4 y^2 & 2& \ldots \cr
                  \sqrt{2} & 2 & 2 & 2+9y^2& \ldots \cr
                  \vdots & \vdots & \vdots & \vdots  & \ddots\cr}~.
\label{10.43}
\ee
Note that the usual Peccei-Quinn case corresponds to the upper-left 
$1\times 1$ matrix, leading to the expected result ${\cal M}^2= m^2_{\rm PQ}$.
Thus, the additional rows and columns reflect the extra KK states, 
and their physical effect is to pull the lowest eigenvalue of this matrix
away from $m^2_{\rm PQ}$.

Deriving the condition for the eigenvalues of this matrix is straightforward.
Let us denote the eigenvalues of this matrix as $\lambda^2$
rather than $\lambda$ because this is a (mass)$^2$ matrix.
We then find that the eigenvalues are the solutions to the  
transcendental equation
\be
         {\pi \tilde \lambda\over y} \, 
            \cot\left( {\pi \tilde \lambda\over y}\right) ~=~
            \tilde \lambda^2~ \ , 
\label{10.44}
\ee
where we have defined the dimensionless eigenvalue
$\tilde \lambda ~\equiv~ \lambda / m_{\rm PQ}$. 
In terms of dimensionful quantities, this transcendental equation
takes the equivalent form\footnote{
Interestingly, this eigenvalue equation is identical to the one that
emerges \cite{ddg2} (see eq.(\ref{10.8}) of the previous 
paragraph) 
when the right-handed neutrino $\nu_R$ is placed in the bulk,
with the mass scale $m_{\rm PQ}$ in the axion case 
corresponding to the Dirac coupling $m$ in
the neutrino case. Remarkably, this correspondence exists
even though the axion and right-handed neutrino have different spins, 
and even though the mechanisms for mass generation are completely
different in the two cases.}  
\be
      \pi R \lambda \,\cot(\pi R\lambda) ~=~  {\lambda^2\over m_{\rm PQ}^2 } ~.
\label{10.46}
\ee
We can check that (\ref{10.46}) makes sense as $R\to 0$.
In this limit, the KK states become infinitely heavy and decouple, and
we are left with the lightest eigenvalue $\lambda= m_{\rm PQ}$.
As $R$ increases, the effect of the additional large dimension is felt
through a reduction of this lowest eigenvalue and, as a result, the mass 
of the lightest axion decreases \cite{ddg3}.

One important consequence, easy to check plotting (\ref{10.46}), is that the
lightest axion mass eigenvalue $m_a$ is strictly bounded by the radius
\be
          m_a ~\le~ {1 \over 2} \,R^{-1}~.
\label{10.47}
\ee
This result holds {\it regardless}\/ of the value of $m_{\rm PQ}$.
Thus, in higher dimensions, when $m_{\rm PQ} \ge 1/2R$, 
the size of the axion mass is set by the 
radius $R$ and not by the Peccei-Quinn scale $f_{\rm PQ}$.  
and therefore the Peccei-Quinn scale essentially {\it decouples}\/ from the axion mass.
Indeed, as long as $m_{\rm PQ} \ge 1/2R$, we see that $m_a \le 1/2R$
{\it regardless}\/ of the specific values of $m_{\rm PQ}$ or 
$\Lambda_{\rm QCD}$. 

This observation has a number of interesting implications.
First, an axion mass 
in the allowed range (\ref{10.24}) is already achieved for 
$R$ in the submillimeter range, independently of $m_{\rm PQ}$.
Second, surprisingly $m_{\rm PQ}$ can 
still be lowered or raised at will without violating the 
constraint (\ref{10.24}), provided $m_{\rm PQ} \ge 1/2R$. 
In other words, having already satisfied the axion {\it mass}\/
constraints by appropriately choosing the value of $R$,  
we are now essentially free to tune $m_{\rm PQ}$ 
(or equivalently the fundamental Peccei-Quinn symmetry breaking
scale $f_{\rm PQ}$)  in such a way as to 
weaken the axion couplings to matter 
and make the axion sufficiently invisible. 
This may therefore provide a new method of obtaining an invisible 
axion.
\section{Low-scale string predictions for accelerators}

One of the main motivations for low-scale string theories comes from the
possibility of testing them at
future colliders. Indeed, virtual string (oscillator) states appear in
all string amplitudes, and in particular in tree-level Veneziano-type
amplitudes, and give deviations from the field-theory amplitudes for
energies $E \le M_I$.
In addition, there are effects of gravitational Kaluza-Klein states 
\cite{grw} both via their direct production and via indirect (virtual)
effects in various cross-sections.  
This paragraph is devoted to the direct evaluation in the Type I string of the
relevant amplitudes, that were estimated in a field-theory context in various
papers \cite{grw}. We will argue, using the results obtained in 
\cite{dm2} (see also \cite{cpp}), that the full
string amplitudes contain some new features that are relevant for the future 
accelerator searches. 

An important notion that appear in string computations is that of {\it
form factor}. In our case we are interested
in the form factor in the brane-brane-bulk vertex and we shall consider,
for definiteness, bulk gravitons. Let us 
assume for the moment that the form factor can be described by the local lagrangian  
\be
{\cal L}_{int} = \int d^4 x d^{\delta} y \ h_{\mu \nu}(x,{\bf y}) B({\bf y}) T^{\mu \nu} (x) \ , \label{12.01}
\ee
where $h_{\mu \nu}$ denotes the graviton fluctuations, $T^{\mu \nu}$
denotes the matter energy-momentum tensor and $B({\bf y})$
(which could also contain derivatives and could even be a nonlocal function)
describes the brane ``thickness''. Defining the form factors $g_{\bf
m}$ and the graviton Kaluza-Klein modes
$h_{\mu \nu}^{\bf m}$ as (for simplicity here we compactify on circles)
\be
B({\bf y}) = \sum_{\bf m} e^{- i {\bf m} {\bf y}} g_{\bf m} \ , \ 
h_{\mu \nu} (x,{\bf y}) = \sum_{\bf m} e^{i {\bf m} {\bf y}} \ h_{\mu \nu}^{(\bf m)}(x) \ , \label{12.02}
\ee
from the KK expansion, we find 
\be
{\cal L}_{int} = \sum_{\bf m} \int d^4 x  \ g_{\bf m} \ h_{\mu
\nu}^{(\bf m)}(x) T^{\mu \nu} (x) \ . \label{12.03}
\ee
The ``thin brane'' approximation
$g_{\bf m} \equiv g= {\rm cst}$,
or equivalently $B(\bf y) \sim \delta (\bf y)$ is widely used in the
phenomenological literature. This however leads to UV
divergences in virtual processes for
$\delta \ge 2$ coming from KK states of very large mass. A typical
procedure to deal with this difficulty, justified by
field-theory considerations \cite{bkny} or by the analogy with heterotic form
factors \cite{antoniadis}, is to suppress the interactions 
with heavy KK gravitons introducing a form factor $g_{\bf m}$ or,
equivalently, a the brane thicknes $B(\bf y)$, of the form
\be
g_{\bf m} \sim e^{-{a {\bf m}^2 \over R^2 M_I^2}} \ , \ B({\bf y}) \sim 
({\pi R^2M_I^2 \over a})^{\delta \over 2} 
e^{-{\delta R^2M_I^2 {\bf y}^2 \over a}} \ , \label{12.04}
\ee    
where $a$ is a (possibly dependent on the string coupling) constant whose
value depends on the model.
One of the main purposes of this Section is to compute the D-brane
string analog of form factors. It will be shown
in particular that (\ref{12.04}) reproduces only the {\it on-shell}
string form factor. We will find that its
off-shell extension is nonlocal and has a different form (see (\ref{12.17})
below), that does not regularize the divergences of virtual
gravitational exchange. The resolution of this apparent puzzle, that will
be described in the second part of this section, is that
in the Type I string this divergence is actually an IR divergence 
and {\it not} an UV one. Therefore,
string theory does not regulate these divergences, that should instead
be cured by the usual procedures the IR divergences are treated in
field theory.   

\vskip 2mm
- {\large \bf Emission of real gravitons}
\vskip 2mm

In the first part of this Section we discuss tree-level string
amplitudes with three gauge bosons and one internal (massive) excitation
of the graviton. For theories with low string scale and (sub)millimeter
dimensions, this type of process is one of the best signals for future
accelerators and was studied in field theory in
\cite{grw}. Schematically, the amplitude for the emission of one 
massive graviton in field theory is Planck suppressed $1/M_P$. The
inclusive cross section for the emission of gravitons of mass less than
the characteristic energy scale of the process $E$ is then proportional to
\be
\sigma_{FT} \sim {1 \over M_P^2} \sum_{m_i=0}^{RE} 1 \sim 
{ E^{ \delta} \over M^{2+ \delta}} \ , \label{12.05} 
\ee 
where in the last line we have used the relation $M_P^2 \sim R^{\delta}
M^{2+\delta}$. In (\ref{12.05}) $M$ is the effective Planck scale
of the higher-dimensional theory, used in most phenomenological papers
on the subject, whose relation to the string scale $M_I$ in toroidal
compactifications is
\be
M/M_I = (1 / \pi)^{1/8} \alpha^{-1/4} \ , \label{12.005}
\ee
where $\alpha=g^2 /(4 \pi)$ and $g$ is the gauge coupling. Therefore,
taking the electromagnetic and
the strong coupling as extreme values, we find $ 1.6 \le M/M_I \le 3$.       
The main point of (\ref{12.05}) is that in the inclusive
cross section the Planck mass suppression for the emission of each 
massive graviton is compensated by the large number $(RE)^{\delta}$
of gravitons kinematically accessible. As a consequence, for energies 
$E$ close to $M$, this process could provide an experimental test/signal
of models with a low string scale.  

A full string formula is needed, however, for energies close to the
string scale $M_I$, where string effects are important and the
amplitude (\ref{12.05}) violates unitary. Moreover, as shown in
\cite{grw}, the signal of graviton emission dominates over the
Standard Model background for energies $E \ge (0.5-0.3) M$, so
that the interesting case is $E \ge M_I$, where string effects play an important role in
the experimental signal and therefore cannot be
ignored\footnote{Recently, the mixing between Higgs and massive gravitons was
proposed as a signal with very small string corrections, provided a
term of the form $\xi R H^2$ term exist in the Lagrangian, where $R$ is
the scalar curvature tensor and $H$ is the Higgs scalar \cite{grw2}. The
string computation of this operator involves one bulk and two brane
fields and can be done along the lines of those performed in this paragraph.}.

The string amplitude involves the correlation function of three gauge vertex
operators, of polarisations $\e_i$ and momenta 
$p_i$ ($i=1,2,3$), and of a
massive winding-type graviton of polarisation $\e_4$ and momentum $p_4$
(see Figure 4).
\begin{figure}
\vspace{4 cm}
\special{hscale=60 vscale=60 voffset=0 hoffset=120
psfile=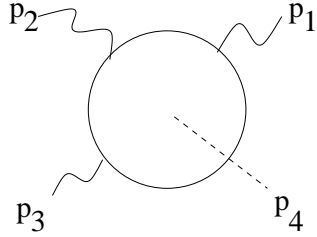}
\caption{The disk amplitude with three open string particles and one
closed string particle.}
\end{figure} 
Defining for convenience the Mandelstam variables
\be
s=-(p_1+p_2)^2,\quad t=-(p_1+p_4)^2,\quad u=-(p_1+p_3)^2 \ , \label{12.9}
\ee
the kinematics of the amplitude is summarized by the equations
\ba
s &=& -2p_1 p_2 = -2p_3 p_4+w^2 \ , \ t=-2p_2 p_3=-2p_1 p_4+w^2 \ , \nonumber\\
u &=& -2p_1 p_3=-2p_2 p_4+w^2 \ , \ s+t+u=w^2 \ . \label{12.10}
\ea
The details of the
computation are given in \cite{dm2}, \cite{cpp}. As shown there, the
final result can be cast in the form
\be
A_4\!=\!{1 \over \sqrt{\pi}} \ 2^{-{w^2 \over M_I^2}} \ 
{ \Gamma\left(-{w^2 / 2M_I^2} \!+\! {1 / 2} \right)
\Gamma\left(-{s / 2M_I^2}\!+\!1 \right)
\Gamma\left(-{t / 2M_I^2}\!+\!1 \right)
\Gamma\left(-{u / 2M_I^2}\!+\!1 \right) \over
\Gamma\left({(s-w^2) / 2M_I^2}\!+\!1 \right)
\Gamma\left({(t-w^2) / 2M_I^2}\!+\!1\right)
\Gamma\left({(u-w^2) / 2M_I^2}\!+\!1\right)} \ A_4^{FT} \ , \label{12.11}  
\ee
where $A_4^{FT}$ is the field-theory amplitude
\cite{grw}. Eq. (\ref{12.11}) obviously reduces to
the field theory result in the limit of low energy ($s,t,u << M_I^2$) and small 
graviton mass ($w^2 << M_I^2$). This string
amplitude has poles for $s,t,u=(2n-2) M_I^2$, with $n$ a
positive integer, corresponding to tree-level open string state
exchanges in the $s$, $t$ and $u$ channels. 
Moreover, there are poles for graviton
winding masses $w^2=(2n-1)M_I^2$, to be interpreted as tree-level mixings between the
graviton and the gauge singlets present at odd levels in the open string spectrum,
which then couple to the gauge fields. The amplitude $A_4$ has also
zeroes for very heavy gravitons $w^2=s+2n M_I^2$, or similar conditions
obtained by the replacements 
$s \rightarrow t,u$. These give interesting selection rules and 
display a typical behavior, not shared by any other field theory process
with missing energy. 

An important question raised by (\ref{12.11}) concerns the string deviations
from the field theory result $A_4^{FT}$. In the s-channel, the energy
corresponding to the first string resonance is
$s=2M_I^2$, and similarly for $t$ and $u$. This supports the natural
expectation that field theory 
breaks down for energies above $M_I$. For energies well below
this value ($s,t,u,w^2 << M_I^2$), it is easy to obtain the corrections to 
the field-theory computation from a power-series expansion 
of (\ref{12.11}).
The first corrections are of the form
\be
A_4 = (1+ {\zeta(2) \over 4} {w^4 \over M_I^4}
+{\zeta(3)\over 4}{stu+w^6\over M_I^6}+
\cdots )  A_4^{FT} \ \label{12.12} 
\ee
and, after T-duality become
\be
A_4 = (1+ {\pi^2 \over 24} {m^4 \over (R_{\perp}M_I)^4}+ \cdots )  A_4^{FT} 
\ . \label{12.13}  
\ee 
Notice that the first correction to the amplitude with 
a massless graviton (of fixed energy) is of order\footnote{This is
probably related to the underlying ${\cal N}=4$ supersymmetry of 
this toroidal compactification.} $E^6/M_I^6$,
and therefore the deviation from the field theoretical result is first expected to
arise from massive gravitons.

A more useful way to define deviations from field
theory is the integrated cross-section $\sigma$, obtained summing
over all graviton masses, up to the available energy $E$  
\be
\sigma = \sum_{m_1 \cdots m_6=0}^{R_{\perp}E} |A_4|^2 \quad , \quad 
\sigma^{FT} = \sum_{m_1 \cdots m_6=0}^{R_{\perp}E} |A_4^{FT}|^2
\ , \label{12.15}
\ee 
where $\sigma^{FT}$ is the corresponding
field theory value. Surprisingly enough, terms of order $E^2$ are absent
in (\ref{12.15}) and therefore at low energies
the string corrections are smaller than expected, of order 
\be
{\sigma-\sigma^{FT} \over \sigma^{FT}} \sim {E^4 \over M_I^4} \ . \label{12.16}
\ee
However, as mentioned above, strong deviations emerge for
$E^2 \sim 2 M_I^2$, where the first string resonance appears and the field
theory approach breaks down.
  
Another interesting quantity is the form-factor 
for two gauge bosons and one winding (KK mode ${\bf m}$ after
T-dualities) graviton, or, in a more phenomenological language, of the
bulk/brane/brane couplings,
which were already used in previous sections to discuss neutrino (and
axion) masses.  A direct on-shell computation can easily be done
\cite{hk}, and the result turns out to be the same in the bosonic string and in
the superstring \cite{dm2}. A partly off-shell expression for the form factor 
can however be obtained factorizing the three gauge bosons -- one
massive graviton amplitude.
Indeed, using (\ref{12.11}) and the two gauge bosons -- one 
massive graviton
amplitude \cite{hk,dm2}, it is possible to find the form factor in
the case where one of the  gauge bosons and the graviton are
off-shell\footnote{The expression (\ref{12.17}) corrects a factor 2 misprint
in the eqs. (1.9) and (4.16) of \cite{dm2}.} \cite{dm2}
\be
g(p_1,p_2,p) = {1 \over M_P \sqrt{\pi}} \ 2^{p^2 \over M_I^2}{\Gamma
(p^2/2M_I^2+1/2) \over
\Gamma (p_1p_2/M_I^2+1)} \ . \label{12.17}
\ee
Notice that, at energies much smaller than the string scale ($p^2,p_1p_2
<< M_I$), this form factor is close to $1/M_P$ for all winding
states, a result that was used in the field theory approach to
the brane/brane/bulk couplings in Section 11. 

{}From (\ref{12.17}) we can deduce a form factor characterizing the
emission of a heavy graviton ($-p^2 >> M_I^2$)
\be
g(p^2) \sim 2 \sqrt{2M_I^2 \over \pi p^2} 
(\tan{\pi p^2 \over M_I^2}) e^{{p^2 \over M_I^2} \ln 2} \ , \label{12.18}            
\ee
where for an on-shell graviton $p^2$ is equal to the KK graviton 
mass $-p^2=m^2/R_{\perp}^2$. It is transparent from this result that we
qualitatively recover, aside from an 
oscillatory factor accounting
for the string resonances, the field-theory exponential form factor 
(\ref{12.04}), but only for {\it on-shell} particles.
Indeed, the off-shell result (\ref{12.17}) contains, as expected, the 
exponential damping of string amplitudes at high-energy \cite{cpp} in
the fixed angle limit, irrespective of the graviton KK mass. 
\vskip 2mm
- {\large \bf Virtual exchange of string and gravitational-type states}
\vskip 2mm

One of the main possible experimental signatures for String Theory is the
tree-level exchange of string oscillators (Regge particles), encoded in
the four-particle Veneziano amplitude
\be
A (s,t) = { \Gamma (1-s/M_I^2) \Gamma (1-t/M_I^2) \over \Gamma (1-s/M_I^2 -
t/M_I^2) } \ , \label{12.010}
\ee
and in similar expressions $A(t,u)$, $A(u,s)$, that have poles
corresponding to massive open string states. They manifest themselves as
deviations from field theory amplitudes for energies close to the string
scale $M_I$. Moreover, they produce an exponential damping $\exp
(-s/M_I^2)$ of the amplitudes at high energy $s >>M_I^2$, for scatterings
at fixed angle.
  
The corresponding one-loop diagrams have a dual interpretation
as tree-level virtual exchanges of gravitational-type particles. In a field-theory
approach, these contributions have the problem that for a
number of compact dimensions
$d \ge 2$ the corresponding KK field theory summations diverge in the
ultraviolet (UV), and therefore the field-theory computation is unreliable.
Indeed, let us consider a 
four-fermion interaction of particles confined to a D3 brane mediated by
KK gravitational excitations orthogonal to it. The 
amplitude for the process reads
\be
A \ = \ {1 \over M_P
^2} \sum_{m_i} {1 \over -s +
{m_1^2 + \cdots m_{\d}^2 \over R_{\perp}^2}} \ , \label{12.1}
\ee
where for simplicity we considered equal radii denoted by $R_{\perp}$ 
and $s=-(p_1+p_2)^2$ is the squared center of mass energy\footnote{With our conventions
$s$ is negative in Euclidean space.}.
The summation clearly diverges for $\d \ge 2$. 
In this cas, the traditional attitude
is to cut the sums for masses heavier than a cutoff
$\Lambda >> R_{\perp}^{-1}$, of the order of the fundamental scale 
$M_I$ in string theory \cite{grw}. This can be implemented in a
proper-time representation of the amplitude
\be
A \ = \ {1 \over M_P^2} \sum_{m_i} \int_{1 / \Lambda^2}^{\infty} dl \ e^{-l
(-s + {m_1^2 + \cdots m_{\d }^2 \over R_{\perp}^2})}
\ = \  {1 \over M_P^2} \int_{1 / \Lambda^2}^{\infty} dl \ e^{sl}
\ \theta_3^{\d } (0,{il \over \pi R_{\perp}^2}) \ , \label{12.2}
\ee 
where $\theta_3 (0,\tau)=\sum_k exp(i \pi k^2 \tau)$ is one of the Jacobi
functions. We shall be interested in the following in the region of
parameter space $-R_{\perp}^2 s >>1$, $R_{\perp} \Lambda >>1$ and  
$-s << \Lambda^2$, in which the available energy is smaller than (but not far
from) the UV cutoff $\Lambda$, but is much bigger than the (inverse)
compact radius $R_{\perp}^{-1}$, of submilimeter size. In this case, the
amplitude can be evaluated and is
\be
A \ = \ {\pi^{\d} R_{\perp}^{\d} \over M_P^2} \int_{1 /
\Lambda^2}^{\infty} {dl \over l^{\d \over 2}} \ e^{sl} \ 
\theta_3^{\d } (0,{i \pi R_{\perp}^2 \over l}) \ \simeq \ 
{2 \pi^{\d \over 2} \over \d-2} {R_{\perp}^{\d} \Lambda^{\d-2} \over
M_P^2} \ = \ {4 \pi^{\d \over 2} \over \d-2} \alpha_{YM}^2 {\Lambda^{\d-2} \over
M_I^{\d+2}} \ , \label{12.3}
\ee
where in the last step we used the relation
$M_P^2=(2/\alpha_{YM}^2)R_{\perp}^{\d} M_I^{2+\d}$, valid for Type I strings,
where $\alpha_G=g_{YM}^2/(4 \pi)$ and $g_{YM}$ is the Yang-Mills
coupling on our brane. The cutoff $\Lambda$ is equivalent
to computing the field-theory diagram using a form factor of the type
(\ref{12.04}) with $\Lambda = M_I / \sqrt{a}$. As shown at the beginning
of this section, however, (\ref{12.04}) is an on-shell form factor in
string theory, whereas virtual particle exchanges ask for an off-shell
form factor. The off-shell form factor (\ref{12.17}), on the other hand,
depends only on the momentum of the massive  gravitons and {\it not} on its
mass. This therefore raises doubts on the way string theory regulates
the divergent sum (\ref{12.1}).
In addition, the high sensitivity of the result (\ref{12.3}) to the
cutoff $\Lambda$ asks 
for a more precise computation in a full string context. As explained
below, string theory does not cut the
divergent sum (\ref{12.05}). The solution to the puzzle is that the
divergent sum is not an UV divergence in string theory, but  
an IR divergence, which has the same status in string and in field 
theory, and asks for resummation of graphs with soft particle
emissions\footnote{Another interesting, related example of this type of
divergence, arises in the one-loop
effective action of toroidal compactifications \cite{bk}. Indeed,
there are $F^4$ terms on the Type I and on the heterotic $SO(32)$ side that match using
the Type I-heterotic duality relations (\ref{1.4}). However, the
coefficient of the corresponding terms is proportional to $\sum_{\bf m \not=0}
(1/{\bf m}^2)$, where ${\bf m}=(m_1 \cdots m_6)$. This sum is divergent
due to the contribution of very heavy KK states, 
as the amplitude (\ref{12.1}). The Type I-heterotic duality check
performed in \cite{bk} then holds with an appropriate identification of the IR cutoffs  
on the two sides.}.

The relevant string diagram is actually one loop and is a priori subdominant with
respect to tree-level Veneziano amplitudes. However, deviations from Newton's
law come precisely from this one-loop diagram, and therefore a precise
evaluation is necessary.

The computation reviewed below was done for the $SO(32)$ Type I 10D 
superstring compactified to 4D on a six-dimensional torus. However,
we will argue later that the result holds for a large class of
orbifolds, including ${\cal N}=2$ and ${\cal N}=1$ 
supersymmetric vacua.
The Type I string diagram that in the low-energy limit contains the 
gravitational exchange mentioned above is the nonplanar cylinder diagram,
in which for simplicity we prefer
to insert gauge bosons rather than fermions in the external lines. This
diagram has two dual interpretations \cite{dua} 
a) tree-level exchange of closed-string
states, if time is chosen to run horizontally (see Fig. 5)
b) one-loop diagram of open strings, if time is chosed to run vertically 
(see Fig. 6). 
\begin{figure}
\vspace{4 cm}
\special{hscale=60 vscale=60 voffset=0 hoffset=100
psfile=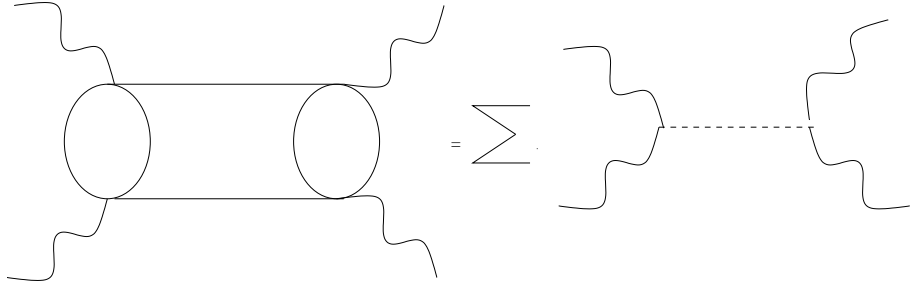}
\caption{The closed channel representation of the amplitude.}
\end{figure}
\begin{figure}
\vspace{4 cm}
\special{hscale=60 vscale=60 voffset=-5 hoffset=100
psfile=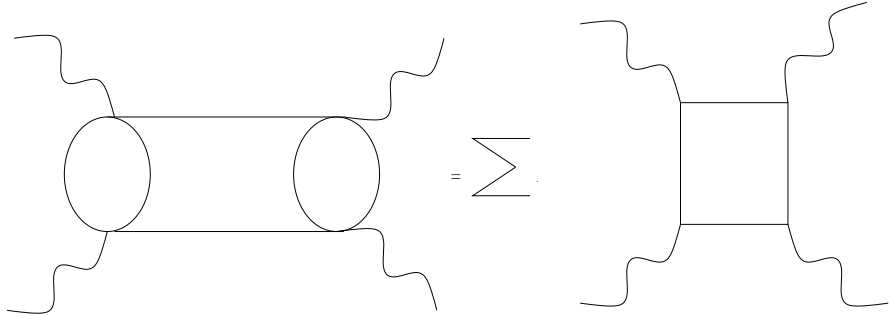}
\caption{The open channel representation of the amplitude.}
\end{figure}
In the two dual representations, the
nonplanar amplitude reads symbolically
\ba
A &=& \sum_n \int_0^{\infty} \ dl \sum_{n_i} \ A_2 (l,n_1 \cdots n_d,n)
\nonumber \\
&=&   \sum_{k_1 \cdots k_4} \int_0^{\infty} \ d\tau_2 \tau_2^{d/2-2} 
\sum_{m_i} \ A_1 (\tau_2,m_1 \cdots m_d,k_1 \cdots k_4) \ , \label{12.4} 
\ea
where $l$ denotes the cylinder parameter in the tree-level channel and
$\tau_2=1/l$ is the one-loop open string parameter\footnote{$\tau_2$ was called
 $t$ in previous chapters. In this paragraph, however, we reserve the
symbol $t$ for one of the Mandelstam variables.}. In the first
representation, the amplitude is interpreted as tree-level exchange of
closed-string particles of mass $(n_1^2+ \cdots n_d^2) R^2M_I^4 + n
M_I^2$, where $n_1 \cdots n_d$ are winding quantum numbers and $n$ is
the string oscillator number. In particular the $n=0$ term reproduces
the field-theory result (\ref{12.1}), and therefore the full expression
(\ref{12.4}) is its string regularization. In the second representation, the
amplitude is interpreted as a sum of box diagrams with particles of
masses $(m_1^2+\cdots m_d^2)/R^2+k_i M_I^2$ ($i=1 \cdots 4$) in 
its four propagators.

The UV limit ($l \rightarrow 0$) of the gravitational 
tree-level diagram is related to the IR limit ($\tau_2
\rightarrow \infty$) of the box diagram.
In particular, in four dimensions, when an IR regulator
$\mu$ is introduced in
the box diagram, the divergence in the Kaluza-Klein summation in the 
gravitational-exchange diagram disappears. 
The final result for the
nonplanar cylinder amplitude in the low energy limit $E/M_I <<1$ ($E$ is
a typical energy scale) in four-dimensions is \cite{dm2}
\ba
A &=& -{1 \over \pi M_P^2 s} + {2 g_{YM}^4 \over \pi^2 } \ [ \ {1 \over st} 
\ln {-s \over
4 \mu^2} \ln {-t \over 4 \mu^2}+ {\rm perms.}]  \nonumber \\
&-& {g_{YM}^4 \over 3 M_I^4} 
\ [ \ \ln {s \over t}  \ln {s t \over \mu^4} + \ln {s \over u} \ln {s u
\over \mu^4} ]+ \cdots \ , \label{12.5}
\ea
where ``perms.'' denotes two additional contributions coming from
permutations of $s,t$, $u$ and ``$\cdots$'' denote terms of higher order
in the low energy expansion. Notice the absence in (\ref{12.5}) of the
contact term (\ref{12.3}), that in the string result is replaced by the
leading string correction, given by the second line in (\ref{12.5}).
The string correction in (\ref{12.5}) is indeed of the same order of
magnitude as (\ref{12.3}) for $\Lambda \sim M_I$, but has an
explicit energy dependence coming from the logarithmic terms.
 
In order to find the appropriate interpretation of (\ref{12.5}) in terms
of field-theory diagrams, it is  convenient to separate the integration region
in (\ref{12.4}) into two parts, introducing an arbitrary parameter $l_0$ and writing
\be
A = \sum_n \int_{l_0}^{\infty} \ dl \sum_{n_i} \ A_2 \ +
 \sum_{k_1 \cdots k_4} \int_{1/l_0}^{\infty} \ d\tau_2 \tau_2^{d/2-2} 
\sum_{m_i} \ A_1 \ . \label{12.6}
\ee  
This has the effect of fixing an UV cutoff $\Lambda=M_I/\sqrt{l_0}$ in the tree-level
exchange diagram, similar to the
one introduced in (\ref{12.2}) and (\ref{12.3}),
as well as a related UV cutoff $\Lambda'=M_I \sqrt{l_0}=M_I^2/\Lambda$ in
the one-loop box diagram described here by $A_1$. 
Computing the low-energy limit of $A_1$ and $A_2$, in 4d we find \cite{dm2}
\ba  
A_1 &=& {2 g_{YM}^4 \over \pi^2} \ [ \ {1 \over st} \ln {-s \over
4 \mu^2} \ln {-t \over 4 \mu^2}+ {\rm perms.}] - 
{g_{YM}^4 \over 3 M_I^4} 
\ [ \ \ln {s \over t}  \ln {s t \over \mu^4} + \ln {s \over u} \ln {s u
\over \mu^4} + {6 \over l_0^2}] + \cdots \nonumber \\
A_2 &=&  -{1 \over \pi M_P^2 s}+ {2g_{YM}^4 \over M_I^4}
\ [ \ {1 \over l_0^2} + \cdots + O({s^2 \over M_I^4})+ \cdots] \  . \label{12.7}
\ea
The $g_{YM}^4$ term in $A_1$ describes a box diagram with four light
particles (of mass $\mu$) circulating in the loop, while the
$g_{YM}^4/M_I^4$ term is the first string correction coming from box
diagrams with one massive particle (of mass $M_I$) and three light
particles of mass $\mu$ in the loop. It also contains the $l_0$ dependent
part of the box diagram with four light particles in the loop. 
The $1/M_I^4l_0^2$ term in $A_2$ can be written as $\Lambda^4 /
M_I^8$ and therefore reproduces the field theory result (\ref{12.3})
in the case of six compact dimensions. However, as expected, a similar 
term with an opposite
sign appears in $A_1$, while the $l_0$ dependent terms cancel. 
In $A_2$ the  $O({s^2 / M_I^4})$ term is $l_0$ independent,
and is actually the first correction to the tree-level graviton exchange.
We emphasize, however, that the only physically meaningful 
amplitude is the full expression (\ref{12.5}). The leading
string correction is therefore the second line of (\ref{12.5}), coming
from box diagrams $A_1$ involving one massive particle in the loop.
 
Strictly speaking, this result is valid for toroidal
compactifications of the $SO(32)$ 10D Type I string. For a general ${\cal
N}=1$ supersymmetric 4D Type I vacuum the amplitude $A$ has
contributions from sectors with various numbers of supersymmetries
\be
A= A^{{\cal N}=4}+ A^{{\cal N}=2}+ A^{{\cal N}=1} \ , \label{12.8}
\ee
where the ${\cal N}=4$ sector contains the six-dimensional compact KK
summations, the ${\cal N}=2$ sectors contain two-dimensional compact KK
summations and the ${\cal N}=1$ sectors contain no KK summations. From the
tree-level ($A_2$) viewpoint, the ${\cal N}=2$ sectors give logarithmic
divergences that in the one-loop box ($A_1$) picture
correspond to additional
infrared divergences associated to wave-functions or vertex corrections,
absent (by nonrenormalization theorems) in the ${\cal N}=4$ theory. 
Similarly, the ${\cal N}=1$ sectors give no KK divergences. As the important 
(power-like) divergences come from the gravitational ${\cal N}=4$
sector, the toroidally compactified Type I superstring
contains the relevant information for our purposes.
Moreover, even if we confine our attention to the Type I
superstring, the formalism can be easily adapted to Type II
strings and to their D-branes. This can be done exchanging some of the
Neumann boundary conditions in the compactified Type I string with the
Dirichlet ones appropriate for the D-branes \cite{hk,myers}. As can be
easily seen, the basic results and conclusions of this Section are unchanged.  

An interesting observation was made recently concerning models where
some of the Standard Model fermions live on
a brane, while others live on a distant brane. In this case, all
amplitudes for processes involving fermions on the two
branes have an exponential damping factor depending on the distance
between the branes \cite{ags}, that would produce spectacular
effects in accelerator experiments.  
  
\section{Conclusions}

The last years had a dramatic effect on our understanding of string
physics and of its possible implications for low energy physics. In particular,
there is a real hope to experimentally test scenarios with a low string
scale, large compactification (TeV) radii and possibly (sub)millimeter
gravitational dimensions. Some of the relevant issues (gauge coupling
unification, supersymmetry breaking, gauge hierarchy) were already
analyzed at the string level using quasirealistic string models, while other
issues (flavor physics, for example) were mainly studied at the field theory
level, so that more detailed string studies would be very useful. 
This review does not cover cosmological issues (see, for example, 
\cite{cosmology} and references therein) and the
recent scenarios related to warped compactifications \cite{rs} (for the
role of warped compactifications in strings, see for example \cite{dm4,warped}). In particular,
the last scenarios provide the first phenomenological models of Anti-de-Sitter
compactifications, which led to the famous AdS/CFT conjecture
\cite{malda} with interesting, nonperturbative results, for the gauge
theory dynamics. 
 
 It is however important to keep in mind that, despite the beautiful new
ideas dealing with large (or infinite) extra dimensions which appeared
recently, the good old picture of the ``desert'' between the weak scale
and a large (of the order of $10^{16}$ GeV) unification scale is still
a viable possibility. Since String Theory at the present time offers no compelling reason in
favor of any of the new scenarios, only new experimental results can provide a hint 
for the real value of the string scale or, more generally, for the
correct picture of physics beyond the Standard Model. 

\vskip 16pt
\begin{flushleft}
{\large \bf Acknowledgments}
\end{flushleft}

\noindent I am grateful to  C. Angelantonj, I. Antoniadis,
C. Bachas, P. Bin{\'e}truy, G. D'Appollonio, C. Deffayet, K.R. Dienes,
T. Gherghetta, C. Grojean, J. Mourad, S. Pokorski, P. Ramond, A. Riotto,
A. Sagnotti and C.A. Savoy for enjoyable collaborations and illuminating 
discussions over the last
years and to L.E. Ib{\'a}{\~n}ez, C. Kounnas, M. Perelstein, M. Peskin and
G. Veneziano for helpful discussions and comments. Special thanks are 
due to Augusto Sagnotti for a a detailed reading of the manuscript and
many suggestions which improved substantially the content of this review. 
\appendix

\section{Jacobi functions, lattice sums and their properties}

For the reader's convenience, in this Appendix we collect the
definitions, transformation properties and some identities for the
modular functions used in the text (for more formulae and
properties of modular functions, see for example \cite{kiritsis}). The 
Dedekind function is defined by the usual product formula (with $q=e^{2\pi i\tau}$)
\be
\eta(\tau) = q^{1\over 24} \prod_{n=1}^\infty (1-q^n)\ , \label{a1}
\ee
whereas the Jacobi $\vartheta$-functions with general characteristic and
arguments  are
\be
\vartheta [{\a \atop \b }] (z,\tau) = \sum_{n\in Z}
e^{i\pi\tau(n-\a)^2} e^{2\pi i(z- \b)(n-\a)} \ . \label{a2}
\ee
The corresponding product representation of the Jacobi functions is
\be
\vartheta [{\a \atop \b }] (z,\tau) = e^{2 \pi i \a (\b-z)} q^{{\a^2
\over 2} } \prod_{n=1}^\infty
  (1-q^n) [1 + q^{n-\a -{1 \over 2}} e^{2\pi i (z-\b) }]
[ 1+ q^{n + \a -{1 \over 2}} e^{-2\pi i (z-\b)}] \ . \label{a02}
\ee
For completeness, we give also the product formulae for the four special 
$\vartheta$-functions with half-integer characteristics
\ba 
\vartheta_1(z,\tau) &\equiv & \vartheta \left[{{1\over 2} \atop {1\over
2} } \right] (z,\tau) = 2q^{1/8}{\rm sin}\pi z\prod_{n=1}^\infty
  (1-q^n)(1-q^ne^{2\pi i z})(1-q^ne^{-2\pi i z}) \ , \nonumber \\
\vartheta_2(z,\tau) &\equiv & \vartheta \left[{{1\over 2} \atop 0 }\right]
  (z,\tau) = 2q^{1/8}{\rm cos}\pi z\prod_{n=1}^\infty
  (1-q^n)(1+q^ne^{2\pi i z})(1+q^ne^{-2\pi i z}) \ , \nonumber \\
\vartheta_3(z,\tau) &\equiv & \vartheta \left[{0 \atop 0 }\right]
  (z,\tau) = \prod_{n=1}^\infty
  (1-q^n)(1+q^{n-1/2}e^{2\pi i z})(1+q^{n-1/2}e^{-2\pi i z}) \ ,
\nonumber \\
\vartheta_4(z,\tau) &\equiv & \vartheta \left[{0 \atop {1\over 2} }\right]
(z,\tau) = \prod_{n=1}^\infty
(1-q^n)(1-q^{n-1/2}e^{2\pi i z})(1-q^{n-1/2}e^{-2\pi i z}) \ . \label{a3}
\ea
The modular properties of these functions are described by
\be
\eta(\tau+1) = e^{i\pi/12}\eta(\tau)\ \ , \ \
\vartheta \left[{\a \atop {\b}}\right] \left({z} ,
  {\tau+1}\right)=
e^{-i\pi\a(\a-1)}\vartheta 
\left[{\a \atop {\a+\b-{1\over 2}}}\right] \left({z} \ ,
  {\tau}\right) \nonumber 
\ee
\be
\eta(-1/\tau) = \sqrt{-i\tau}\; \eta(\tau)\ \ , \ \ 
\vartheta \left[{\a \atop {\b}}\right] \left({z \over \tau}, {-1 \over \tau}\right)=
\sqrt{-i \tau} \ e^{2 i \pi \a \b +{i \pi z^2 / \tau}} \ 
\vartheta \left[{{\b} \atop - \a}\right] (z, \tau ) \ . \label{a4}
\ee

The relevant Kaluza-Klein and winding lattice summations appearing in the text are
\be
\sum_m P_{m+a} (\tau) \equiv \sum_m q^{\pi \alpha' (m+a)^2 \over R^2}
\quad , \quad
\sum_n W_{n+b} (\tau) \equiv \sum_n q^{{\pi \over 4 \alpha'} {(n+b)^2
R^2}} \ , \label{a5}
\ee
with $q=e^{2 \pi i \tau}$, where in the summations relevant to the D9 
branes, for example,
$\tau=it/2$ in $P_{m+a}$ and $\tau=il$ in $W_{n+b}$. A Poisson formula
used frequently in order to pass from the one loop open-string channel
to the tree-level closed string channel is
\be
\sum_m e^{- \pi \alpha' t {(m+a)^2 \over R^2}} = ({R^2 l \over 2
\alpha'})^{1 \over 2} \sum_n  e^{2\pi i a n} e^{- {\pi l \over 2
\alpha'} n^2 R^2} \ , \label{a6}
\ee
where $t/2=1/l$. 
\section{Glossary}

- {\bf Chan-Paton factors}: quantum numbers which sit at the end of
  open strings, which give rise to the gauge group and the charged matter
  quantum numbers.  

- {\bf critical dimension}: spacetime dimension (10 for superstrings )
  in which the 2d Weyl anomaly cancels and the world-sheet theory has
  a background which is Poincar\'e invariant.

- {\bf Dp-brane}: dynamical surface which spans $p+1$ spacetime
  dimensions, containing  gauge group and charged
  matter, on which open strings (with Dirichlet boundary conditions) can
  end. D-branes are embedded in an underlying space of dimension ten for
  superstrings. 
  
- {\bf Green-Schwarz mechanism}: gauge and gravitational anomaly cancellation
  in 10d due to the nonlinear gauge transformation of the antisymmetric
  tensor field.

- {\bf GSO projection}: projection of physical states which enforces
  modular invariance.

- {\bf modular invariance}: invariance of the one-loop partition function of
  closed strings under global reparametrizations of the torus.

- {\bf no-scale model}: supergravity models with (tree-level) zero
  vacuum energy and broken supersymmetry, having flat directions in the
  scalar potential, along which the size of supersymmetry breaking is
  classically undetermined.

- {\bf orbifolds}: compact spaces of the type $M/G$, where $G$ is a discrete
  group, on which string propagation can be exactly solved. 
  Supersymmetry is generically partly or completely broken, such that
  the method can generate chiral models in four-dimensions.
   
- {\bf orientifolds}: String models constructed by gauging world-sheet 
  symmetries $H$ (for example the world-sheet parity $\Omega$), 
  such that physical states are $H$ invariant. The topological
  expansion in this case involves non-orientable surfaces (e.g. Klein bottle
  or M\"obius strip).

- {\bf orientifold (O) plane}: fixed (non-dynamical) surface under 
  the orientifold projection, carrying Ramond-Ramond charges.

- {\bf partition function}: the vacuum-energy at a given order in the 
  topological expansion.

- {\bf Ramond-Ramond fields}: Antisymmetric tensor fields of different
  rank present in the closed string spectrum, which couple to
  orientifold planes and D-branes.
   
- {\bf Scherk-Schwarz mechanism}: breaking of supersymmetry due to
  boundary conditions in the compact space, different for bosons and fermions.

- {\bf vertex operators}: operators constructed out of world-sheet degrees
  of freedom, used in constructing string correlations functions
  for external on-shell particles.

- {\bf Wilson line}: gauge field in compact space with vanishing
  field strength. The wave function of charged states acquires a
  phase after a closed loop in the compact direction. The corresponding
  states become massive and break the gauge group to a subgroup.
   
- {\bf winding state}: massive state coming from closed strings wrapping
  a compact coordinate. A string wrapping $n$ times a circle of radius $R$
  gives a mass $n \ R M_s^2$.  


\end{document}